\UseRawInputEncoding

\documentclass[a4paper,aps,prl,amsmath,amssymb,floatfix,preprint,citeautoscript,noeprint,superscriptaddress,12pt]{revtex4-2}

\usepackage{dsfont}
\usepackage{lipsum} 
\usepackage{bibentry}
\usepackage[english]{babel}
\usepackage[dvipsnames]{xcolor}
\usepackage{graphicx}
\usepackage[caption=false]{subfig} 
\usepackage{amsmath,amssymb,bm}
\usepackage[version=3]{mhchem}
\usepackage{verbatim}
\usepackage{multirow}
\usepackage{dcolumn}
\usepackage{float}
\usepackage{nicefrac}
\usepackage{siunitx}
\DeclareSIUnit{\kjmol}{\kilo\joule\per\mole}
\DeclareSIUnit{\calorie}{cal}
\DeclareSIUnit{\kcalmol}{\kilo\calorie\per\mole}
\usepackage{booktabs}
\usepackage{chemformula}
\usepackage{wrapfig}
\usepackage{enumitem}  
\usepackage{transparent}
\usepackage[colorlinks,allcolors=black,citecolor=blue,urlcolor=blue]{hyperref}
\usepackage{tcolorbox}
\usepackage{relsize}
\hypersetup{
    citecolor=purple,
    colorlinks=true,
    linkcolor=black,    
    urlcolor=purple,
    }
\usepackage{etoolbox}

\newcommand{\etal}{\emph{et al.}}

\usepackage{sansmathfonts}
\usepackage[justification=centerlast, font={sf}, labelfont={bf}]{caption}
\makeatletter
\def\@fnsymbol#1{%
  \ifcase#1\or *\or *\or *\or *\or *\else\@ctrerr\fi}
\makeatother


\begin{document}

\author{Benjamin X. Shi}
\email{mail@benjaminshi.com}
\affiliation{Initiative for Computational Catalysis, Flatiron Institute, New York, NY 10010, USA}

\author{Timothy C. Berkelbach}
\email{t.berkelbach@columbia.edu}
\affiliation{Initiative for Computational Catalysis, Flatiron Institute, New York, NY 10010, USA}
\affiliation{Department of Chemistry, Columbia University, New York, NY 10027, USA}%

\title{Practical and accurate density functionals for transition-metal heterogeneous catalysis}
\date{\today}

\begin{abstract}
Density functional theory (DFT) underpins modern atomistic simulations of transition-metal surfaces.
It can predict key properties linked to catalytic performance, such as adsorption energies and barrier heights, enabling new paradigms in rational catalyst design.
These applications require reliable density functionals, however achieving transition-metal chemical accuracy (\SI{13}{\kjmol}) on these properties remains challenging.
We introduce a framework for designing new functionals tailored to catalytic processes on transition-metal surfaces, building on recent non-self-consistent approaches.
Within this framework, we develop a hybrid and a double-hybrid functional that achieve unprecedented accuracy, with the latter reaching transition-metal chemical accuracy on average across 39 experimental adsorption reactions.
In addition, both functionals demonstrate balanced performance for 17 barrier heights and correct qualitative failures of standard functionals, including CO adsorption on Pt(111) and graphene on Ni(111).
They are computationally efficient, readily integrated into existing DFT codes, and supported by open-source workflows to facilitate adoption.
More broadly, this framework provides a systematic route towards improved functionals for heterogeneous catalysis and complex materials.
\end{abstract}

\maketitle

\section{Introduction}
Our atomic-level understanding of catalytic reactions on transition-metal surfaces has advanced in recent years due to density functional theory (DFT)~\cite{norskovComputationalDesignSolid2009c}.
Complex catalytic processes can be studied through kinetic models~\cite{xuFormationActiveSites2023} derived from barrier heights~\cite{borodinQuantumEffectsThermal2022} and adsorption energies computed from DFT~\cite{jiangPredictiveModelDiscovery2025,lorenzuttiMicroenvironmentEffectsElectrochemical2025}.
More generally, DFT has been used to reveal fundamental scaling relationships between properties such as the adsorption energy on a surface and its catalytic activity or selectivity~\cite{michaelidesIdentificationGeneralLinear2003a,bligaardBronstedEvansPolanyi2004a,wangBreakingLinearScaling2025}.
With this framework, the complexity of heterogeneous catalysis can be simplified into easy-to-compute quantities, opening avenues for rational design~\cite{greeleyAlloyCatalystsDesigned2004a,studtDiscoveryNiGaCatalyst2014a,zhaoTheoryguidedDesignCatalytic2019a}.
For such applications, the effectiveness of DFT lies in an accurate determination of these central quantities~\cite{calle-vallejoFindingOptimalSurface2015,hannaganFirstprinciplesDesignSingleatomalloy2021a},
and inaccurate prediction~\cite{dilibertoUniversalPrinciplesRational2022,xuRevisitingUniversalPrinciple2024a} can readily mislabel a promising catalyst as an inert one or vice versa.

The practical application of DFT requires density functional approximations~\cite{perdewJacobLadderDensity2001b} (DFAs), which span increasing sophistication from the local density approximation to generalized-gradient approximations (GGAs), meta-GGAs, hybrids, and double hybrids~\cite{goerigkLookDensityFunctional2017d}.
Although semilocal DFAs such as GGAs and meta-GGAs are the workhorse of computational materials science, they exhibit systematic failures for adsorption on transition metal surfaces~\cite{schimkaAccurateSurfaceAdsorption2010}.
For example, achieving errors below \SI{20}{\kjmol} across broad transition-metal adsorption energy datasets has proved difficult~\cite{mallikarjunsharadaAdsorptionTransitionMetal2019a,goltlComparingPerformanceDensity2020a}, despite the fact that errors below \SI{4}{\kjmol} are attainable for insulating surfaces~\cite{r.rehakIncludingDispersionDensity2020a,shiAccurateEfficientFramework2025b}.
Beyond quantitative errors, semilocal DFAs also fail to qualitatively describe the adsorption mechanism of molecules, most famously the CO adsorption puzzle~\cite{feibelmanCOPt111Puzzle2001}:
across important transition-metal surfaces, semilocal DFAs consistently predict the wrong adsorption site for CO, incorrectly favoring the hollow (FCC) site over the (on-)top site.

The next level above semilocal DFAs are hybrid DFAs,
which incorporate a fraction of exact exchange (EXX), and while their significantly higher cost is justified by clear improvements for insulators~\cite{borlidoExchangecorrelationFunctionalsBand2020a}, the benefits for transition-metal surfaces are not clear.
For example, the popular HSE06 DFA, a screened (range-separated) hybrid, has been applied to several transition-metal surface databases of experimental values, where its accuracy for both adsorption energies~\cite{mallikarjunsharadaAdsorptionTransitionMetal2019a} and barrier heights~\cite{mallikarjunsharadaSBH10BenchmarkDatabase2017} is found to be worse than semilocal DFAs.
Similarly, HSE06 and other common hybrid DFAs such as PBE0 and B3LYP show mixed success for the CO adsorption puzzle: while they perform well for some transition-metal surfaces~\cite{huExactTreatmentExchange2007d} [Cu(111) and Rh(111)], the canonical case of Pt(111) remains a persistent challenge~\cite{stroppaShortcomingsSemilocalHybrid2008}.
By contrast, embedded cluster studies~\cite{araujoAdsorptionEnergiesTransition2022d} have shown that the hybrid M06 can be applied to small clusters to locally correct PBE+D3 estimates from periodic slabs, reporting some of the most accurate predictions to date.
However, the transferability of these approaches to global properties~\cite{janthonBulkPropertiesTransition2014} (e.g., surface energies, monolayer adsorption, or liquid/metal interfaces~\cite{ogasawaraStructureBondingWater2002a}) remains unclear.
More fundamentally, the direct use of unscreened hybrid functionals to periodic metal slabs is questionable, due to their incorrect description of the band structure and their convergence challenges in the self-consistent field (SCF) cycle~\cite{mihmShortcutThermodynamicLimit2021c,yuOptimizationRandomPhase2026}.

Double hybrids occupy the top of the DFA hierarchy and incorporate a fraction of correlation energy from many-body methods, typically second-order M{\o}ller-Plesset perturbation theory (MP2).
The application of MP2-based double hybrids to periodic transition-metal surfaces has been limited by their substantially increased cost and the formal divergence of the MP2 correlation energy for metallic systems~\cite{kellerRegularizedSecondorderCorrelation2022,bystromSizeConsistentAdiabaticConnection2026}.
Such limitations can be sidestepped again via embedded cluster methods~\cite{libischEmbeddedCorrelatedWavefunction2014c,sauerInitioCalculationsMolecule2019b}, which have been used to apply MP2~\cite{huExactTreatmentExchange2007d} and the XYG3 double-hybrid DFA~\cite{chenAccurateDescriptionsMoleculesurface2023}.
Similar approaches~\cite{zhaoRevisitingUnderstandingElectrochemical2021a,caoQuantumManybodySimulations2025a,fantaResolutionSelectivitySteps2025} have also been pursued with other high-level methods besides MP2, but they come at even higher costs, which limit them to model systems~\cite{carboneCOAdsorptionPt1112024}.
The most widely used many-body method is the random-phase approximation (RPA) ~\cite{olsenRandomPhaseApproximation2013,oudotReactionBarriersMetal2024}.
However, while successful for selected systems~\cite{garridotorresAdsorptionEnergiesBenzene2017,sheldonAdsorptionCH4Pt1112021,weiIntroducingEmbeddedRandom2023a}, including the CO adsorption puzzle~\cite{renExploringRandomPhase2009a,schimkaAccurateSurfaceAdsorption2010}, broader applications of RPA have exposed limitations in its accuracy~\cite{schmidtBenchmarkDatabaseTransition2018,szaroBenchmarkingAccuracyDensity2023}.
These observations have motivated efforts to empirically optimize RPA~\cite{kimAssessingExchangeCorrelationFunctionals2024,yuOptimizationRandomPhase2026} and to develop RPA-based double hybrids~\cite{mezeiConstructionApplicationNew2015a,grimmeComputationallyEfficientDouble2016}, although the latter have not yet been applied to solid-state systems.

In recent years, there has been significant interest in non-self-consistent (NSC) DFAs~\cite{simImprovingResultsImproving2022a}, in which orbitals and densities are obtained self-consistently with one functional and the energy is evaluated in a one-shot manner using another.
This approach has been successful for molecular systems~\cite{songExtendingDensityFunctional2023}, where the meta-GGA SCAN~\cite{sunStronglyConstrainedAppropriately2015c} or r$^2$SCAN functional evaluated with Hartree-Fock orbitals [(r$^2$)SCAN@HF] has shown strong performance for noncovalent interactions~\cite{songExtendingDensityFunctional2023,kaplanHowDoesHFDFT2024,kimExtendingDensityCorrectedDensity2025} and reaction barriers~\cite{kaplanUnderstandingDensityDrivenErrors2023}.
More recently, this accuracy has also been demonstrated for solid transition-metal oxides by combining hybrid variants of r$^2$SCAN with separate exchange admixtures (r$^2$SCANX@r$^2$SCANY)~\cite{gopidiReducingSelfInteractionError2025}.
NSC-DFAs remain largely unexplored for surfaces, particularly catalytic transition-metal surfaces.
The only reported examples are CO adsorption, where PBE@PBE+U~\cite{patraRethinkingCOAdsorption2019a} was shown to reproduce earlier DFT+U results correcting its site preference~\cite{kresseSignificanceSingleelectronEnergies2003,soiniDFT+UmolMethodIts2014}, and O$_2$ dissociation on Al(111), where HSE03-1/3x@RPBE was able to reproduce self-consistent HSE03-1/3x at reduced computational cost~\cite{gerritsDensityFunctionalTheory2020b,breeDissociativeChemisorptionO22024}.

In this work, we extend the NSC-DFA framework to catalytic transition-metal surfaces.
We introduce a hybrid and double-hybrid NSC-DFA within this framework that addresses key deficiencies of (self-consistent) semilocal DFAs.
Unlike prior applications to insulators that used HF densities, we employ GGA DFAs, which are more appropriate for metallic systems while being computationally efficient.
Specifically, we generate these densities (and orbitals) from the BEEF-vdW~\cite{wellendorffDensityFunctionalsSurface2012a,studtCOCO2Hydrogenation2013,medfordAssessingReliabilityCalculated2014} DFA, a dispersion-corrected GGA designed for metal surfaces.
From the BEEF-vdW orbitals, we construct a hybrid DFA evaluated non-self-consistently (hBEEF-vdW@BEEF-vdW) with one additional free parameter for the fraction of screened (short-range) exact exchange.
We also construct a double-hybrid extension (dhBEEF-vdW@BEEF-vdW) with a second free parameter for the fraction of RPA correlation (see Methods).
The parameters of these NSC-DFAs are empirically optimized on a small curated set of seven experimental adsorption energies (Supplementary Section~\ref{si-sec:opt_process}).

The resulting two NSC-DFAs surpass conventional RPA in accuracy while maintaining a highly economical computational cost.
The hybrid NSC-DFA, hBEEF-vdW@BEEF-vdW, matches the performance of RPA for 39 adsorption energies---far beyond its limited tuning set---and improves upon RPA for 17 surface reaction barriers.
Notably, it resolves the CO adsorption puzzle, correctly predicting the preferred binding configuration of CO on Pt(111) that has challenged previous self-consistent semilocal (and even hybrid) DFAs.
Our double-hybrid DFA, dhBEEF-vdW@BEEF-vdW, yields substantial accuracy improvements over RPA, achieving a mean absolute deviation of \SI{11.8}{\kjmol} (within ``transition-metal chemical accuracy''~\cite{deyonkerQuantitativeComputationalThermochemistry2007} of \SI{13}{\kjmol}) for adsorption energies and outperforming RPA for barrier heights, while retaining correct behavior for CO adsorption.

Both NSC-DFAs are practical, cost-efficient, and compatible with existing DFT codes.
We provide easy-to-modify recipes for the QuAcc automated workflow library~\cite{rosenQuaccQuantumAccelerator2024}, enabling immediate application to systems of interest.
Beyond improved accuracy, they offer key advantages over their self-consistent counterparts.
In particular, both DFAs avoid a hybrid SCF iteration, which significantly lowers cost and precludes convergence issues. 
Here, the use of screened exact exchange and RPA correlation (as opposed to MP2) also ensures a well-defined description of metallic solids.
As a result, our hybrid NSC-DFA, hBEEF-vdW@BEEF-vdW, requires the equivalent of only one SCF cycle of a hybrid DFA, which is about 20 times more expensive than the GGA SCF calculation for the systems studied in this work (Supplementary Table~\ref{si-tab:ce39_cost_ratio}).
In addition, our double-hybrid NSC-DFA, dhBEEF-vdW@BEEF-vdW, is designed to be only twice as expensive than hBEEF-vdW@BEEF-vdW.
Finally, for both functionals, the use of screened exact exchange~\cite{heydHybridFunctionalsBased2003a}, instead of global exact exchange, helps make our approach more practical due to the low cost and fast convergence with vacuum spacing and Brillouin zone sampling, while enabling cost-saving composite approximations using BEEF-vdW (see Methods and Supplementary Section~\ref{si-sec:practicalities}).

\section{Results}
\subsection{Accurate adsorption energy predictions}

We first validate our proposed NSC-DFAs against the CE39 dataset by \citet{wellendorffBenchmarkDatabaseAdsorption2015}, which consists of experimental adsorption energies for 39 diverse reactions corrected for zero-point vibrational effects.
It covers chemisorption and physisorption of key catalytic reactants such as hydrogen, alkanes, alcohols, water, and oxygen on technologically-relevant transition-metal surfaces, and spans a large energy range of over \SI{200}{\kjmol} per product.
This dataset is a standard benchmark for computational catalysis, and it has been a prevailing challenge~\cite{mallikarjunsharadaAdsorptionTransitionMetal2019a,kothakondaChemicalAccuracyChemi2026} to design DFAs that (1) achieve errors below \SI{20}{\kjmol}, let alone the \SI{13}{\kjmol} target that defines transition-metal chemical accuracy, and (2) give a balanced description of both physisorption and chemisorption.

\begin{figure}[h]
    \includegraphics[width=\textwidth]{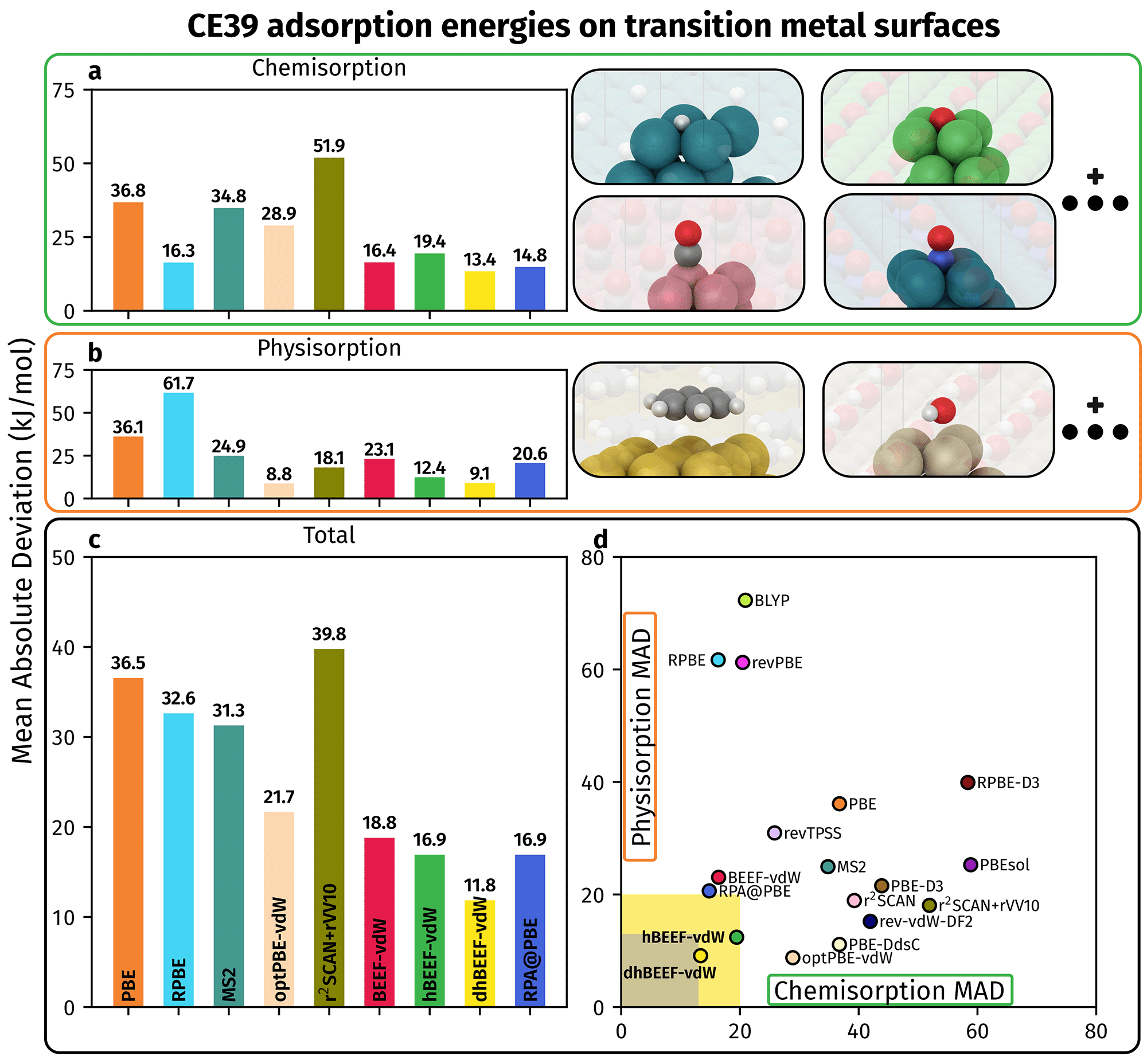}
    \caption{\label{fig:ce39_dftxc}\textbf{Accurate and balanced performance for adsorption on transition metal surfaces.} Comparison of hBEEF-vdW@BEEF-vdW and dhBEEF-vdW@BEEF-vdW against common (self-consistent) density functional approximations for adsorptions energies of the CE39 dataset~\cite{wellendorffBenchmarkDatabaseAdsorption2015}. We give mean absolute deviations (MAD per product formed) w.r.t.\ experimental values from Ref.~\citenum{wellendorffBenchmarkDatabaseAdsorption2015} for the \textbf{a} chemisorption and \textbf{b} physisorption subsets. We also compare the \textbf{c} total performance and \textbf{d} show the correlation between the physisorption and chemisorption subset MADs. In \textbf{d}, we highlight MAD values within \SI{20}{\kjmol} (yellow) and \SI{13}{\kjmol} (gray) for both the physisorption and chemisorption subsets, thresholds achieved only by hBEEF-vdW@BEEF-vdW and dhBEEF-vdW@BEEF-vdW, respectively.}
\end{figure}

We benchmark the performance of hBEEF-vdW@BEEF-vdW and dhBEEF-vdW@BEEF-vdW on the CE39 dataset against several common (self-consistent) DFAs and the RPA in Figure~\ref{fig:ce39_dftxc}; we evaluate the RPA correlation energy with PBE orbitals (RPA@PBE), which is the most common choice in the literature.
With the double-hybrid dhBEEF-vdW@BEEF-vdW, we obtain the best accuracy to date for the full CE39 dataset, leading to a mean absolute deviation (MAD) of \SI{11.8}{\kjmol}, surpassing transition-metal chemical accuracy.
Unlike many other DFAs and the RPA, the accuracy of dhBEEF-vdW@BEEF-vdW is also well-balanced, with an MAD of \SI{13.4}{\kjmol} and \SI{9.1}{\kjmol} on the chemisorbed and physisorbed subsets, respectively.
The overall performance is about $\SI{5}{\kjmol}$ more accurate than RPA, primarily due to a significant reduction of error on the physisorbed subset of the CE39 dataset, for which RPA has a MAD of \SI{20.6}{\kjmol}.
Our hybrid hBEEF-vdW@BEEF-vdW improves upon the base BEEF-vdW DFA, bringing the MAD down from \SI{18.8}{\kjmol} to \SI{16.9}{\kjmol}, such that it rivals RPA (also MAD of \SI{16.9}{\kjmol}) at lower costs.

As shown in Figure~\ref{fig:ce39_dftxc}d, we have benchmarked a broad set of DFAs and the RPA on the CE39 dataset and none have been able to reach errors below \SI{20}{\kjmol} on the chemisorption and physisorption subsets simultaneously (BEEF-vdW and RPA are the closest).
Both hBEEF-vdW@BEEF-vdW and dhBEEF-vdW@BEEF-vdW achieve this goal, with dhBEEF-vdW@BEEF-vdW being the first DFA to achieve close to transition-metal chemical accuracy of \SI{13}{\kjmol} or better across both subsets of the CE39 dataset.
The DFAs we benchmark are all widely employed in computational catalysis, and dhBEEF-vdW@BEEF-vdW represents the best performing for the chemisorption subset while rivaling the best performing DFA (optPBE-vdW at \SI{8.8}{\kjmol}) for the physisorption subset, with hBEEF-vdW@BEEF-vdW also performing strongly at slightly lower cost.

\clearpage

\subsection{Reliable description of CO and graphene adsorption}

Beyond quantitative improvements, our proposed NSC-DFAs also give qualitative improvements over standard self-consistent DFAs.
As mentioned in the Introduction, predicting the correct CO adsorption site, particularly on Pt(111), has represented the canonical puzzle that has challenged GGAs, meta-GGAs and hybrid DFAs~\cite{stroppaShortcomingsSemilocalHybrid2008}.
Similarly, the adsorption of graphene on Ni(111) has also been contentious~\cite{janthonTheoreticalAssessmentGraphenemetal2013,christianAdsorptionGrapheneNickel2017a}, due to a delicate competition between physisorption far from the surface and chemisorption close to the surface.

\begin{figure}[h]
    \includegraphics[width=\textwidth]{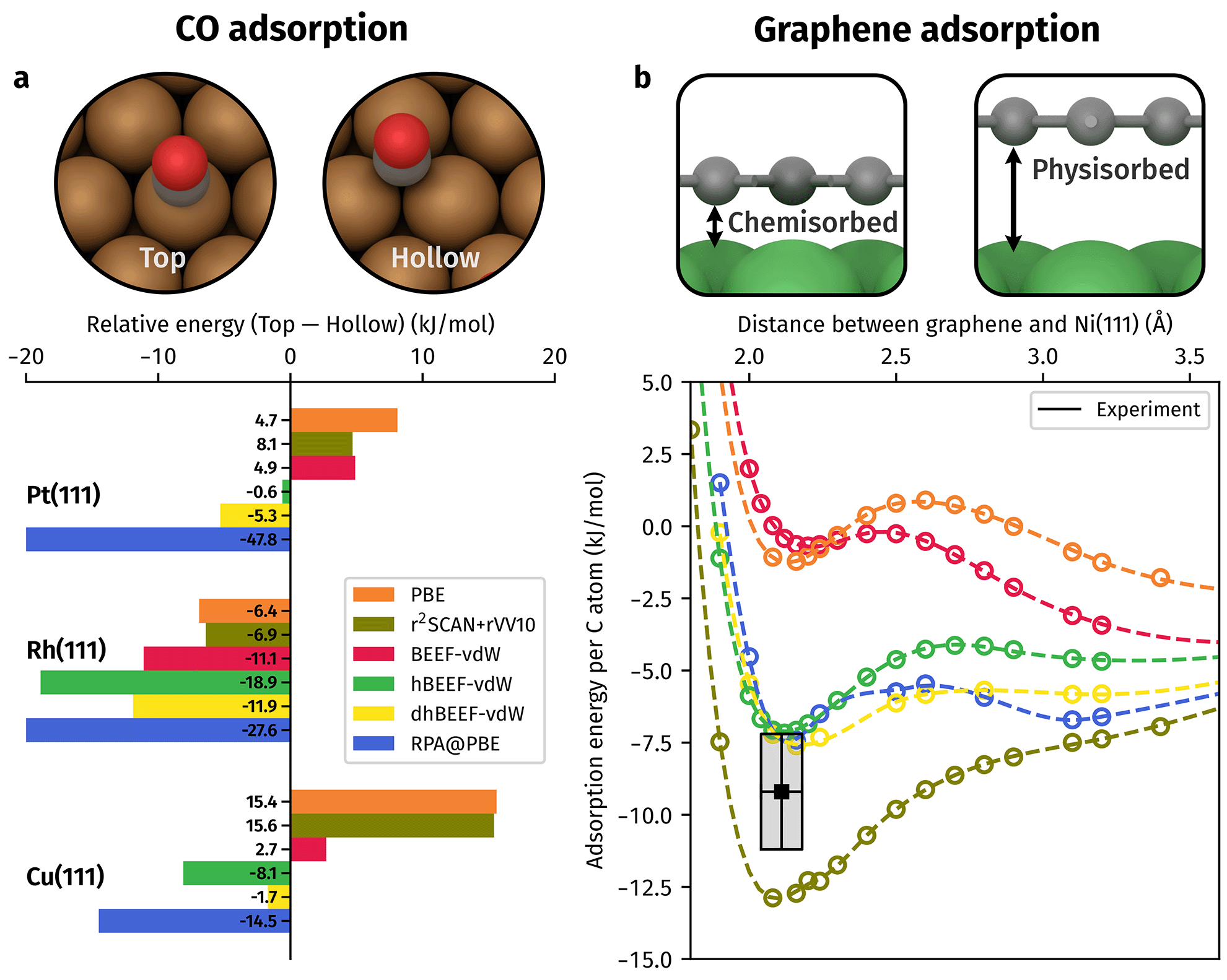}
    \caption{\label{fig:co_gr_ads}\textbf{Overcoming qualitative failures of self-consistent DFAs for adsorption.} Comparison of hBEEF-vdW@BEEF-vdW and dhBEEF-vdW@BEEF-vdW against common (self-consistent) density functional approximations for \textbf{a} CO adsorption on Pt(111), Rh(111) and Cu(111) and \textbf{b} graphene adsorption on Ni(111). We give the relative energy between the (on-)top and hollow (FCC) site in \textbf{a}, with a negative value favoring the top site. The experimental value for the adsorption energy and distance in \textbf{b} were taken from Ref.~\citenum{janthonTheoreticalAssessmentGraphenemetal2013}.}
\end{figure}

We compare a set of DFAs to our NSC-DFAs for CO adsorption across three transition metals [Pt(111), Rh(111) and Cu(111)] in Figure~\ref{fig:co_gr_ads}a, where the relative energy between the (on-)top and hollow (FCC) sites is plotted (with a negative value favoring the top site).
Experimental measurements show that CO adsorbs on the top site for all three systems~\cite{haydenInfraredSpectroscopicStudy1985,ogletreeLEEDIntensityAnalysis1986,beutlerAdsorptionSitesCO1997}, but for both Cu(111) and Pt(111), the semilocal DFAs [r$^2$SCAN-rVV10, PBE and BEEF-vdW] predict the hollow site to be most stable.
For Cu(111), the correct ordering is recovered with hBEEF-vdW@BEEF-vdW, which predicts a relative energy of ${-}\SI{8.1}{\kjmol}$, favoring the top site, whereas  BEEF-vdW predicts ${+}\SI{2.7}{\kjmol}$, favoring the hollow site.
For Pt(111), hBEEF-vdW@BEEF-vdW also favors the top site, but only by ${-}\SI{0.6}{\kjmol}$.
The top site is further favored when increasing the percentage of exact exchange from its current value of 17.5\% to higher levels (Supplementary Section~\ref{si-sec:exx_effect}).
The correct behavior is also observed from dhBEEF-vdW@BEEF-vdW, in part due to its higher fraction of exact exchange (the standard 25\%).
All DFAs (including NSC-DFAs) are found to predict CO adsorption on Rh(111) correctly.
The RPA consistently favors the top site (by 15--\SI{48}{\kjmol}) for all three transition-metal surfaces.

We compare the same set of DFAs for the adsorption of graphene on the Ni(111) surface in Figure~\ref{fig:co_gr_ads}b.
Experiments indicate that it forms a chemisorbed state with an adsorption energy per C atom of $-9.2\pm\SI{2.0}{\kjmol}$ and distance of $2.11\pm0.11\,$pm~\cite{janthonTheoreticalAssessmentGraphenemetal2013} above the surface, highlighted in the gray box.
PBE and BEEF-vdW fail to reproduce this result, predicting weak chemisorption stability and favoring the physisorbed configuration.
We also find that this challenge is observed in many other commonly employed DFAs (Supplementary Section~\ref{si-sec:graphene_dft}).
In contrast, hBEEF-vdW@BEEF-vdW and dhBEEF-vdW@BEEF-vdW (and the RPA) predict the chemisorbed state to be the most stable, with an adsorption energy and chemisorption distance that lie within the errors bars of the experimental values.

\subsection{Towards right answers for the right reasons on surface barrier heights}

We now assess the performance of our NSC-DFAs for reaction barrier heights on transition-metal surfaces, using the 17 reaction barriers compiled in the SBH17 dataset~\cite{tchakouaSBH17BenchmarkDatabase2023}.
This dataset covers the dissociation of CH$_4$, H$_2$, and N$_2$ on various surfaces of Ni, Cu, Ru, Ag, Ir, and Pt.
In particular, N$_2$ dissociation on Ru(0001) and Ru(10$\bar{1}$0) exhibits significant charge transfer, making them challenging barriers to predict.
These reactions have been extensively studied, yet a persistent paradox is that PBE performs best~\cite{tchakouaSBH17BenchmarkDatabase2023} for barrier heights despite being among the worst for adsorption energies (as we showed in Figure~\ref{fig:ce39_dftxc}).

\begin{figure}[h]
    \includegraphics[width=\textwidth]{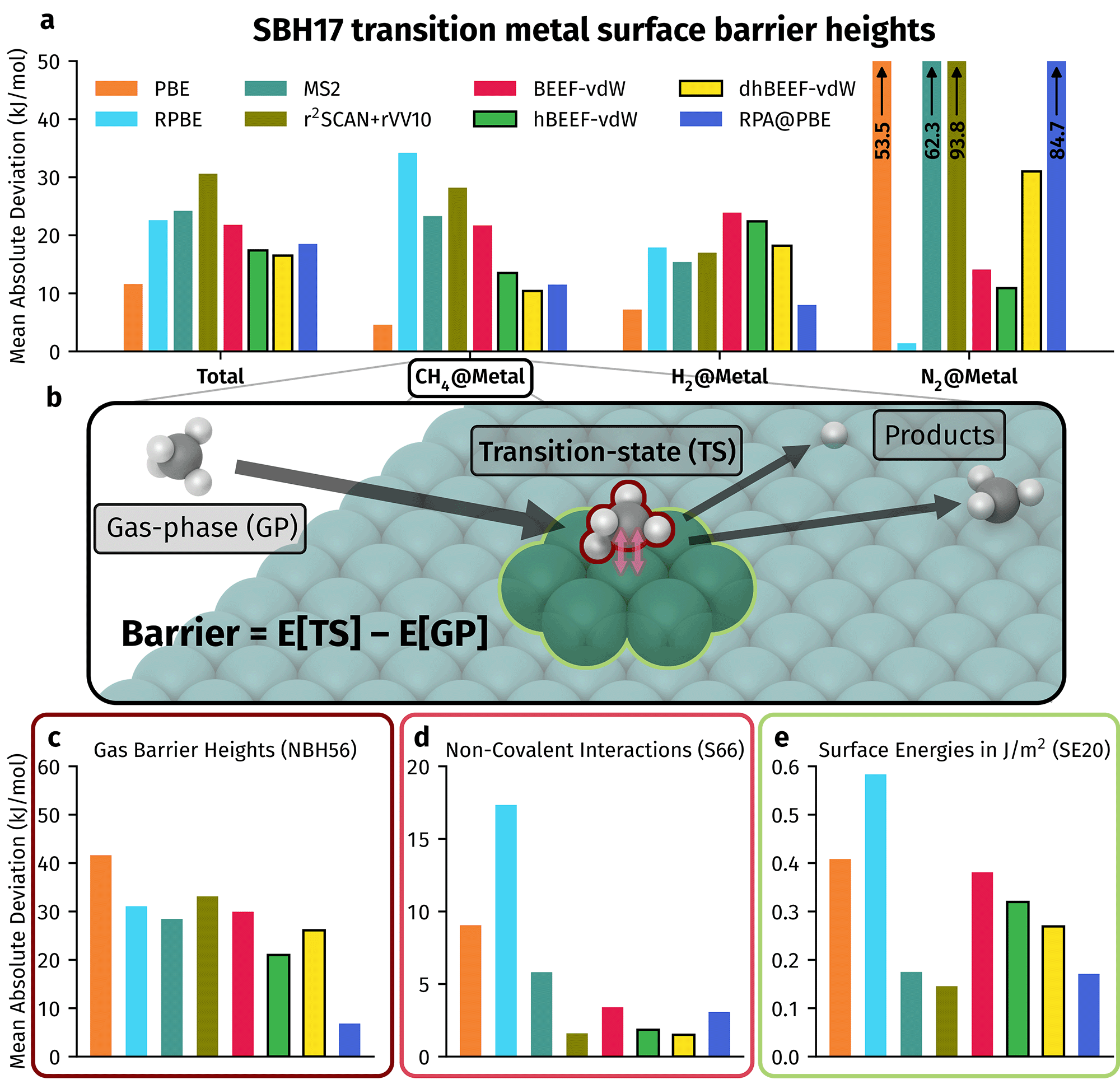}
    \caption{\label{fig:sbh17}\textbf{Improving barrier heights for the right reasons on transition metal surfaces.} Comparison of hBEEF-vdW@BEEF-vdW and dhBEEF-vdW@BEEF-vdW against common (self-consistent) density functional approximations for the \textbf{a} SBH17 dataset~\cite{tchakouaSBH17BenchmarkDatabase2023}. We provide mean absolute deviations (MADs) against experimental values for the total SBH17 dataset as well as its CH$_4$ [visualized in \textbf{b}], H$_2$ and N$_2$ dissociation subsets. Further comparison is given for \textbf{c} gas-phase barrier heights (NBH56; the neutral subset of the BH46 and DBH22 datasets in Ref.~\citenum{liangGoldStandardChemicalDatabase2025}), \textbf{d} gas-phase dimer non-covalent interactions (S66 dataset~\cite{rezacS66WellbalancedDatabase2011}) and \textbf{e} metal surface energies (SE20 dataset) from Ref.~\citenum{lundgaardMBEEFvdWRobustFitting2016}.}
\end{figure}

Figure~\ref{fig:sbh17}a compares the performance of hBEEF-vdW@BEEF-vdW and dhBEEF-vdW@BEEF-vdW against several DFAs for the SBH17 dataset.
While BEEF-vdW performs strongly for the CE39 dataset, its performance for the SBH17 dataset is weaker, with an MAD of \SI{21.8}{\kjmol}.
Both hBEEF-vdW@BEEF-vdW and dhBEEF-vdW@BEEF-vdW improve over BEEF-vdW, reaching an MAD of \SI{17.4}{\kjmol} and \SI{16.5}{\kjmol}, respectively.
PBE remains the best, with an MAD of \SI{11.6}{\kjmol} but its performance for the N$_2$ dissociation subset is poor, with an MAD of \SI{53.5}{\kjmol}.
These large errors are expected for systems with significant charge transfer (Supplementary Section~\ref{si-sec:ct_barrier}) but are masked in the SBH17 dataset, which includes only two N$_2$ dissociation reactions, previously noted as a qualitative limitation of the dataset~\cite{oudotReactionBarriersMetal2024}.
In fact, this failure is also inherited and exacerbated in RPA@PBE, leading to an MAD of \SI{84.7}{\kjmol} for the N$_2$ dissociation subset.
Our two new DFAs, hBEEF-vdW@BEEF-vdW and dhBEEF-vdW@BEEF-vdW, give relatively consistent performance across all three sets of systems, with the large errors from RPA for N$_2$ dissociation partially mitigated within dhBEEF-vdW@BEEF-vdW.

Improving barrier heights for the right reasons requires accurately describing all relevant components: (1) gas-phase reaction barriers, (2) noncovalent interaction energies between molecules and metal surfaces, and (3) surface energies.
In Figures~\ref{fig:sbh17}b-d, we assess the DFAs across these various components using a neutral subset of 56 reactions (dubbed the NBH56 dataset) taken from the DBH22 and BH46 dataset of gas-phase barrier heights~\cite{liangGoldStandardChemicalDatabase2025}  for (1), the S66 dataset~\cite{rezacS66WellbalancedDatabase2011} of dimer binding energies for (2), and surface energies for 20 metals~\cite{lundgaardMBEEFvdWRobustFitting2016} (dubbed the SE20 dataset in this work) for (3).
For all of these datasets, PBE is one of the worst performing.
Its good performance for CH$_4$ and H$_2$ barrier heights on metal surfaces likely arises from fortuitous error cancellation, where the underestimated gas-phase barriers is offset by weaker binding energies that destabilize the adsorbed transition state.
On the other hand, hBEEF-vdW@BEEF-vdW and dhBEEF-vdW@BEEF-vdW are much better than PBE and consistently improve upon BEEF-vdW across all three datasets.
There is still significant room for improvement, particularly for surface energies, where both hBEEF-vdW@BEEF-vdW and dhBEEF-vdW@BEEF-vdW perform worse than the r$^2$SCAN-rVV10 DFA or the RPA.
Additionally, RPA performs much better than our NSC-DFAs for gas-phase reaction barriers, although this may be due to error cancellation~\cite{gouldSimpleSelfinteractionCorrection2019} that does not transfer to heterogeneous reactions.

\section{Discussion}

We discuss here the origins of the qualitative success of hBEEF-vdW@BEEF-vdW for the CO adsorption puzzle, particularly on Pt(111).
Although self-consistent hybrid DFAs correct semilocal DFAs for Rh(111) and Cu(111), a comprehensive analysis of previous studies~\cite{stroppaShortcomingsSemilocalHybrid2008} showed that the issue persists for Pt(111) for standard global (or screened) hybrids.
As discussed in Supplementary Section~\ref{si-sec:co_ads}, this reflects two competing effects: bonding to the hollow site is weakened due to a raised CO 2$\pi^*$ level that suppresses electron back-donation from the metal, but its bonding is strengthened by the reduced steric repulsion as hybrids localize charge on metal atoms.
As hBEEF-vdW@BEEF-vdW builds on top of the GGA BEEF-vdW density, the decreased steric repulsion does not occur.
On the other hand, we show in Supplementary Section~\ref{si-sec:co_ads} that the 2$\pi^*$ level continues to be raised.
In fact, it increases as a function of the admixture of exact exchange, leading to a corresponding monotonic increase in the stability of the top site.

We now discuss the cost of both hBEEF-vdW@BEEF-vdW and dhBEEF-vdW@BEEF-vdW.
Besides its qualitative and quantitative improvements, hBEEF-vdW@BEEF-vdW has a significantly lower cost relative to self-consistent hybrid DFAs.
By building on top of self-consistent GGA orbitals, the expensive EXX energy must be evaluated only once, in comparison to the tens to hundreds of evaluations that may be required to converge the SCF iteration with a self-consistent hybrid DFA.
This leads to affordable computational costs of 900\,CPU core-hours on average for each system in the CE39 dataset, which is about 20 times more expensive than a self-consistent BEEF-vdW calculation.
In addition to the single EXX evaluation, dhBEEF-vdW@BEEF-vdW has an additional cost to evaluate the RPA correlation energy.
Although RPA is traditionally more expensive than self-consistent hybrid DFAs~\cite{szaroBenchmarkingAccuracyDensity2023}, we have taken several practical measures (detailed in the Methods and Supplementary Section~\ref{si-sec:practicalities}) to ensure that it is on average less than twice the cost of a single EXX evaluation (i.e., the cost of hBEEF-vdW@BEEF-vdW), making it significantly cheaper to evaluate than self-consistent hybrids.
Briefly, (1) the preceding BEEF-vdW results can be reused to overcome finite-size and vacuum effects and (2) relaxed electronic-structure settings can be used for the RPA correlation energy as its contribution is scaled to only 15\% in dhBEEF-vdW@BEEF-vdW.

This work highlights that significant improvements can be obtained from only one or two empirically tuned parameters in hBEEF-vdW@BEEF-vdW and dhBEEF-vdW@BEEF-vdW, respectively.
We adopt simple corrections, adding fixed fractions of exact exchange or RPA correlation to a computationally affordable, self-consistent GGA energy.
This makes the approach practical and cost-effective to implement in existing DFT codes (see Supplementary Section~\ref{si-sec:workflow}).
However, this choice also limits the level of accuracy that can be achieved.
In particular, the performance on gas-phase datasets in Figure~\ref{fig:sbh17} is still lacking compared to modern empirically parametrized hybrid DFAs.
Our proposed non-self-consistent DFA framework allows flexibility in both the hybrid DFA energy evaluation and the GGA used to generate the orbitals, offering opportunities for further improvement by modifying either choice.

\section{Conclusions}

In summary, we introduce a framework for generating accurate and affordable density functional approximations to study heterogeneous catalysis on transition-metal surfaces, where a hybrid or double-hybrid DFA is evaluated non-self-consistently with GGA orbitals.
We have proposed two DFAs within this framework: the hybrid hBEEF-vdW@BEEF-vdW and the double-hybrid dhBEEF-vdW@BEEF-vdW.
We show that hBEEF-vdW@BEEF-vdW resolves qualitative challenges within standard self-consistent hybrid density functionals, predicting the correct CO adsorption site on Pt(111) while providing improved performance on 39 adsorption energies and 17 barrier heights over BEEF-vdW, considered the current workhorse DFA for metallic surfaces.
We find that dhBEEF-vdW@BEEF-vdW can achieve transition metal chemical accuracy on this large dataset of adsorption energies, together with a further improvement in barrier heights.
These improvements to the reaction barriers on transition metal surfaces are demonstrated to be for the right reasons, yielding better gas-phase barrier heights, non-covalent interactions, and surface energies.

Their practical utility and low computational cost make these DFAs well suited for studying more complex adsorption reactions~\cite{silbaughEnergiesFormationReactions2016,campbellExperimentalEnergiesFormation2025a} or to compare against accurate experimental barrier heights~\cite{nitzThermalRatesHightemperature2024,nitzExperimentalBenchmarkBarrier2025}.
These qualities, together with its high accuracy, also makes them ideal for training reliable machine-learning interatomic potentials.
To promote adoption and further development, this framework has been implemented as a user-friendly recipe in the QuAcc computational workflow library~\cite{rosenQuaccQuantumAccelerator2024}, enabling high-throughput calculations across diverse metal surfaces.
More broadly, this formalism opens a new avenue for DFA development for transition-metal surfaces, long dominated by GGAs and meta-GGAs, with substantial flexibility for further improvement.

\section{Acknowledgements}

The Flatiron Institute is a division of the Simons Foundation.

\section{Competing Interests Statement}
The authors declare no competing interests.

\section{Methods}
\subsection{Non-self-consistent density functional approximations for metals}

We describe our two non-self-consistent (NSC) density functional approximations (DFAs) in this section, building on a general framework where the density is evaluated with one functional, while the energy is evaluated with a different functional.
Although this framework has been motivated by a formal partitioning of DFT errors into functional-driven and density-driven contributions, its success has also been attributed to favorable error cancellation between these contributions~\cite{kanungoUnconventionalErrorCancellation2024}.
In molecules and insulators, good performance is obtained~\cite{simImprovingResultsImproving2022a,kanungoUnconventionalErrorCancellation2024} by using Hartree-Fock or high-exact-exchange hybrids for the density and SCAN-based meta-GGAs for the energy.
In this work, we adapt the framework to metallic solids by using semilocal DFAs (e.g., BEEF-vdW) to generate densities while evaluating hybrid or double-hybrid functionals non-self-consistently for the energy.
Owing to the delocalized character of metals, GGAs such as BEEF-vdW are expected to produce more reliable densities~\cite{wellendorffDensityFunctionalsSurface2012a,paierWhyDoesB3LYP2007a,gaoApplicabilityHybridFunctionals2016} than hybrids, while hybrid or double-hybrid evaluations on these densities address can help address functional-driven errors.

For our hybrid NSC-DFA, we use the exchange (BFx) and correlation (BFc) functionals from BEEF-vdW (BF-vdW) in the form:
\begin{equation}
\begin{split}
E_\text{xc}^\text{hBF-vdW} &= a_h \cdot \left ( E_x^{\text{HF,SR}}[n^\text{BF-vdW}] + E_x^{\text{BFx,LR}}[n^\text{BF-vdW}] \right) + \left(1 - a_h\right) \cdot E_x^{\mathrm{BFx}}[n^\text{BF-vdW}]  \\
&\hspace{1em} + E_c^{\text{BFc}}[n^\text{BF-vdW}] + E_c^\text{NL}[n^\text{BF-vdW}],
\end{split}
\end{equation}
where $n^\text{BF-vdW}$ is the (self-consistent) electron density from BF-vdW, $a_h$ is the fraction of screened, short-range Hartree-Fock (HF) exact exchange, and $E_c^\text{NL}$ is the non-local van der Waals correlation contribution.
For our double-hybrid NSC-DFA, we also mix in a fraction of RPA correlation (RPAc) energy:
\begin{equation}
\begin{split}
E_\text{xc}^\text{dhBF-vdW} &= a_{dh} \cdot \left( E_x^{\text{HF,SR}}[n^\text{BF-vdW}] + E_x^{\text{BFx,LR}}[n^\text{BF-vdW}] \right) + \left(1 - a_{dh}\right) \cdot E_x^{\mathrm{BFx}}[n^\text{BF-vdW}] \\ 
    &\hspace{1em} + b_{dh} \cdot E_c^\text{RPAc}[n^\text{PBE}] + \left( 1-b_{dh} \right) \cdot E_c^{\text{BFc}}[n^\text{BF-vdW}] + E_c^\text{NL}[n^\text{BF-vdW}].
\end{split}
\end{equation}
The range separation is performed with the error function, and, in practice, we use $E_x^{\text{BFx,LR}}[n^\text{BF-vdW}] \approx E_x^{\text{PBEx,LR}}[n^\text{BF-vdW}]$ for simplicity when using VASP.
Note that the RPA correlation energy is evaluated on PBE orbitals, which we find gives non-negligible improvements over BEEF-vdW orbitals (Supplementary Section~\ref{si-sec:beef_orbitals}).

To determine the mixing parameters $a_h, a_{dh}, b_{dh}$ and the range-separation parameter $\omega$, we evaluate the performance on seven adsorption energies taken from the CE39 dataset (Supplementary Section~\ref{si-sec:opt_process}). 
We select the range-separation parameter $\omega = 0.3$~\AA$^{-1}$, due to its balanced performance and cost-saving implications, as well-established by the HSE family of functionals~\cite{heydHybridFunctionalsBased2003a}.
We find $a_h=0.175$ and $a_{dh}=0.25, b_{dh}=0.15$ to be optimal on our test set, and these values define our NSC-DFAs.

\subsection{Practical advantages of NSC-DFA framework}

Our NSC-DFA framework addresses the limitations of directly applying (global) hybrid, double-hybrid DFAs, or RPA to molecules on metal surfaces, which face practical and cost disadvantages compared to semilocal functionals.
Specifically, we highlight a well-known challenge~\cite{schaferCeriumOxidesRole2021b,schaferUnderstandingDiscrepanciesNoncovalent2025,yuOptimizationRandomPhase2026} in Supplementary Section~\ref{si-sec:gas_challenge}.
With periodic boundary conditions, the exact exchange (EXX) energy of an isolated molecule converges slowly with unit cell volume, due to the long-range nature of the Coulomb interaction and the associated integrable divergence in the EXX energy expression.
For similar reasons, it is also challenging to converge the EXX energy to the thermodynamic limit (i.e., dense $k$-point mesh) when studying metallic slabs (see Supplementary Section~\ref{si-sec:slab_practical}).
These challenges are eliminated in our NSC-DFAs due to the use of a short-range, screened EXX energy. 
As a result, hBEEF-vdW@BEEF-vdW and dhBEEF-vdW@BEEF-vdW exhibit convergence with respect to unit cell volume and $k$-point sampling similar to semilocal DFAs such as BEEF-vdW (just like for HSE functionals~\cite{heydHybridFunctionalsBased2003a}).
Beyond their faster convergence relative to global hybrids or RPA, their similar convergence behavior to BEEF-vdW enables practical and cost-effective correction schemes for the cell volume and $k$-point sampling, as discussed in Supplementary Section~\ref{si-sec:practicalities}.

In the double-hybrid dhBEEF-vdW@BEEF-vdW NSC-DFA, the modest 15\% RPA correlation contribution enables further practical advantages.
As shown in Supplementary Section~\ref{si-sec:dhrpa_conv}, fully converging the RPA correlation energy requires strict electronic structure settings.
However, because it uses only a small fraction of the RPA correlation energy, dhBEEF-vdW@BEEF-vdW can be evaluated with a lower energy cutoff and coarser fast Fourier transform grids without introducing significant errors.

\subsection{Modeling procedure for adsorption energy and barrier heights}

All adsorption energy calculations were modeled with four-layer (4L) slabs with a subsequent correction towards a larger number of layers (5L or 6L).
The adsorption energy for DFT calculations have the form:
\begin{equation}
    E_\text{ads} = E[\text{Mol@Slab}] - E[\text{Mol}] - E[\text{Slab}] + \Delta E_\text{ads}^\text{layer}.
\end{equation}
Here, $E[\text{Mol}]$, $E[\text{Slab}]$ and $E[\text{Mol@Slab}]$ are the total electronic energies of the molecule (Mol), surface slab and the adsorbed complex, respectively.
The geometries for these systems were optimized using the BEEF-vdW DFA.
The molecule calculations were performed in a large unit cell box with \SI{15}{\angstrom} of vacuum in each direction.
The $\Delta E_\text{ads}^\text{layer}$ accounts for the effects of missing slab layers between a 4L slab with respect to an infinite layer slab and is approximated by subtracting the adsorption energy for the 4L slab from a 6L slab using BEEF-vdW.
For the RPA calculations, we took a slightly different approach, where:
\begin{equation}
    E_\text{ads} = E[\text{Mol@Slab}] - E[\text{Mol}//\text{Slab}] - E[\text{Slab}] + \Delta_\text{vac} + \Delta E_\text{ads}^\text{layer}.
\end{equation}
Here, the molecule is placed in the same unit cell as the slab or adsorbed complex, as indicated by Mol//Slab, together with the same $k$-point mesh.
As shown in Supplementary Section~\ref{si-sec:gas_challenge}, the construction of hBEEF-vdW@BEEF-vdW and dhBEEF-vdW@BEEF-vdW leads to vacuum and $k$-point convergence behavior similar to BEEF-vdW.
This allows vacuum effects to be accounted for at the BEEF-vdW level through the $\Delta_\text{vac}$ term, which accounts for the change in energy between the molecule in the Mol//Slab unit cell and a unit cell with $20\,$\AA{} of vacuum.

The geometries for the SBH17 dataset were taken from Ref.~\citenum{tchakouaSBH17BenchmarkDatabase2023}, which were optimized using the PBE DFA, and already contain a sufficient number of layers.
There are two structures for each system: the molecule in the gas phase placed more than \SI{10}{\angstrom} away from the slab (dubbed GP+Slab), and the adsorbed molecule in its transition-state on the slab (TS+Slab).
For the DFT calculations, we calculated the barrier height as follows:
\begin{equation} \label{eq:full_eb}
E_\text{b} = E[\text{TS+Slab}] - E[\text{GP+Slab}].
\end{equation}
The resulting unit cell (and thus number of plane-waves) is very large to accommodate vacuum around both the gas-phase molecule and the slab in GP+Slab.
This can make the EXX and RPA calculations quite costly.
For these contributions, we instead generated structures truncated down to 4 layers with a \SI{15}{\angstrom} vacuum (w.r.t. the slab) indicated by ``4L,small''.
The molecule (from the gas-phase geometry) is placed within the same unit cell in a separate calculation, and the barrier height is calculated as:
\begin{equation} \label{eq:eb_small}
E_\text{b} = E[\text{TS+Slab}_\text{4L,small}] - E[\text{Mol}//\text{GP+Slab}_\text{4L,small}] - E[\text{Slab}_\text{4L,small}].
\end{equation}
We find that barrier heights computed in Equation~\ref{eq:eb_small} closely approximated the original Equation~\ref{eq:full_eb} to better than \SI{3}{\kjmol} on average across the SBH17 dataset.

\subsection{Density functional theory calculation details}

Most DFT calculations presented in this work were performed in the Vienna \textit{Ab Initio} Simulation Package (VASP)~\cite{kresseEfficiencyAbinitioTotal1996a,kresseEfficientIterativeSchemes1996a,kresseUltrasoftPseudopotentialsProjector1999b}.
We used the standard projector augmented wave (PAW) potentials for the DFT calculations and the associated GW-optimized PAW potentials for the RPA calculations.
For the Ir and Ru elements, smaller core GW-optimized PAW potentials were used for the RPA calculations because corresponding versions for the standard valence core were not available.
The DFT calculations were performed with a \SI{550}{e\volt} energy cutoff, and we used a $9\times9\times1$ and $6\times6\times1$ $k$-point mesh for the $2\times2$ and $3\times3$ slabs within the CE39 dataset and CO adsorption calculations, respectively.
This $k$-point mesh was increased to $12\times12\times1$ and $8\times8\times1$ for the EXX and RPA components of the hybrid and double-hybrid calculations.
We used a $9\times9\times1$ $k$-point mesh for all components for the SBH17 dataset and $12\times12\times1$ $k$-point mesh for the adsorption of graphene on Ni(111).
The RPA calculations were performed with a smaller $310\,$eV cutoff and extrapolated to the basis set limit.
The DFT calculations were performed with a smearing width of \SI{0.10}{e\volt} and Methfessel-Paxton smearing.
Both the EXX and RPA calculations were performed with the same smearing width, but with Fermi-Dirac smearing, where we have used the finite-temperature formalism~\cite{kaltakMinimaxIsometryMethod2020} of RPA.
We used {\tt Normal} fast Fourier transform grids for the EXX energy, while the {\tt Fast} settings were used in evaluating the RPA correlation energy.

\subsection{Gas-phase random phase approximation calculation details}

RPA calculations were performed in the MRCC~\cite{kallayMRCCProgramSystem2020a} molecular quantum chemistry code for the gas-phase NBH56 and S66 datasets in Figure~\ref{fig:sbh17}.
The complete basis set limit was reached using a two point extrapolation~\cite{neeseRevisitingAtomicNatural2011a} of the aug-cc-pVQZ and aug-cc-pV5Z pair of basis sets for the separate EXX and correlation energy components.

\section{Data Availability}

See the supplementary information for a detailed compilation of the obtained results as well as further data and analysis to support the points made throughout the text. The input and output files associated with this work and all analysis will be made available on Github and Zenodo upon publication.

\section{Code Availability}

The scripts and workflow for performing the non-self-consistent density functional approximations will be made available on Github and Zenodo upon publication.

\bibliography{references.bib}

\end{document}


\author{Benjamin X. Shi}
\email{mail@benjaminshi.com}
\affiliation{Initiative for Computational Catalysis, Flatiron Institute, New York, NY 10010, USA}

\author{Timothy C. Berkelbach}
\email{t.berkelbach@columbia.edu}
\affiliation{Initiative for Computational Catalysis, Flatiron Institute, New York, NY 10010, USA}
\affiliation{Department of Chemistry, Columbia University, New York, NY 10027, USA}

\title{Supplementary Information: Practical and accurate density functionals for transition-metal heterogeneous catalysis}

\date{\today}

\maketitle
\tableofcontents

\newpage 
\section{\label{sec:workflow}Performing hBEEF-vdW@BEEF-vdW and dhBEEF-vdW@BEEF-vdW}

Although our NSC-DFAs are formally defined in the Methods section of the main text, in this work we use VASP (interfaced to the {\tt LIBXC} library~\cite{lehtolaRecentDevelopmentsLibxc2018b}), which requires several calculations to evaluate our functionals.
Both hBEEF-vdW@BEEF-vdW and dhBEEF-vdW@BEEF-vdW follow the same set of calculations, with dhBEEF-vdW@BEEF-vdW incorporating a subsequent set of RPA calculations.
This results in 4 calculations for hBEEF-vdW@BEEF-vdW and 7 calculations for dhBEEF-vdW@BEEF-vdW (or 8 if the total RPA energy is wanted).
This may seem like a lot of calculations but within the Github data repository (to be shared upon publication), we provide bash scripts to automate the entire process in VASP.
Furthermore, we also provide a pre-built recipe for the QuAcc comptuational workflow library that will allow for automated high-throughput calculations with hBEEF-vdW@BEEF-vdW or dhBEEF-vdW@BEEF-vdW.

The set of calculations for hBEEF-vdW@BEEF-vdW are shown below. In each calculation, the energy can be obtained as the last column (ENERGY) from the line containing ``{\tt energy  without entropy=     NUMBER  energy(sigma->0) =     ENERGY}'' --- this is the last line indicated by where there is no spacing between the ``entropy'' and ``='' sign:
\begin{enumerate}
    \item \textbf{Initial BEEF-vdW} --- Perform a BEEF-vdW calculations (versions prior to 6.4.3 require copying the {\tt vdw\_kernel.bindat} file containing the van der Waals kernel) with the following {\tt INCAR} commands:
    \begin{verbatim}
ENCUT    = 550
LASPH    = .TRUE.
ISMEAR   = -1
SIGMA    = 0.10
GGA      = LIBXC
LIBXC1   = GGA_XC_BEEFVDW
LUSE_VDW = .TRUE.
ZAB_VDW  = -1.8867
ALGO     = Normal
ISTART   = 0
LWAVE    = .TRUE.
LCHARG   = .FALSE.
\end{verbatim}
    \item \textbf{Non-local van der Waals correlation} --- Obtain the vdW correlation energy by performing a BEEF (no vdW) calculation on the BEEF-vdW orbitals (kept in {\tt WAVECAR} calculated from the previous calculation):
    \begin{verbatim}
ENCUT    = 550
LASPH    = .TRUE.
ISMEAR   = -1
SIGMA    = 0.10
GGA      = LIBXC
LIBXC1   = GGA_XC_BEEFVDW
ALGO     = Eigenval
ISTART   = 1
LWAVE    = .FALSE.
LCHARG   = .FALSE.
\end{verbatim}
    \item \textbf{BEEF exchange and correlation} --- Calculate the exchange energy using the BEEF-vdW orbitals.
    \begin{verbatim}
ENCUT    = 550
LASPH    = .TRUE.
ISMEAR   = -1
SIGMA    = 0.10
GGA      = LIBXC
LIBXC1   = GGA_X_BEEFVDW
ALGO     = Eigenval
ISTART   = 1
LWAVE    = .FALSE.
LCHARG   = .FALSE.
\end{verbatim}
    \item \textbf{Exact exchange} --- Calculate the sum of the short-range exact exchange energy and the long-range PBE exchange energy evaluated on the BEEF-vdW orbitals:
    \begin{verbatim}
ENCUT    = 550
LASPH    = .TRUE.
ISMEAR   = -1
SIGMA    = 0.10
GGA      = PE
LHFCALC  = .TRUE.
AEXX     = 1.0
HFSCREEN = 0.3
ALGO     = Eigenval
ISTART   = 1
LWAVE    = .FALSE.
LCHARG   = .FALSE.
\end{verbatim}
\end{enumerate}

An energy is associated with each calculation (where we will call $E_1$ as the energy obtained from the first calculation) and the final hBEEF-vdW@BEEF-vdW energy is given by the following equation:
\begin{equation}
E_\text{hBEEF-vdW@BEEF-vdW} =
\underbrace{0.175\,E_4}_{\text{SR HF + LR PBE exchange}}
+ \underbrace{0.825\,E_3}_{\text{BEEF exchange}}
+ \underbrace{\left( E_2 - E_3 \right)}_{\text{BEEF correlation}}
+ \underbrace{\left( E_1 - E_2 \right)}_{\text{nonlocal vdW}}.
\end{equation}
We have labeled the contributions each term constitutes in the underbraces.
The first term on the right contains both the short-range (SR) HF exchange and also long-range (LR) PBE exchange, where the latter has been used to approximate LR BEEF exchange for simplicity within VASP.
The sum of the first two terms on the right also contain other contributions to the final energy, such as the kinetic and Hartree energy.

Building on top of the first 4 calculations, the RPA energy from PBE orbitals (at smaller energy cutoff) is also required and follows a four-step procedure:
\begin{enumerate}
    \setcounter{enumi}{4}
    \item \textbf{PBE energy and orbitals} --- We start a standard PBE calculation using a smaller energy cutoff.
    \begin{verbatim}
ENCUT    = 310
LASPH    = .TRUE.
ISMEAR   = -1
SIGMA    = 0.10
GGA      = PE
LMAXFOCKAE = 4
ALGO     = Normal
ISTART   = 0
LWAVE    = .TRUE.
LCHARG   = .FALSE.
\end{verbatim}
    \item \textbf{Exact exchange from DFT orbitals} --- Calculate the (unscreened) exact exchange energy evaluated on the PBE orbitals. This is optional as it is not used in the final calculation, but enables an RPA estimate.
    \begin{verbatim}
ENCUT    = 310
LASPH    = .TRUE.
ISMEAR   = -1
SIGMA    = 0.10
GGA      = PE
LMAXFOCKAE = 4
PRECFOCK = Normal
LHFCALC = .TRUE.
AEXX = 1.0
ALGO     = Eigenval
ISTART   = 0
LWAVE    = .FALSE.
LCHARG   = .FALSE.
\end{verbatim}
    \item \textbf{Obtain full set of unoccupied orbitals} --- From the PBE orbitals, perform an exact diagonalization to obtain the full set of unoccupied (and occupied) orbitals needed for the RPA calculation. This requires finding the number of occupied states ({\tt numBANDS}) by searching for the ``{\tt maximum number of plane-waves: numBANDS}'' string within the previous PBE calculation in (5).
    \begin{verbatim}
ENCUT    = 310
LASPH    = .TRUE.
ISMEAR   = -1
SIGMA    = 0.10
GGA      = PE
LMAXFOCKAE = 4
NBANDS   = numBANDS
ALGO     = Exact
ISTART   = 0
LWAVE    = .TRUE.
LCHARG   = .FALSE.
\end{verbatim}
    \item \textbf{Get RPA correlation energy} --- Using the full set of orbitals computed from the preceding calculation, compute the RPA correlation energy. The correlation energy can be obtained from the third column (not the last) of the line containing {\tt converged value}.
    \begin{verbatim}
ENCUT    = 310
LASPH    = .TRUE.
ISMEAR   = -1
SIGMA    = 0.10
LMAXFOCKAE = 4
NBANDS   = numBANDS
ALGO     = ACFDTR
LOPTICS  = .FALSE.
PRECFOCK = Fast
LFINITE_TEMPERATURE = .TRUE.
ISTART   = 0
LWAVE    = .TRUE.
LCHARG   = .FALSE.
\end{verbatim}
\end{enumerate}

The final dhBEEF-vdW@BEEF-vdW energy can then be obtained as follows:
\begin{equation}
E_\text{dhBEEF-vdW@BEEF-vdW} =
\underbrace{0.25\,E_4}_{\text{SR HF + LR PBE exchange}}
+ \underbrace{0.75\,E_3}_{\text{BEEF exchange}}
+ \underbrace{0.15\,E_8}_{\text{RPA correlation}}
+ \underbrace{0.85\,\left( E_2 - E_3 \right)}_{\text{BEEF correlation}}
+ \underbrace{\left( E_1 - E_2 \right)}_{\text{nonlocal vdW}}.
\end{equation}

\clearpage
\section{\label{sec:opt_process}Optimizing the hBEEF-vdW and dhBEEF-vdW coefficients}

As mentioned in the Methods of the main text, there is only one parameter ($a_h$) in hBEEF-vdW and two parameters ($a_{dh}$ and $b_{dh}$) for dhBEEF-vdW@BEEF-vdW.
We used a small set of 7 diverse and representative adsorption systems, dubbed CE7, taken from the CE39 dataset to determine the optimal parameters and ingredients for these NSC-DFAs.
The systems have been selected to represent an assortment of binding mechanisms (including chemisorption and physisorption) on a diverse set of transition metal surfaces.
These 7 reactions are as follows:
\begin{enumerate}
    \item $\text{CO} + \text{Rh}(111) \;\longrightarrow\; \text{CO}/ \text{Rh}(111)$
    \item $\text{NO} + \text{Pd}(111) \;\longrightarrow\; \text{NO}/ \text{Pd}(111)$
    \item $\text{O}_2 + \text{Ni}(100) \;\longrightarrow\; 2 \cdot \text{O}/ \text{Ni}(100)$
    \item $\text{H}_2 + \text{Pt}(111) \;\longrightarrow\; 2 \cdot \text{H}/ \text{Pt}(111)$
    \item $\text{NH}_3 + \text{Cu}(111) \;\longrightarrow\; \text{NH}_3/ \text{Cu}(111)$
    \item $\text{C}_6\text{H}_6 + \text{Pt}(111) \;\longrightarrow\; \text{C}_6\text{H}_6/ \text{Pt}(111)$
    \item $\text{H}_2\text{O} + \text{Pt}(111) \;\longrightarrow\; \text{H}_2\text{O}/ \text{Pt}(111)$
\end{enumerate}

\subsection{Effect of HF screening parameter}
\begin{figure}[h]
    \includegraphics[width=0.7\textwidth]{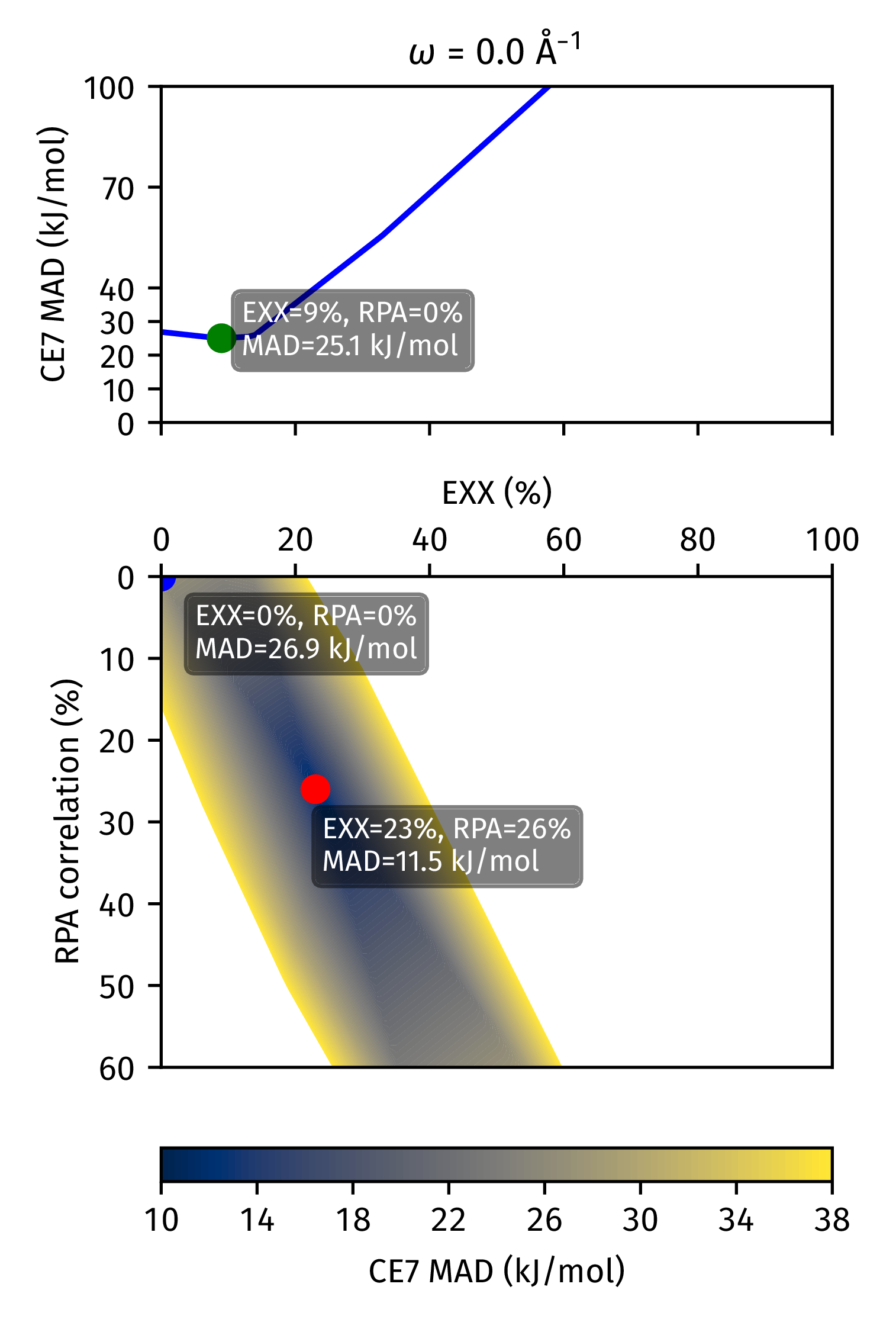}
    \caption{\label{fig:opt_xc_w0}Change in mean absolute deviation (MAD) on the CE7 dataset as a function of global exact exchange and RPA correlation using PBE orbitals for dhBEEF-vdW@BEEF-vdW. The absolute differences computed in the MAD is weighted per product.}
\end{figure}

\begin{figure}[p]
    \includegraphics[width=0.7\textwidth]{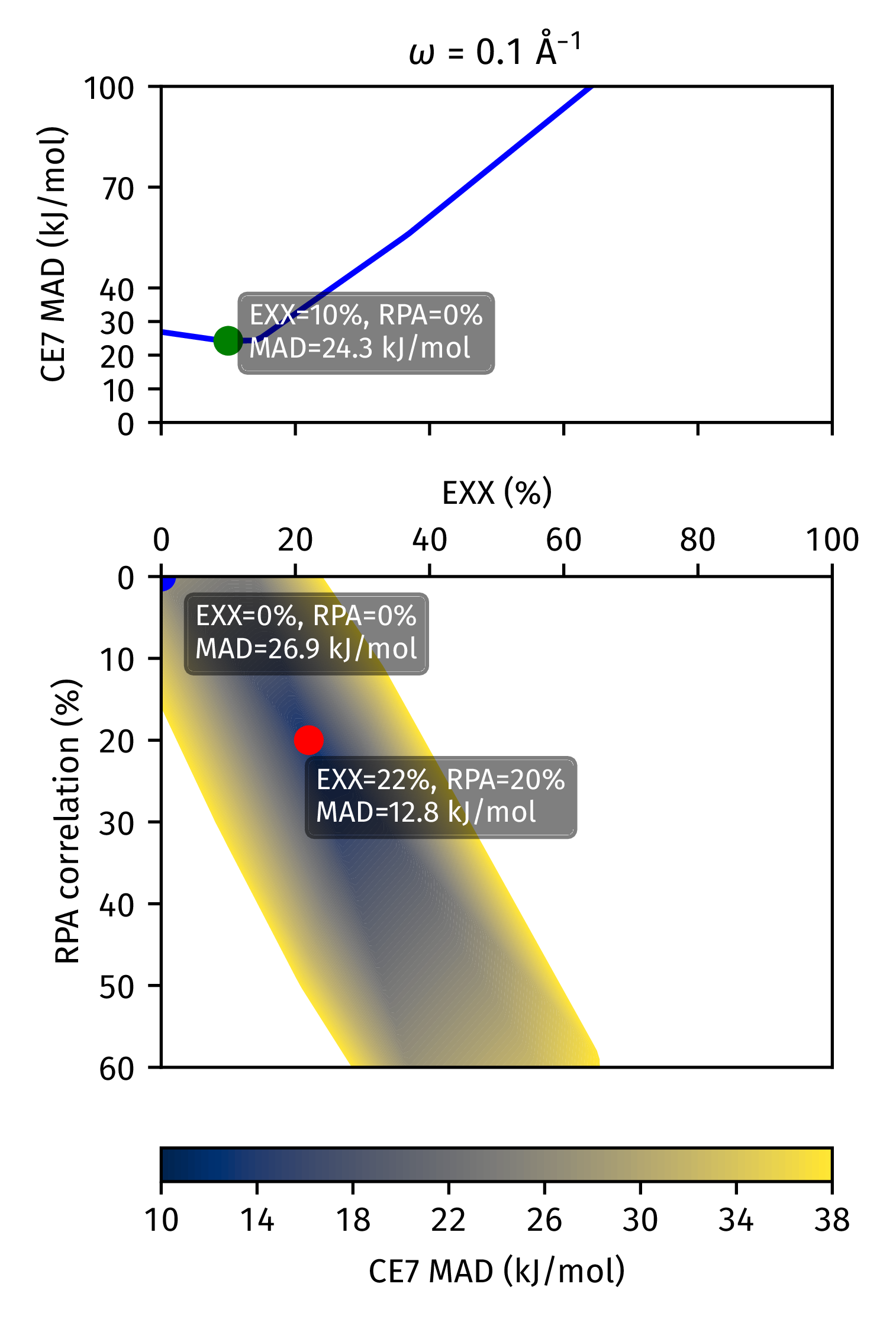}
    \caption{\label{fig:opt_xc_w1}Change in mean absolute deviation (MAD) on the CE7 dataset as a function of screened exact exchange with $\omega=0.1\,$\AA{}$^{-1}$ and RPA correlation using PBE orbitals for dhBEEF-vdW@BEEF-vdW. The absolute differences computed in the MAD is weighted per product.}
\end{figure}

\begin{figure}[p]
    \includegraphics[width=0.7\textwidth]{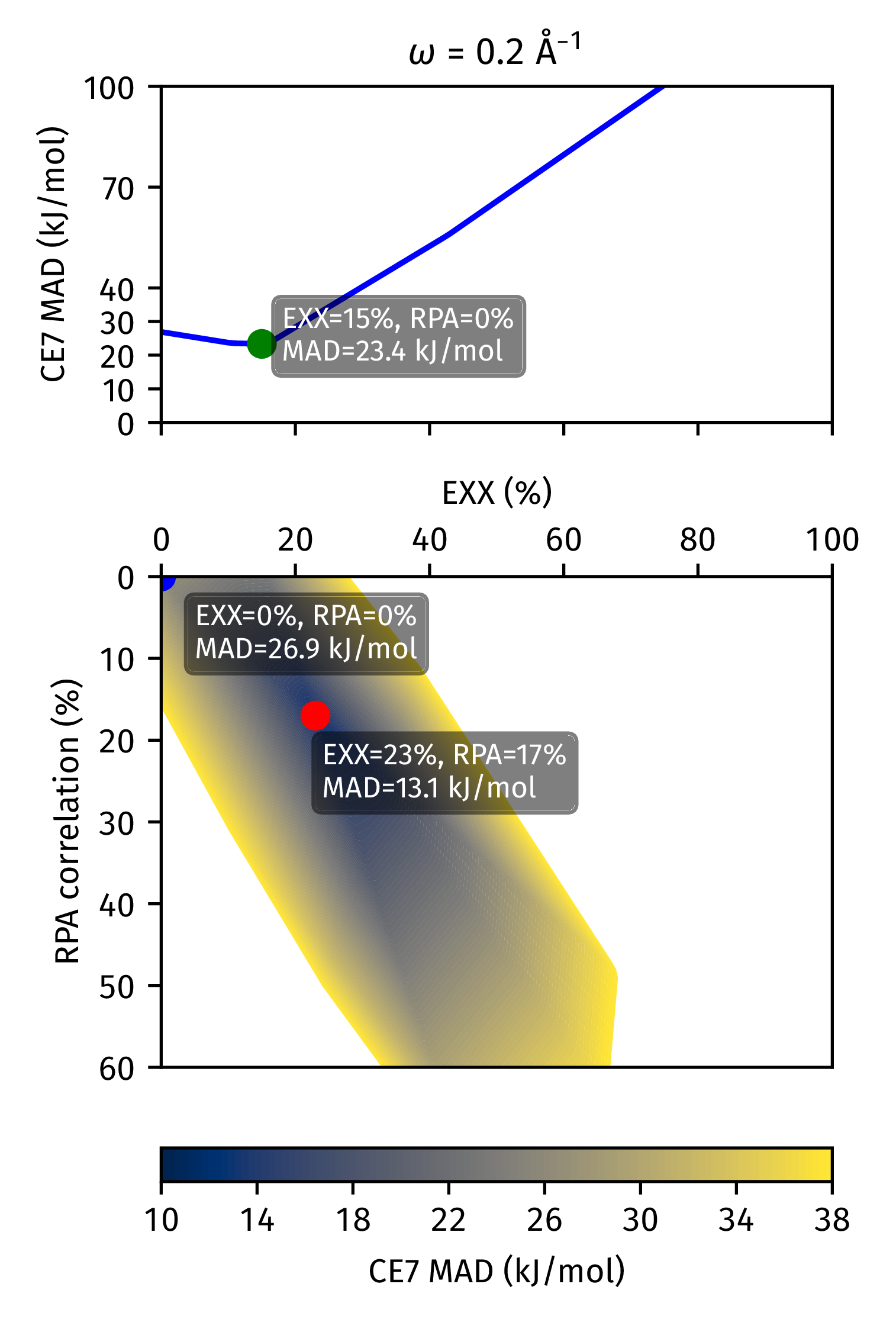}
    \caption{\label{fig:opt_xc_w2}Change in mean absolute deviation (MAD) on the CE7 dataset as a function of screened exact exchange with $\omega=0.2\,$\AA{}$^{-1}$ and RPA correlation using PBE orbitals for dhBEEF-vdW@BEEF-vdW. The absolute differences computed in the MAD is weighted per product.}
\end{figure}

\begin{figure}[p]
    \includegraphics[width=0.7\textwidth]{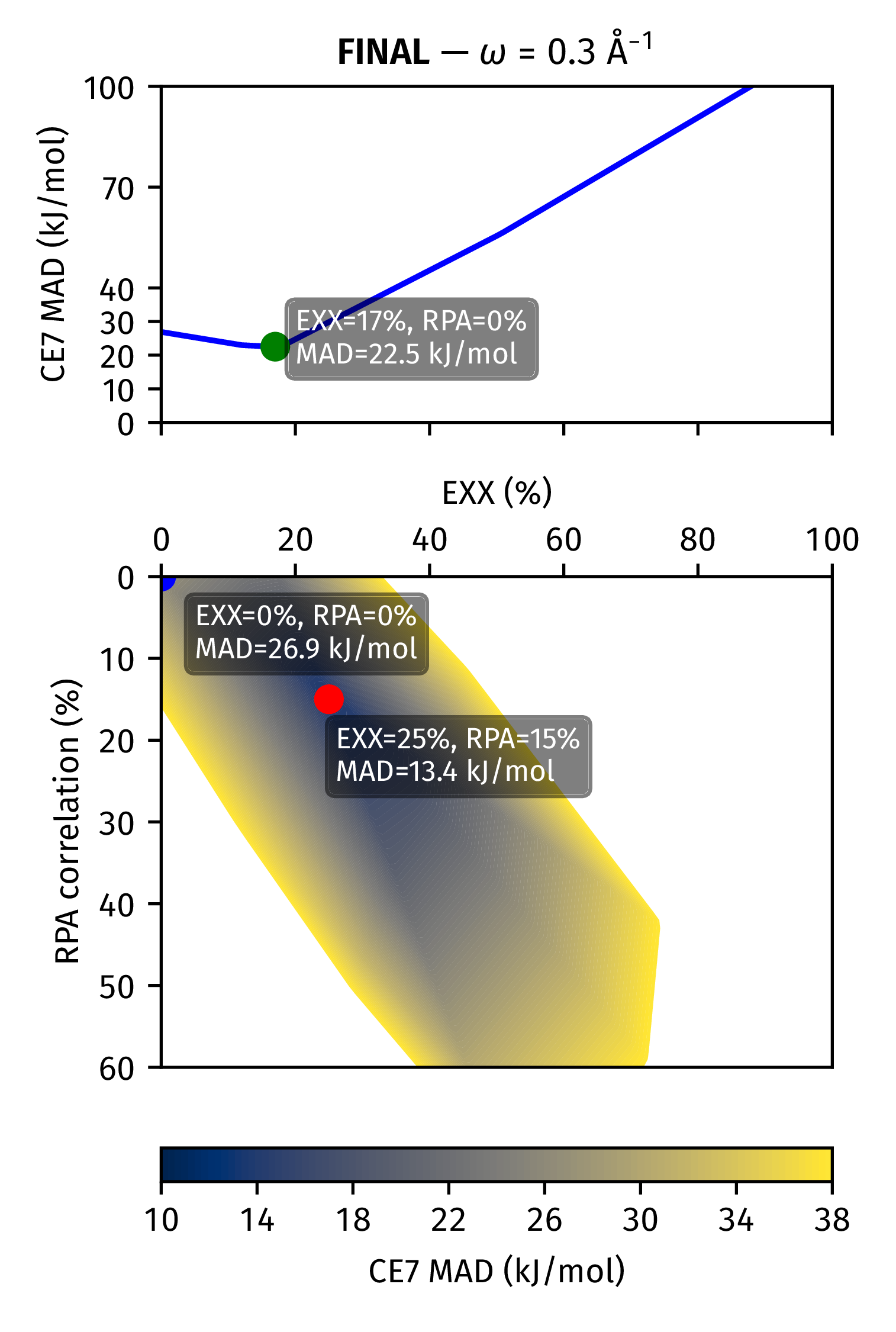}
    \caption{\label{fig:opt_xc_w3}Change in mean absolute deviation (MAD) on the CE7 dataset as a function of screened exact exchange with $\omega=0.3\,$\AA{}$^{-1}$ and RPA correlation using PBE orbitals for dhBEEF-vdW@BEEF-vdW. The absolute differences computed in the MAD is weighted per product.}
\end{figure}

In Figures~\ref{fig:opt_xc_w0} -- \ref{fig:opt_xc_w3}, we highlight the effect of varying the amount of exact exchange (EXX) and RPA correlation on the the mean absolute deviation (weighted per product) on the adsorption energy for the CE7 dataset.
We have compared increasing degree of short range-separation ($\omega$) on the EXX for values of 0.0, 0.1, 0.2 and \SI{0.3}{\per\angstrom}.

It can be seen that the overall effect of adding short-range screening does not modify the MAD significantly.
When simply varying the \% of EXX (with no RPA correlation) in the top panels, increasing the $\omega$ parameter decreases the overall MAD on CE7, from a MAD of \SI{25.1}{\kjmol} for global EXX (i.e., $\omega=\SI{0.0}{\per\angstrom}$) to \SI{22.5}{\kjmol} for $\omega=\SI{0.3}{\per\angstrom}$.
The optimal amount of EXX also increases from 9\% to 17\%, but remains small and in the values typical of standard hybrid DFAs.
Overall, the addition of EXX is beneficial relative to BEEF-vdW, which has an MAD of \SI{26.9}{\kjmol} on the CE7 dataset.

When incorporating RPA correlation, we observe significant decreases in the MAD, down to ${\sim}\SI{13}{\kjmol}$.
There is a smaller dependence on the $\omega$ parameter, going from an optimal MAD of \SI{11.5}{\kjmol} for $\omega=\SI{0.0}{\per\angstrom}$ to \SI{13.4}{\kjmol} for $\omega=\SI{0.3}{\per\angstrom}$.
The optimal EXX also doesn't change, hovering around 24\%, while there is some change in the optimal RPA amount, going from 26\% for $\omega=\SI{0.0}{\per\angstrom}$ to 15\% for $\omega=\SI{0.3}{\per\angstrom}$.

With these observations in mind, in our final NSC-DFAs, we used a range-separation parameter $\omega=\SI{0.3}{\per\angstrom}$ due to its balance at improving the performance of hBEEF-vdW@BEEF-vdW while maintaining a small change on dhBEEF-vdW@BEEF-vdW with respect to full global EXX.
As discussed in the main text, there are also several advantages towards incorporating range-separation, namely (1) faster convergence with $k{-}$point mesh and (2) faster convergence with vacuum amount for the gas-phase molecule.
Based on the observations in Figure~\ref{fig:opt_xc_w3}, the optimal value of EXX was chosen to be 17.5\% for hBEEF-vdW, while we use a value of 25\% for dhBEEF-vdW, together with 15\% RPA correlation for dhBEEF-vdW.

\clearpage

\subsection{\label{sec:beef_orbitals}PBE vs. BEEF-vdW orbitals for RPA correlation}

\begin{figure}[h]
    \includegraphics[width=0.7\textwidth]{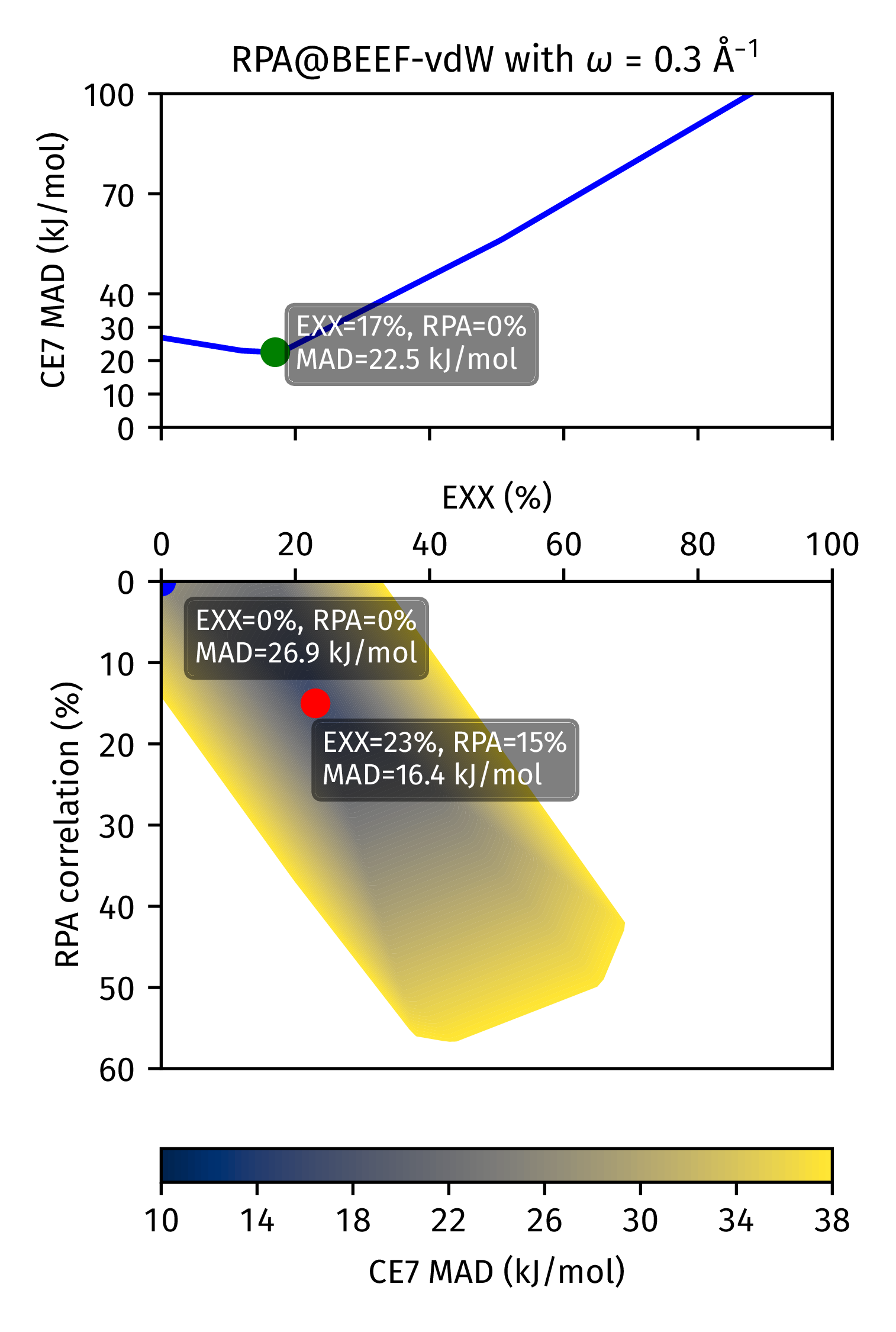}
    \caption{\label{fig:opt_xc_beef}Change in mean absolute deviation (MAD) on the CE7 dataset as a function of screened exact exchange with $\omega=0.3\,$\AA{}$^{-1}$ and RPA correlation using the BEEF-vdW orbitals for dhBEEF-vdW@BEEF-vdW. The absolute differences computed in the MAD is weighted per product.}
\end{figure}

We also originally considered utilizing RPA on BEEF-vdW orbitals, as opposed to PBE orbitals for the RPA correlation component.
As shown in Figure~\ref{fig:opt_xc_beef}, this leads to overall higher MAD of \SI{16.4}{\kjmol} with RPA@BEEF-vdW correlation compared to \SI{13.4}{\kjmol} for RPA@PBE correlation.
Thus, we opted to stick to RPA@PBE correlation.
Practically, this does not add additional costs or steps in dhBEEF-vdW@BEEF-vdW in Section~\ref{sec:workflow} because the initial DFT calculation (whether PBE or BEEF-vdW) had to be re-performed at a smaller energy cutoff and cannot be inherited from the original BEEF-vdW calculations in Step 1.

\clearpage

\section{\label{sec:practicalities}Practical advantages of NSC-DFA framework}

In this section, we aim to highlight the practical advantages of using our final chosen NSC-DFAs settings (i.e., screened exact exchange and small percentage of RPA correlation) from Sec.~\ref{sec:opt_process} over directly applying either a global hybrid or RPA.
The challenge with global hybrids comes from converging the (global) exact exchange (EXX) component, while RPA is affected by both the use of global EXX and with calculating its correlation energy.
Specifically, the two disadvantages are:
\begin{enumerate}
    \item The slow convergence of the EXX energy as a function of vacuum size for the gas-phase molecule, required to calculate the adsorption energy. In this section, this vacuum size corresponds to the unit cell lengths (given the small size of the molecules), so a doubling in the vacuum size relates to a doubling of the unit cell length in all three dimensions, and thus an 8-fold increase in volume.
    \item The slow and erratic convergence on $E_\text{ads}$ with $k$-point mesh for both EXX and RPA correlation, coming from the metallic slab.
\end{enumerate}
For 1, we will use the isolated gas-phase molecule energy of CO and H$_2$O to highlight these challenges and how this is addressed by our NSC-DFA framework.
For 2, we will use the adsorption energy of CO and H$_2$O on the Pt(111) surface, representing a classical example of chemisorption and physisorption, respectively, to highlight the practical advantages of the NSC-DFAs.
For RPA and dhBEEF-vdW@BEEF-vdW, we have used BEEF-vdW orbitals throughout this section, but the behavior is expected to be consistent with using PBE orbitals.

\subsection{\label{sec:gas_challenge}Gas-phase calculation}

The challenges with converging the gas-phase calculation with vacuum size (i.e., unit cell length) is discussed in great detail in Ref.~\citenum{schaferCeriumOxidesRole2021b}.
In brief, the Coulomb interaction is represented as a discrete Fourier sum over reciprocal lattice vectors and $k$-points, recovering the exact $1/|\mathbf{r}-\mathbf{r}'|$ form only in the thermodynamic limit.
For the single $\Gamma$-only sampling, the Coulomb potential becomes artificially periodic in real space, causing intermolecular interactions to be mapped onto intramolecular ones, resulting in spurious interactions.
While this effect is prominent for (unscreened) EXX, DFT exchange decays significantly quicker and thus, the effect is much smaller.
Similarly, adding (short) range-separation to EXX will also lessen this effect because long-range exchange effects are treated at the DFT level.
We note that while this effect can be partly improved for the isolated gas-phase molecule through imposing radial cutoffs~\cite{spencerEfficientCalculationExact2008} to the Coulomb interaction (\texttt{HFRCUT = -1}), this choice must also be applied to the metallic slab, which is expected to make convergence slower, hence we have used the standard probe-charge Ewald summation~\cite{paierPerdewBurkeErnzerhof2005}.

\begin{figure}[p]
    \includegraphics[width=\textwidth]{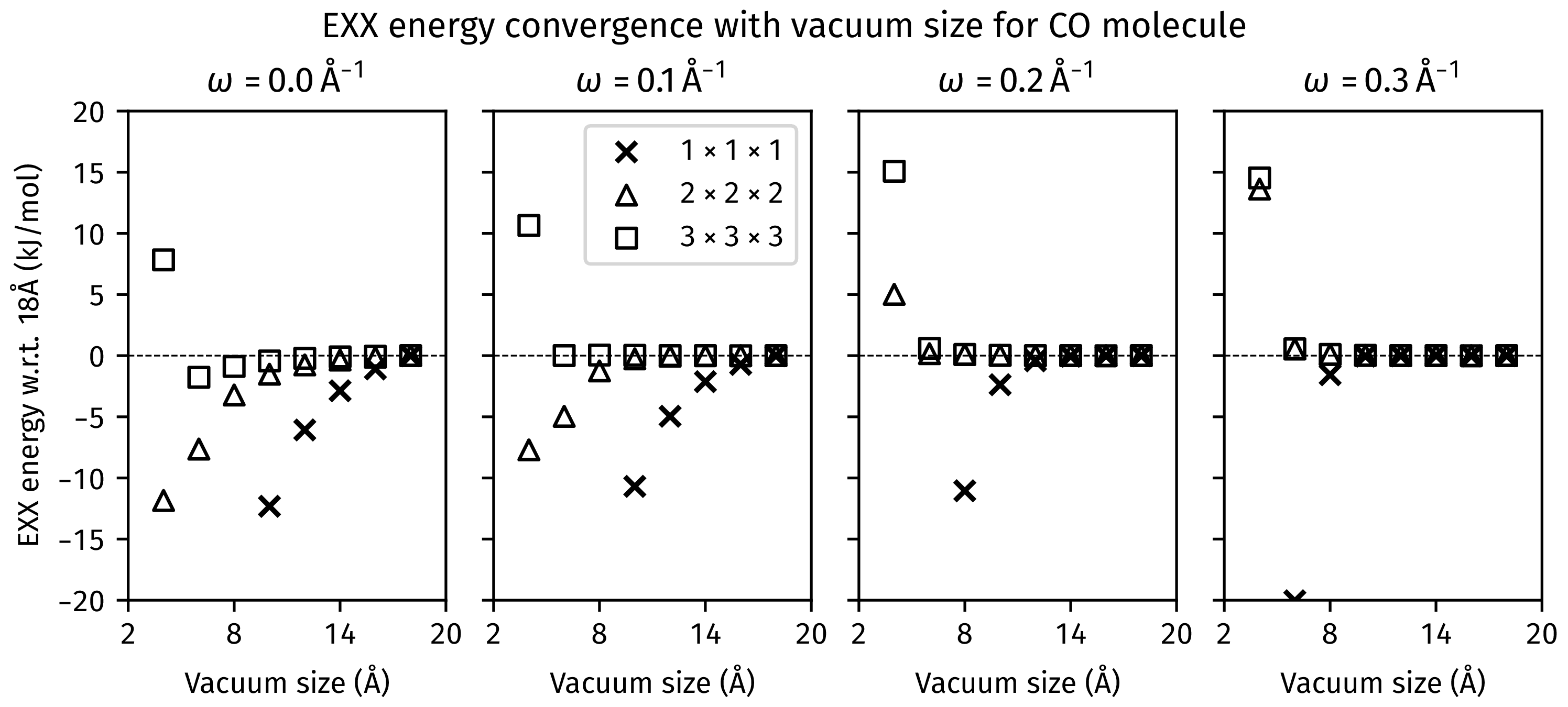}
    \caption{\label{fig:gas_co_kpoints_conv}Convergence of EXX (calculated on BEEF-vdW orbitals) with varying levels of short range-separation for the isolated CO molecule, from global $\omega=\SI{0.0}{\per\angstrom}$ to $\omega=\SI{0.3}{\per\angstrom}$ in steps of $\SI{0.1}{\per\angstrom}$. We plot this as a function of vacuum size and also compare the effect of using denser $k$-point meshes. We used an energy cutoff of \SI{310}{eV} for RPA correlation and \SI{550}{eV} for the BEEF-vdW and EXX components.}
\end{figure}

\begin{figure}[h]
    \includegraphics[width=\textwidth]{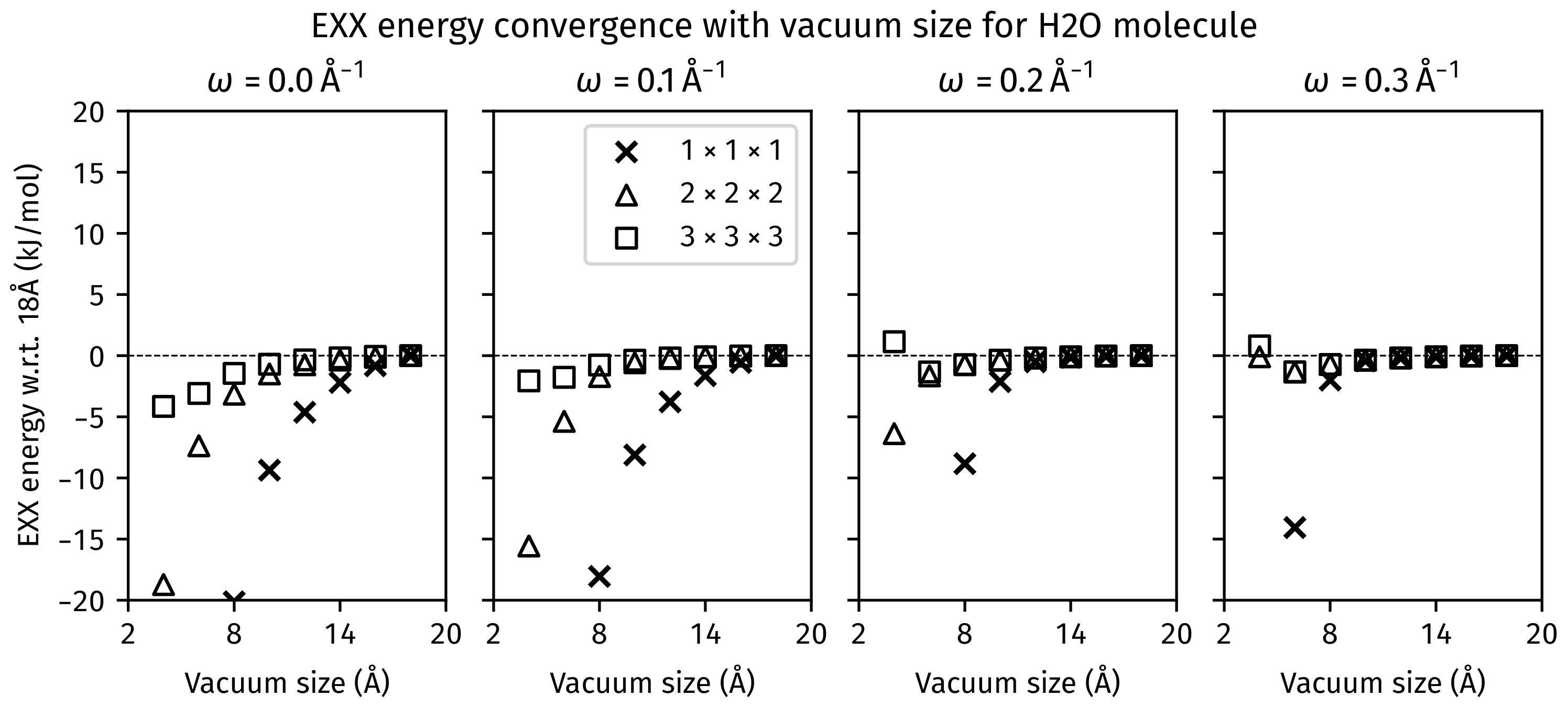}
    \caption{\label{fig:gas_h2o_kpoints_conv}Convergence of EXX (calculated on BEEF-vdW orbitals) with varying levels of short range-separation for the isolated H$_2$O molecule, from global $\omega=\SI{0.0}{\per\angstrom}$ to $\omega=\SI{0.3}{\per\angstrom}$ in steps of $\SI{0.1}{\per\angstrom}$. We plot this as a function of vacuum size and also compare the effect of using denser $k$-point meshes. We used an energy cutoff of \SI{310}{eV} for RPA correlation and \SI{550}{eV} for the BEEF-vdW and EXX components.}
\end{figure}

We highlight the slow convergence with vacuum size for global EXX (corresponding to $\omega=\SI{0.0}{\per\angstrom}$) in Figures~\ref{fig:gas_co_kpoints_conv} and~\ref{fig:gas_h2o_kpoints_conv} for the isolated CO and H$_2$O molecules, respectively.
As the short-range-separation inverse length is increased, used within a complementary error function $\operatorname{erfc}\!\left(\omega \cdot\lvert \mathbf{r} - \mathbf{r}' \rvert \right)$ that multiplies  the Coulomb
interaction, this convergence becomes much faster, plateauing at smaller vacuum sizes for both CO and H$_2$O.
Similarly, this convergence also improves when the $k$-point mesh is increased, further increasing the convergence rate with vacuum size for the short range-separated EXX energy where $\omega=\SI{0.3}{\per\angstrom}$.

\clearpage

\begin{figure}[h]
    \includegraphics[width=\textwidth]{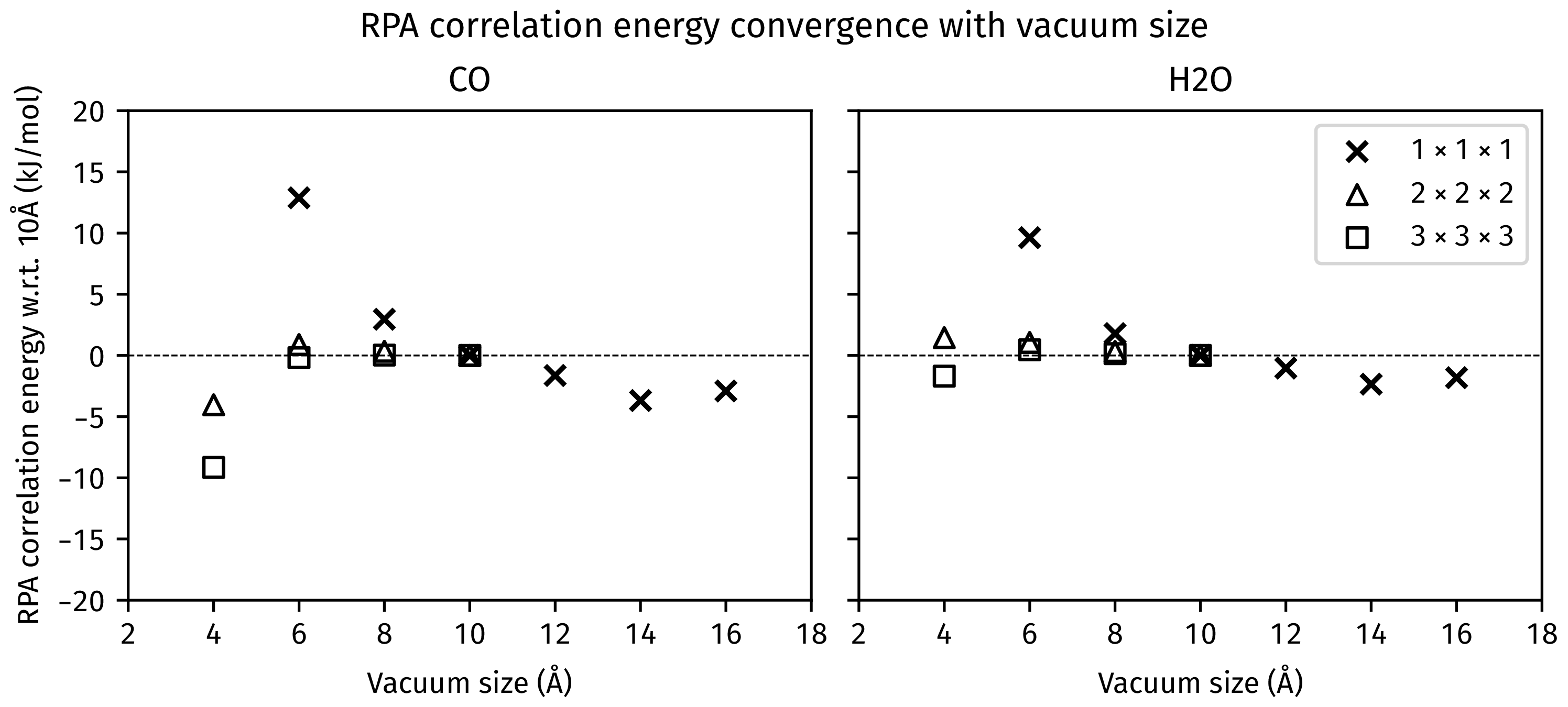}
    \caption{\label{fig:gas_rpa_kpoints_conv}Convergence of RPA correlation energy for the isolated CO (left panel) and H$_2$O (right panel). We plot this as a function of vacuum size and also compare the effect of using denser $k$-point meshes. We used a $10\,$\AA{} box as the reference because that was the largest size that could be performed for the $2 \times 2 \times 1$ and $3 \times 3 \times 1$  $k$-point mesh, although we can go up to $16\,$\AA{} for the $\Gamma$-point calculation. We used an energy cutoff of \SI{310}{eV} for RPA correlation and \SI{550}{eV} for the BEEF-vdW and EXX components.}
\end{figure}

In Figure~\ref{fig:gas_rpa_kpoints_conv}, it can be seen that the slower convergence with vacuum size from utilizing a $\Gamma$-point calculation also pertains to the correlation energy of RPA.
For example, it requires almost 14$\,$\AA{} of vacuum to converge to within \SI{1}{\kjmol} for both the CO and H$_2$O molecule.
On the other hand, convergence of the total energy to within \SI{1}{\kjmol} is already achieved by $6\,$\AA{} with both a $3 \times 3 \times 1$ and $2 \times 2 \times 1$ $k$-point mesh.

\begin{figure}[h]
    \includegraphics[width=\textwidth]{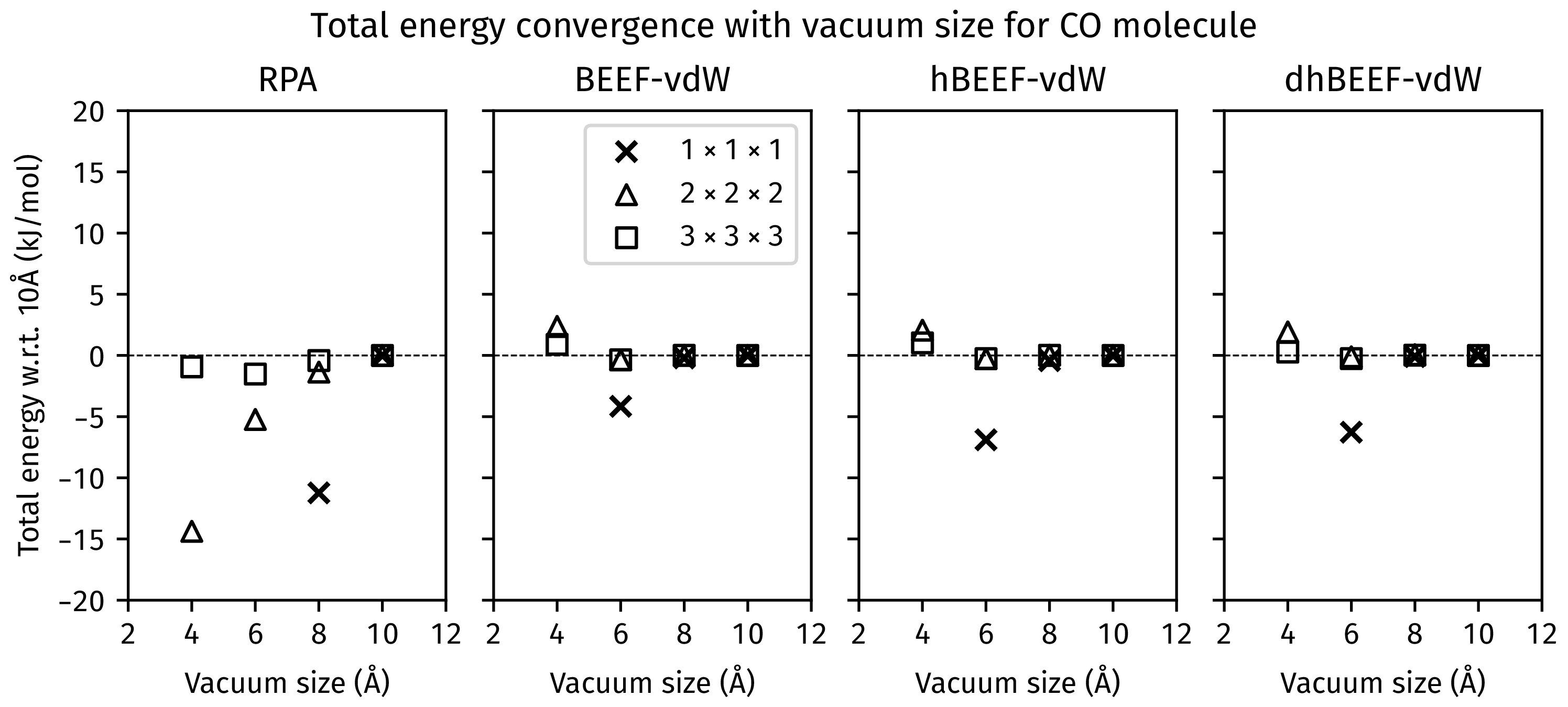}
    \caption{\label{fig:gas_CO_total_kpoints_conv}Convergence of RPA, BEEF-vdW, hBEEF-vdW@BEEF-vdW and dhBEEF-vdW@BEEF-vdW as a function of vacuum size for the isolated CO molecule. We also compare the effect of using denser $k$-point meshes. We used an energy cutoff of \SI{310}{eV} for RPA correlation and \SI{550}{eV} for the BEEF-vdW and EXX components.}
\end{figure}

\begin{figure}[h]
    \includegraphics[width=\textwidth]{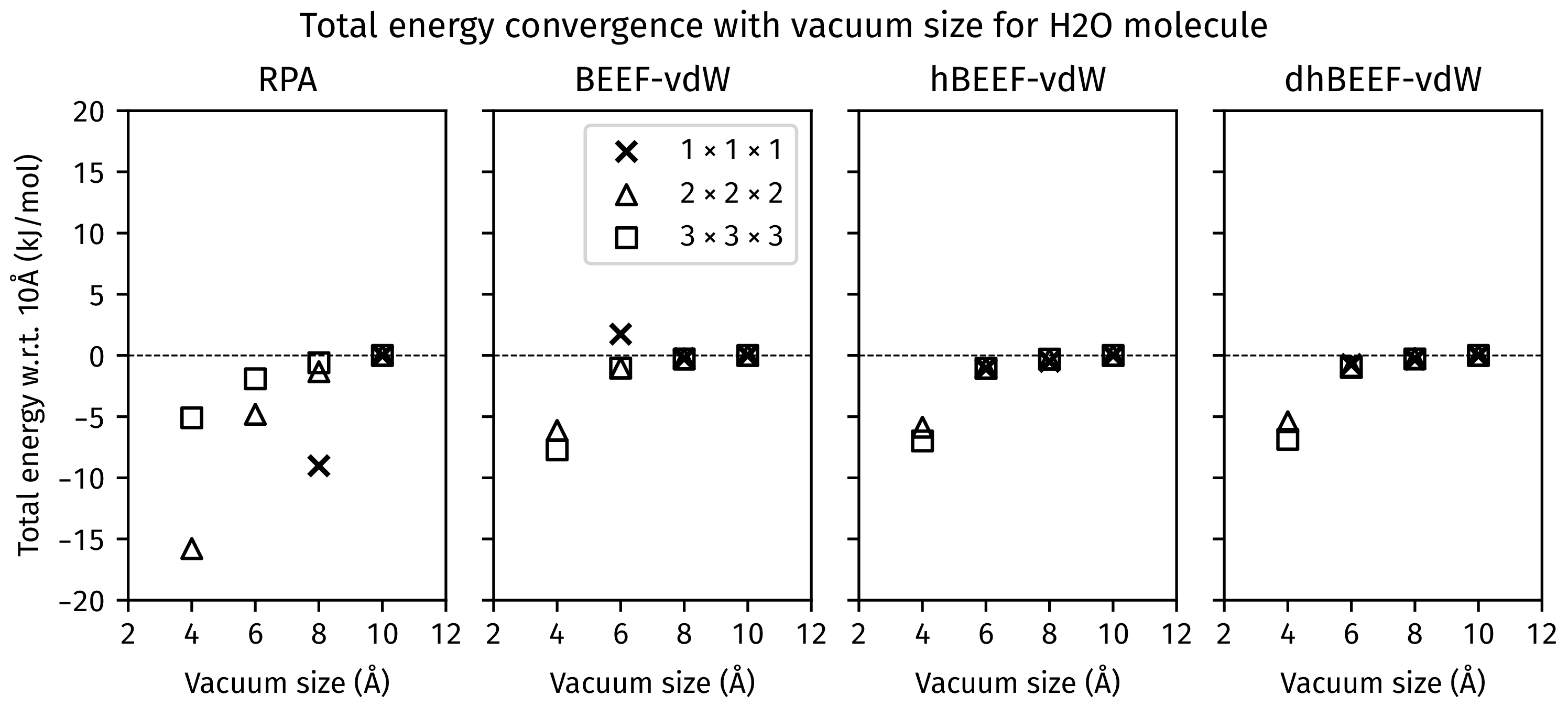}
    \caption{\label{fig:gas_h2o_total_kpoints_conv}Convergence of RPA, BEEF-vdW, hBEEF-vdW@BEEF-vdW and dhBEEF-vdW@BEEF-vdW as a function of vacuum size for the isolated H$_2$O molecule. We also compare the effect of using denser $k$-point meshes. We used an energy cutoff of \SI{310}{eV} for RPA correlation and \SI{550}{eV} for the BEEF-vdW and EXX components.}
\end{figure}

We compare the convergence of the RPA total energy against BEEF-vdW, hBEEF-vdW@BEEF-vdW and dhBEEF-vdW@BEEF-vdW in Figures~\ref{fig:gas_CO_total_kpoints_conv} and~\ref{fig:gas_h2o_total_kpoints_conv} for the isolated CO and H$_2$O molecules.
The RPA total energy is the sum of global EXX and RPA correlation energies, and as both strongly depend on vacuum size with the $\Gamma$-point calculation, there is also a significant dependence in the total energy.
This is contrasted with BEEF-vdW, which achieves rapid convergence with vacuum size, reaching below \SI{1}{\kjmol} for any $k$-point mesh at \SI{8}{\AA} of vacuum.
Both the hBEEF-vdW@BEEF-vdW and dhBEEF-vdW@BEEF-vdW NSC-DFAs also converge rapidly, with a similar rate as BEEF-vdW.
This arises because both uses screened exact exchange with the rapidly converging $\omega=0.3\,$\AA{}$^{-1}$.
In addition, dhBEEF-vdW@BEEF-vdW only uses a small fraction (0.15) of RPA correlation, so its impact on the convergence of this NSC-DFA is small.

\clearpage

As the hBEEF-vdW@BEEF-vdW and dhBEEF-vdW@BEEF-vdW NSC-DFAs converge in a similar manner with vacuum size and $k$-point mesh as BEEF-vdW, we test a composite correction where BEEF-vdW is used to correct the effects from using a small vacuum or small number of $k$-points.
Such a correction $\Delta_\text{vac}$ can enable much more economical calculations, since we find that the RPA calculation (to get the correlation energy in dhBEEF-vdW@BEEF-vdW) can quickly become expensive with the size of vacuum or $k$-point mesh.
For example, the errors for using hBEEF-vdW@BEEF-vdW with a single $\Gamma$-point mesh and \SI{4}{\angstrom} of vacuum (which we will indicate as $E_\text{ads}[\text{hBEEF-vdW}//1\times1\times1//\SI{4}{\angstrom}]$)  can be approximately corrected to $3\times3\times3$ mesh and \SI{10}{\angstrom} of vacuum with BEEF-vdW as follows:
\begin{equation}
    E_\text{ads}[\text{hBEEF-vdW}//3\times3\times3//\SI{10}{\angstrom}] \approx  E_\text{ads}[\text{hBEEF-vdW}//1\times1\times1//\SI{4}{\angstrom}] + \Delta_\text{vac},
\end{equation}
where:
\begin{equation}
    \Delta_\text{vac} = E_\text{ads}[\text{BEEF-vdW}//1\times1\times1//\SI{4}{\angstrom}] - E_\text{ads}[\text{BEEF-vdW}//3\times3\times3//\SI{10}{\angstrom}]
\end{equation}

\begin{figure}[p]
    \includegraphics[width=\textwidth]{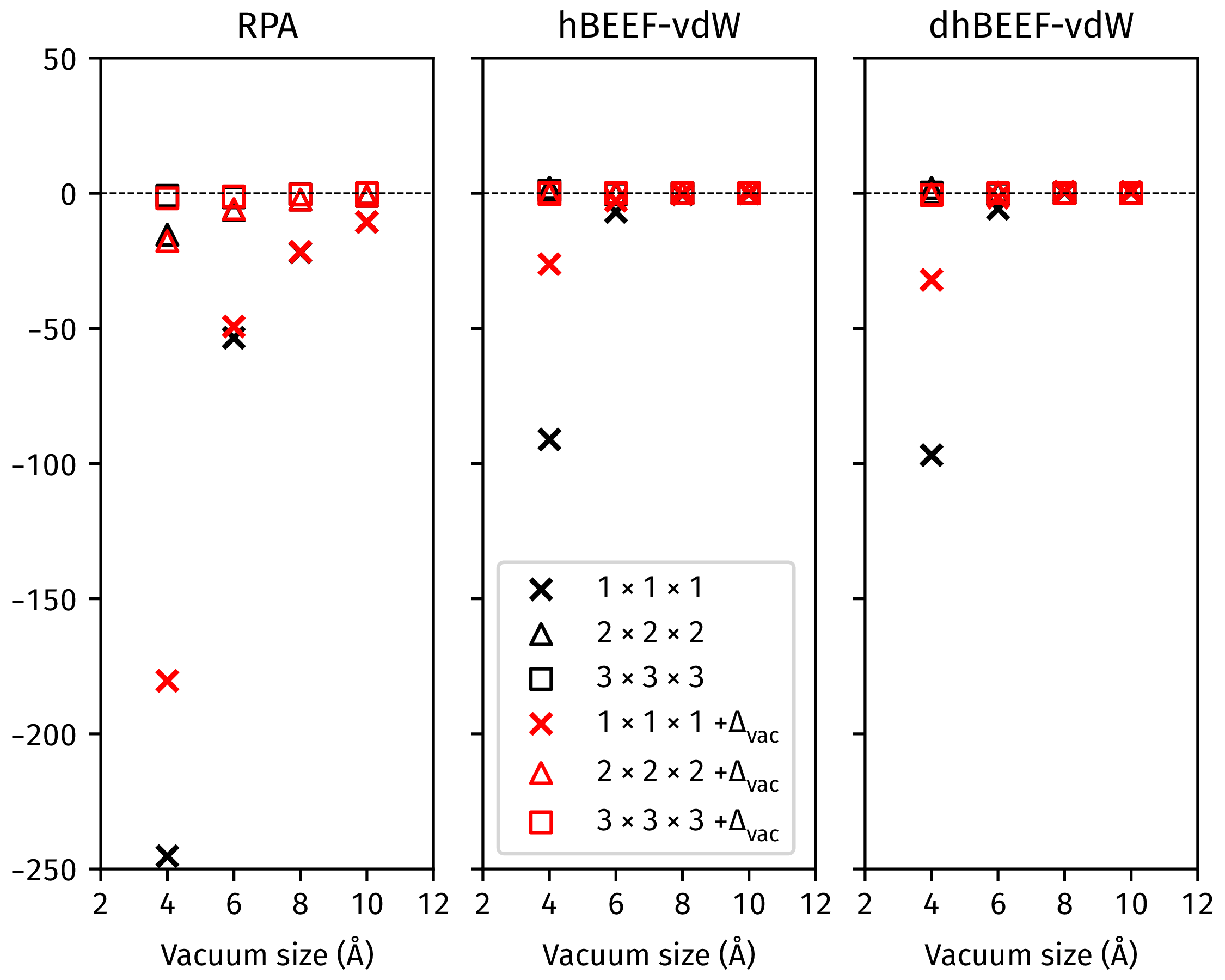}
    \caption{\label{fig:gas_CO_delta_kpoints_conv}Convergence of RPA, hBEEF-vdW@BEEF-vdW and dhBEEF-vdW@BEEF-vdW as a function of vacuum size for the isolated CO molecule with and without a $\Delta_\text{vac}$ that adjusts for lateral interactions with BEEF-vdW. We also compare the effect of using denser $k$-point meshes. We used an energy cutoff of \SI{310}{eV} for RPA correlation and \SI{550}{eV} for the BEEF-vdW and EXX components.}
\end{figure}

\begin{figure}[p]
    \includegraphics[width=\textwidth]{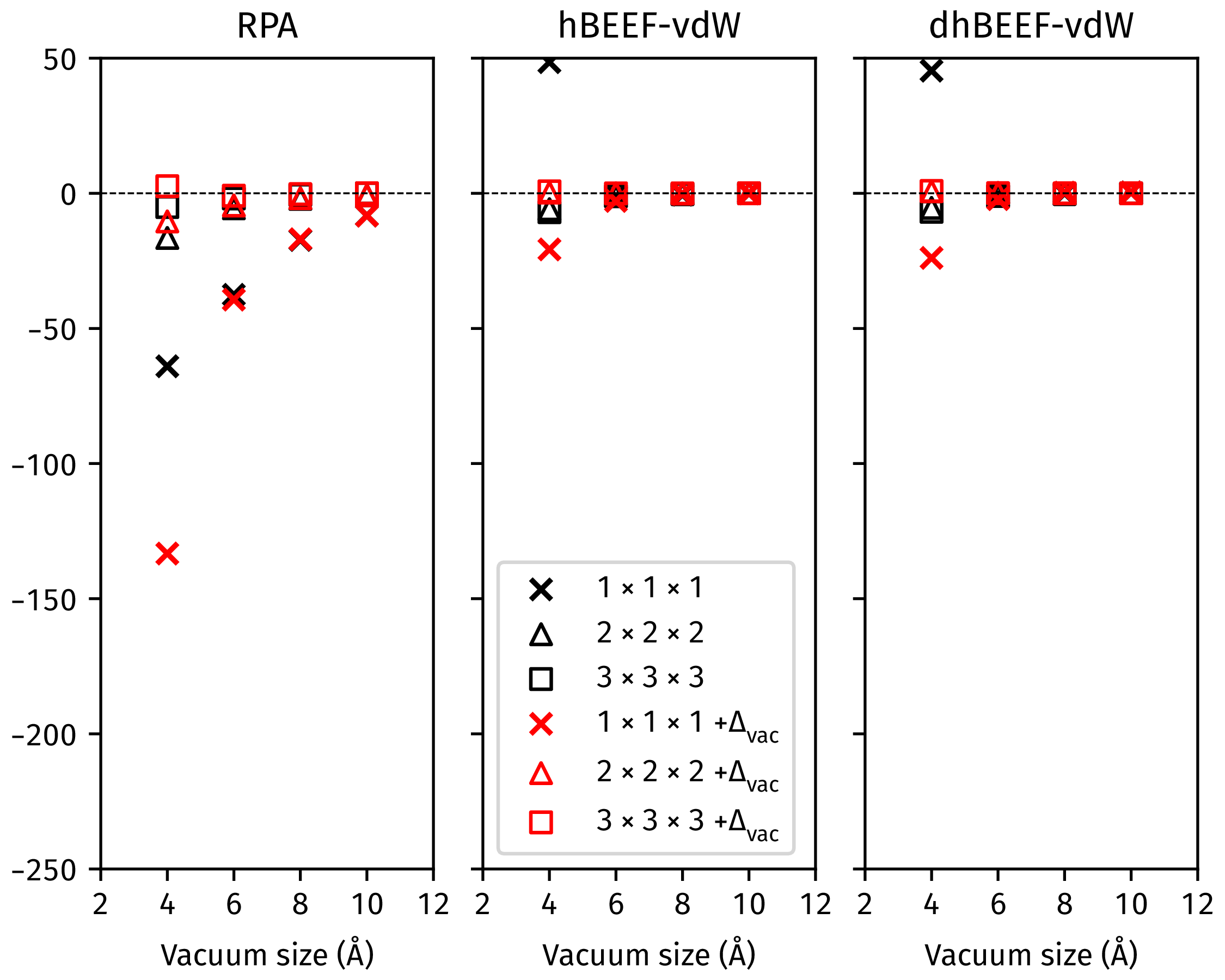}
    \caption{\label{fig:gas_h2o_delta_kpoints_conv}Convergence of RPA, hBEEF-vdW@BEEF-vdW and dhBEEF-vdW@BEEF-vdW as a function of vacuum size for the isolated H$_2$O molecule with and without a $\Delta_\text{vac}$ that adjusts for lateral interactions with BEEF-vdW. We also compare the effect of using denser $k$-point meshes. We used an energy cutoff of \SI{310}{eV} for RPA correlation and \SI{550}{eV} for the BEEF-vdW and EXX components.}
\end{figure}

We test the effect of including $\Delta_\text{vac}$ for the isolated CO and H$_2$O molecules in Figures~\ref{fig:gas_CO_delta_kpoints_conv} and~\ref{fig:gas_h2o_delta_kpoints_conv}, respectively.
For both hBEEF-vdW@BEEF-vdW and dhBEEF-vdW@BEEF-vdW, this correction is highly effective, and allows for errors below \SI{1}{\kjmol} for the $3 \times 3 \times 1$ and $2 \times 2 \times 1$ $k$-point mesh to be achieved at \SI{4}{\AA} of vacuum.
While the error is lowered to \SI{26}{\kjmol} from \SI{91}{\kjmol} for the CO molecule for \SI{4}{\AA}, this becomes less than \SI{3}{\kjmol} for \SI{6}{\AA} of vacuum.
It is important to highlight that there is significant computational savings from this choice because a doubling in the vacuum (e.g., unit cell length from \SI{4}{\AA} to \SI{8}{\AA}) leads to an 8-fold increase in volume and number of plane-waves.
As RPA scales at least cubically with the number of plane-waves~\cite{riemelmoserPlaneWaveBasis2020}, this means that doubling the vacuum can lead to at least a 64-fold increase in the computational cost.
This significant cost decrease and fast convergence both motivate the inclusion of this correction in Equation~4 of the main text.

\clearpage

\subsection{\label{sec:slab_practical}Metallic slab calculation}

\begin{figure}[h]
    \includegraphics[width=\textwidth]{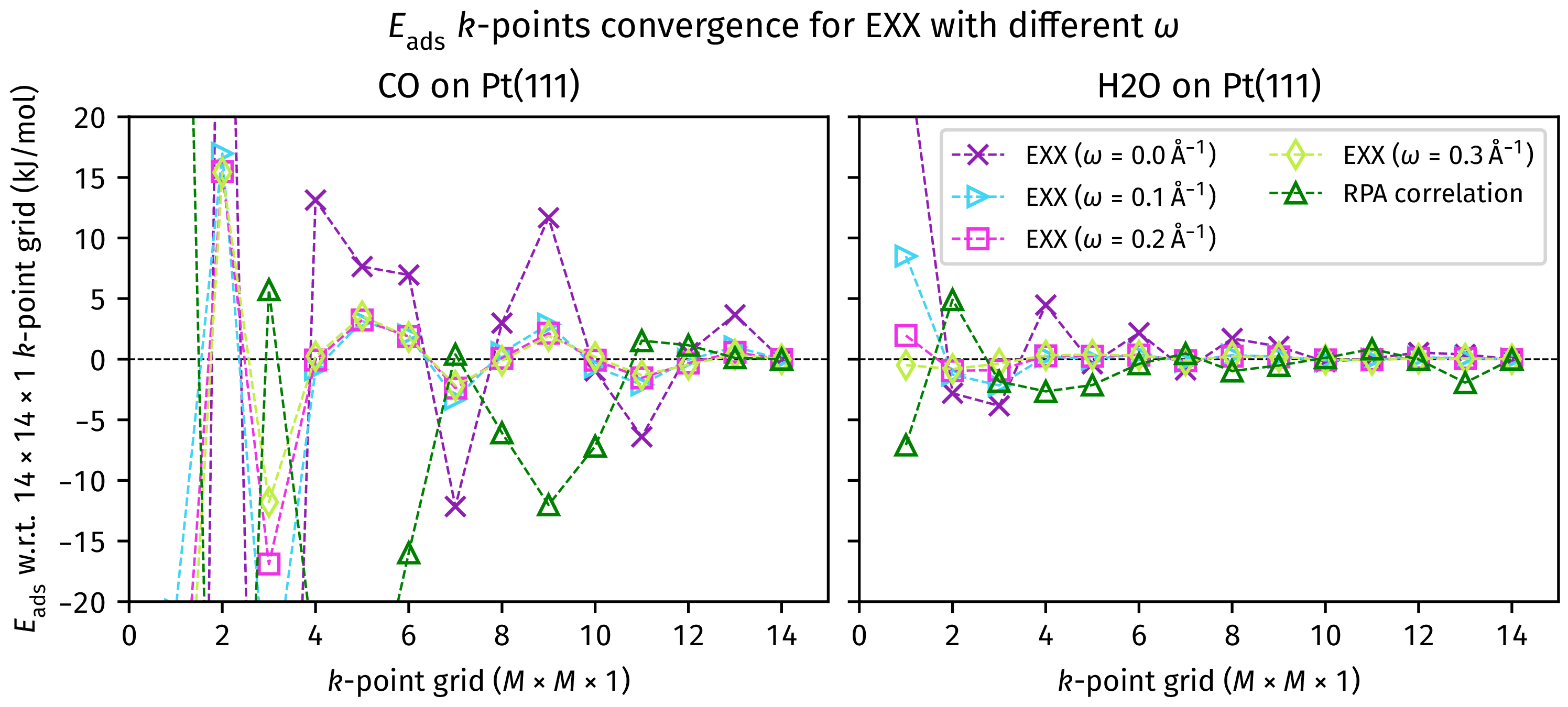}
    \caption{\label{fig:eads_exx_kpoints_conv}Convergence of the adsorption energy of CO (left) and H$_2$O on Pt(111) with $k$-point mesh as a function of global exact exchange (equivalent to range-separation of $\omega=0.0\,$\AA{}$^{-1}$) to a range-separated exact exchange with $\omega=0.3\,$\AA{}$^{-1}$ in steps of $0.1\,$\AA{}$^{-1}$. We used an energy cutoff of \SI{550}{eV} for the EXX components.}
\end{figure}

Besides the isolated gas-phase molecule, there are also well-known challenges with converging metallic calculations, where large numbers of $k$-points are required to account for discontinuous changes in orbital occupation near the Fermi level.
This becomes particularly prominent for the calculation of exact exchange compared to a semilocal DFT exchange, arising from a numerically sharp Coulomb singularity~\cite{paierPerdewBurkeErnzerhof2005} at small momentum transfer.
We demonstrate this in Figure~\ref{fig:eads_exx_kpoints_conv} for the adsorption energy ($E_\text{ads}$) of CO and H$_2$O on Pt(111).
We note here that $E_\text{ads}$ is taken as the definition from Equation~4 of the main text.
We observe that global EXX converges very slowly with $k$-point mesh, particularly for CO on Pt(111) due to its chemisorption nature.
For this system, there are still changes of the order of \SI{5}{\kjmol} between a $13 \times 13 \times 1$ and $14 \times 14 \times 1$ $k$-point mesh.
However, adding range-separation~\cite{heydHybridFunctionalsBased2003a,paierScreenedHybridDensity2006a} resolves these challenges, allowing for a smooth convergence that reaches below \SI{5}{\kjmol} for a $4 \times 4 \times 1$ grid.
We show in Figure~\ref{fig:eads_exx_kpoints_conv} that the RPA correlation energy also converges relatively slowly, reaching below \SI{5}{\kjmol} in convergence only at a $11 \times 11 \times 1$ grid.

\begin{figure}[h]
    \includegraphics[width=\textwidth]{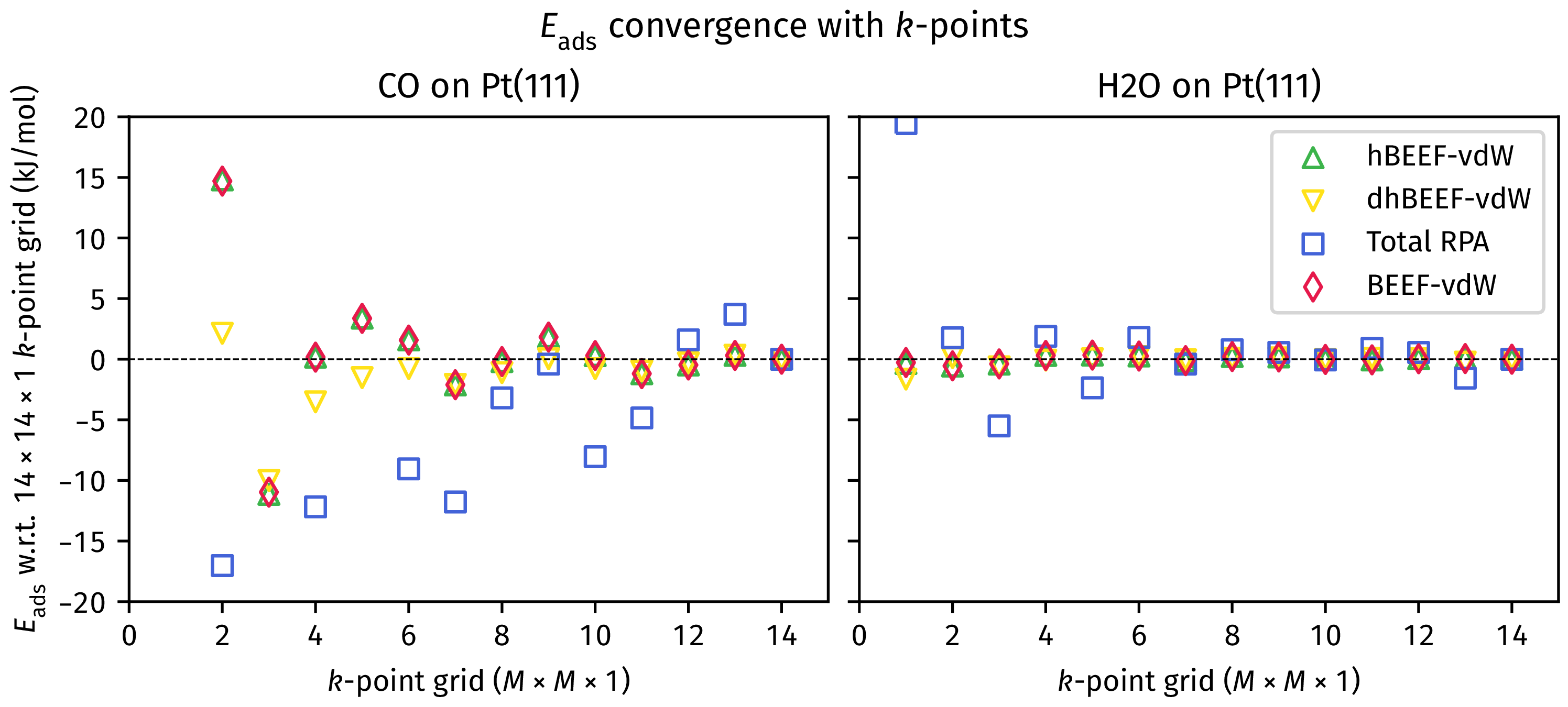}
    \caption{\label{fig:eads_kpoints_conv}Convergence of the adsorption energy of CO (left) and H$_2$O on Pt(111) with $k$-point mesh for BEEF-vdW, hBEEF-vdW@BEEF-vdW, dhBEEF-vdW@BEEF-vdW and RPA. We used an energy cutoff of \SI{310}{eV} for RPA correlation and \SI{550}{eV} for the BEEF-vdW and EXX components.}
\end{figure}

The slow convergence of the global EXX and RPA correlation component also affects the final RPA total energy, which we show in Figure~\ref{fig:eads_kpoints_conv}.
In comparison, BEEF-vdW converges much faster, reaching below \SI{5}{\kjmol} for a $4 \times 4 \times 1$ grid for CO on Pt(111).
We find that the convergence behavior of hBEEF-vdW@BEEF-vdW almost exactly mirrors that of BEEF-vdW.
For dhBEEF-vdW@BEEF-vdW, the convergence is also fast, reaching below \SI{5}{\kjmol} error for a $4 \times 4 \times 1$ grid for CO on Pt(111).
However, its convergence behavior is slightly different from that of BEEF-vdW (and hBEEF-vdW@BEEF-vdW) at earlier $k$-point grids for CO on Pt(111), but start to behave similarly from a $7 \times 7 \times 1$ grid onwards.

\begin{figure}[h]
    \includegraphics[width=\textwidth]{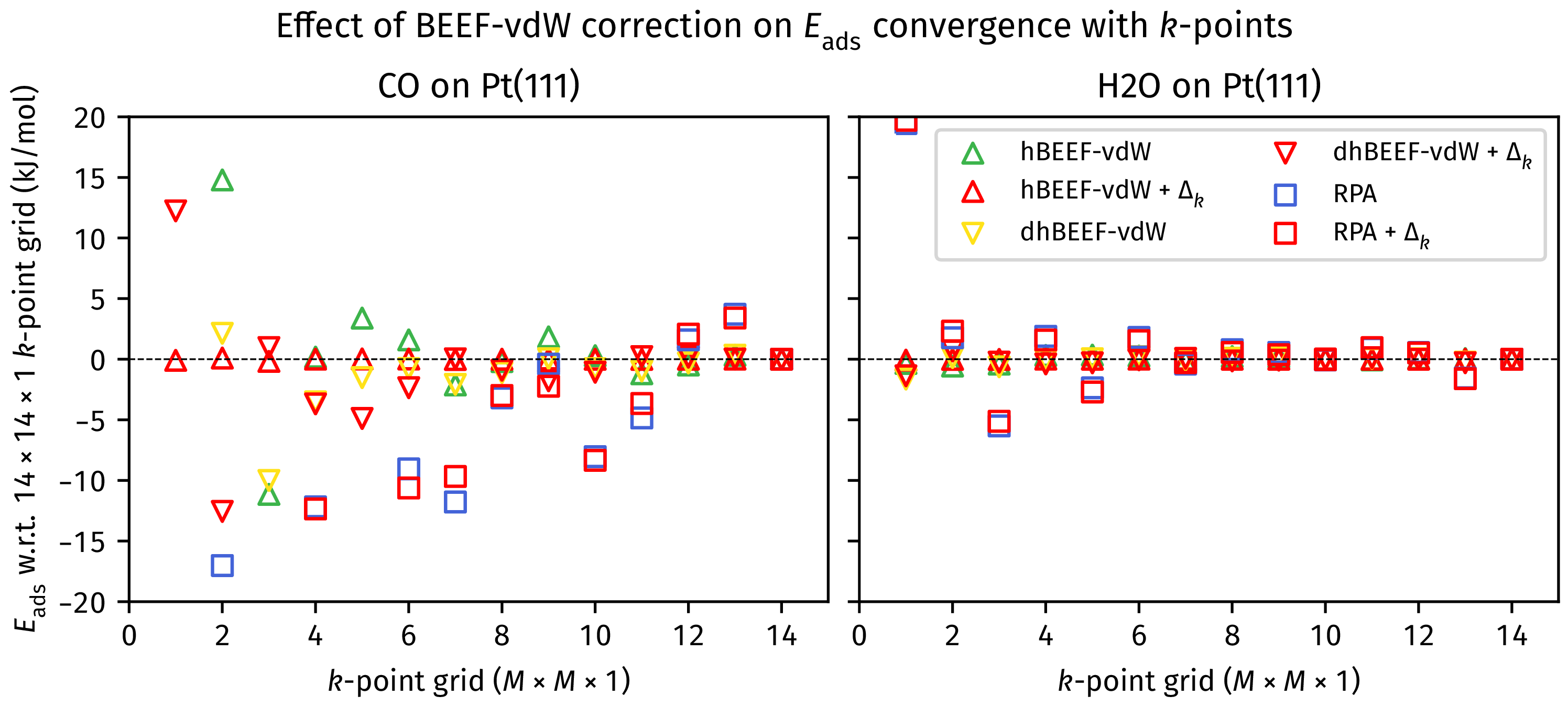}
    \caption{\label{fig:eads_delta_kpoints_conv}Convergence of the adsorption energy of CO (left) and H$_2$O on Pt(111) with $k$-point mesh for hBEEF-vdW@BEEF-vdW, dhBEEF-vdW@BEEF-vdW and RPA. We also plot the same calculations with a $\Delta_k$ correction. We used an energy cutoff of \SI{310}{eV} for RPA correlation and \SI{550}{eV} for the BEEF-vdW and EXX components.}
\end{figure}

Given the very similar $k$-point convergence behavior of hBEEF-vdW@BEEF-vdW and dhBEEF-vdW@BEEF-vdW, we expect that it would be possible to reach the thermodynamic limit (here approximated as a $14 \times 14 \times 1$ grid) from a smaller $M \times M \times 1$ grid through an equation of the form:
\begin{equation}
    E_\text{ads}[\text{hBEEF-vdW}//14 \times 14 \times 1] \approx  E_\text{ads}[\text{hBEEF-vdW}//M \times M \times 1] + \Delta_{k},
\end{equation}
where:
\begin{equation}
    \Delta_{k} = E_\text{ads}[\text{BEEF-vdW} //14 \times 14 \times 1] - E_\text{ads}[\text{BEEF-vdW} //M \times M \times 1].
\end{equation}
We showcase the validity of this protocol in Figure~\ref{fig:eads_delta_kpoints_conv}.
It can be seen that with the $\Delta_k$ correction, hBEEF-vdW@BEEF-vdW can be converged to less than \SI{1}{\kjmol} for CO adsorption on Pt(111) by the single $\Gamma$-point calculation, as opposed to the $12 \times 12 \times 1$ grid without $\Delta_k$.
Similarly, the convergence is faster for dhBEEF-vdW@BEEF-vdW, with errors generally lowered by half when adding $\Delta_k$.
On the other hand, we find that the convergence is not improved when adding $\Delta_k$ to RPA due to their differing convergence behavior.

\subsection{\label{sec:dhrpa_conv}Smaller contribution of RPA in double-hybrids}

\begin{figure}[h]
    \includegraphics[width=\textwidth]{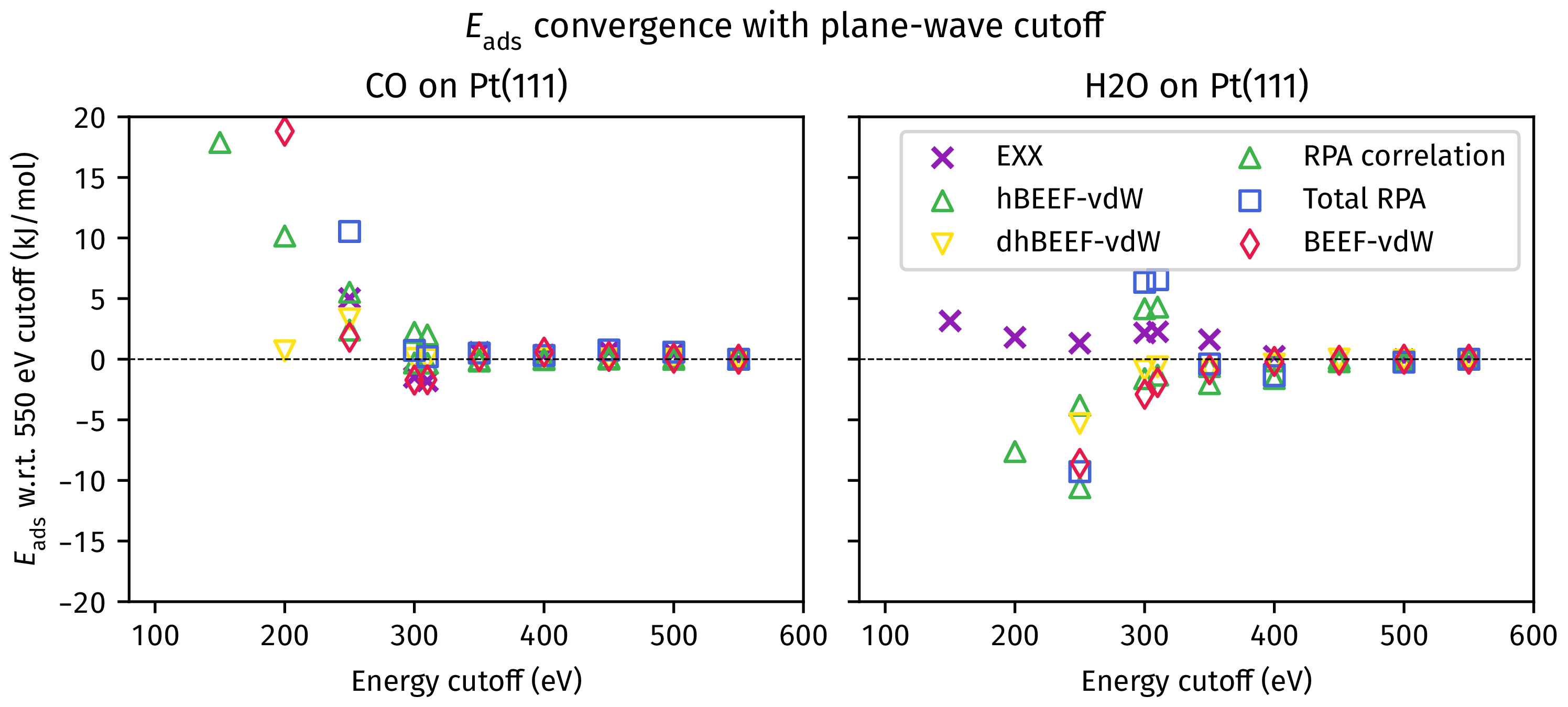}
    \caption{\label{fig:eads_encut_conv}Convergence of the adsorption energy of CO (left) and H$_2$O on Pt(111) with energy cutoff for hBEEF-vdW@BEEF-vdW, dhBEEF-vdW@BEEF-vdW and RPA. We used a $7\times7\times1$ $k$-point mesh. }
\end{figure}

\begin{figure}[h]
    \includegraphics[width=\textwidth]{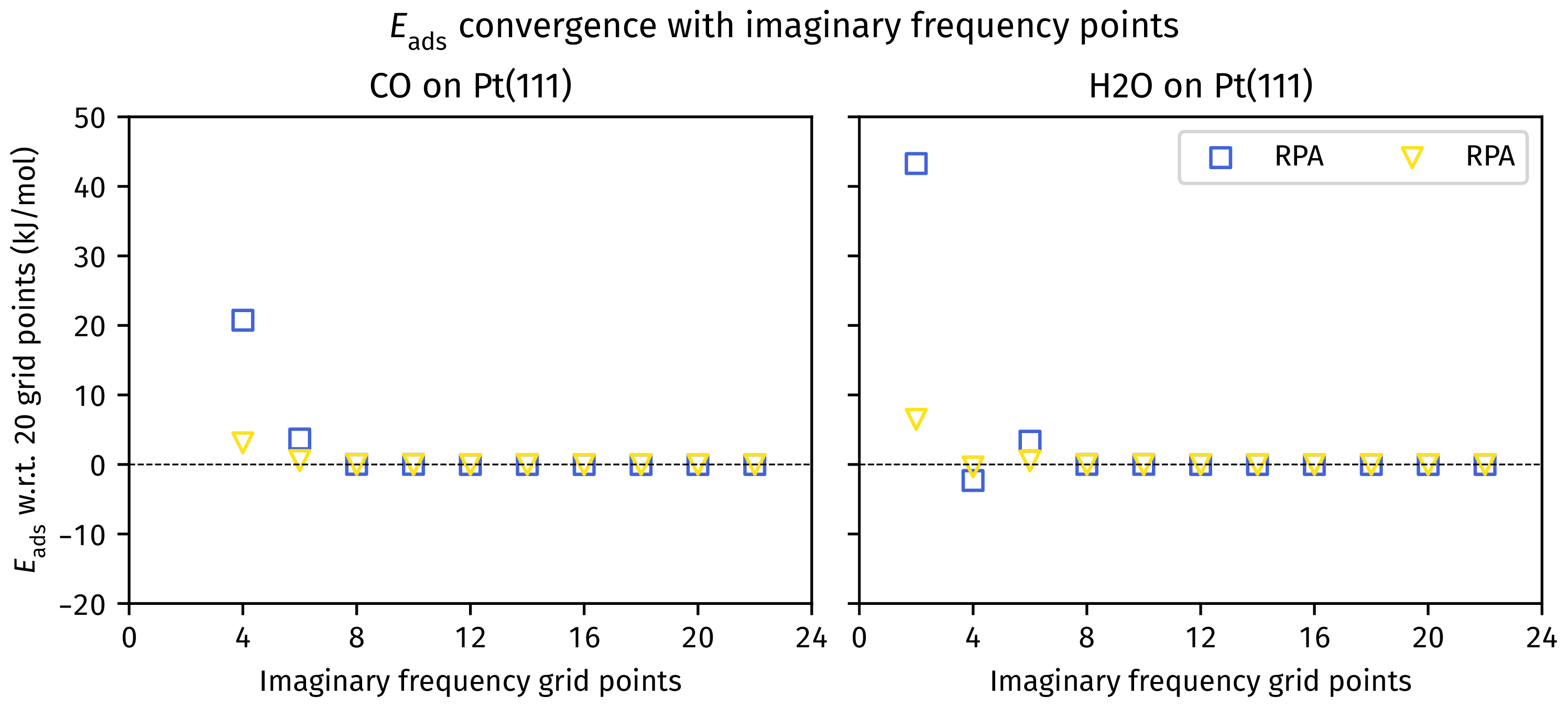}
    \caption{\label{fig:eads_nomega_conv}Convergence of the adsorption energy of CO (left) and H$_2$O on Pt(111) with number of imaginary frequency points in RPA and dhBEEF-vdW@BEEF-vdW. We used a $7\times7\times1$ $k$-point mesh and \SI{310}{eV} energy cutoff.}
\end{figure}

\begin{figure}[h]
    \includegraphics[width=\textwidth]{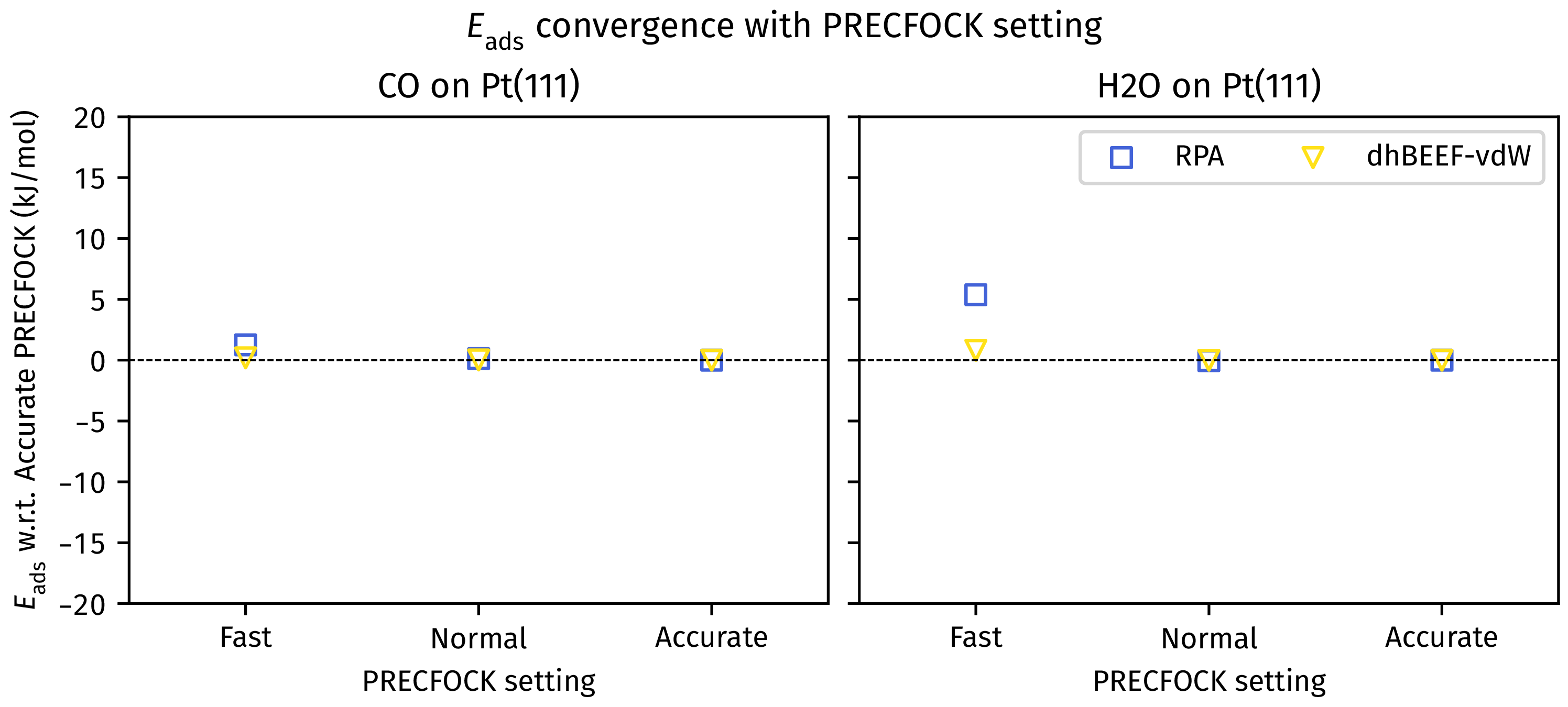}
    \caption{\label{fig:eads_precfock_conv}Convergence of the adsorption energy of CO (left) and H$_2$O on Pt(111) with the fast Fourier transform (FFT) grids used in the exact exchange routines in RPA and dhBEEF-vdW@BEEF-vdW. We used a $7\times7\times1$ $k$-point mesh and \SI{310}{eV} energy cutoff.}
\end{figure}

Besides the $k$-point convergence, we also use the $E_\text{ads}$ of CO and H$_2$O on Pt(111) to highlight other basic convergence parameters in this section.
In Figure~\ref{fig:eads_encut_conv}, we show that there is a relatively smooth convergence with energy cutoff, with hBEEF-vdW@BEEF-vdW and dhBEEF-vdW@BEEF-vdW both converging to below \SI{5}{\kjmol} by our chosen \SI{310}{eV} energy cutoff for both systems.
On the other hand, RPA has not converged to below \SI{5}{\kjmol} at that cutoff for H$_2$O on Pt(111), highlighting the more stringent requirements required for this method.
We also show in Figure~\ref{fig:eads_nomega_conv} that the RPA correlation energy exhibits rapid convergence with respect to the number of imaginary frequency grid points (\texttt{NOMEGA}) employed in the numerical integration over the imaginary frequency axis.
This rapid convergence (achieved by \texttt{NOMEGA}$\,=8$, as opposed to the recommended \texttt{NOMEGA}$\,=16$--24 for metals) probably arises because of error cancellation between the adsorbate-slab complex and metallic slab.
Finally we also consider the convergence as a function of density of the fast Fourier transform (FFT) grids, defined from \texttt{Fast}, \texttt{Normal} to \texttt{Accurate} in order of increasing grid density.
Here, each increase in the FFT grid density is expected to double the execution time and memory requirement.
As can be seen in Figure~\ref{fig:eads_precfock_conv}, for RPA, the use of \texttt{Fast} grids exceeds \SI{5}{\kjmol} error w.r.t.\ the \texttt{Accurate} grids for H$_2$O on Pt(111).
On the other hand, this error is lowered significantly below \SI{1}{\kjmol} when utilized in dhBEEF-vdW@BEEF-vdW.

\clearpage

\section{\label{sec:co_ads}Qualitative improvements on CO adsorption puzzle}

In this section, we discuss briefly the positive impact both exact exchange and RPA correlation have on the the CO adsorption puzzle for Pt(111).
We will aim to highlight why there is the expected improvements for CO on Pt(111) when using hBEEF-vdW@BEEF-vdW by comparing its density of states to BEEF-vdW and GW(RPA) references.

\subsection{\label{sec:exx_effect}Effect of exact exchange}

\begin{figure}[h]
    \includegraphics[width=0.7\textwidth]{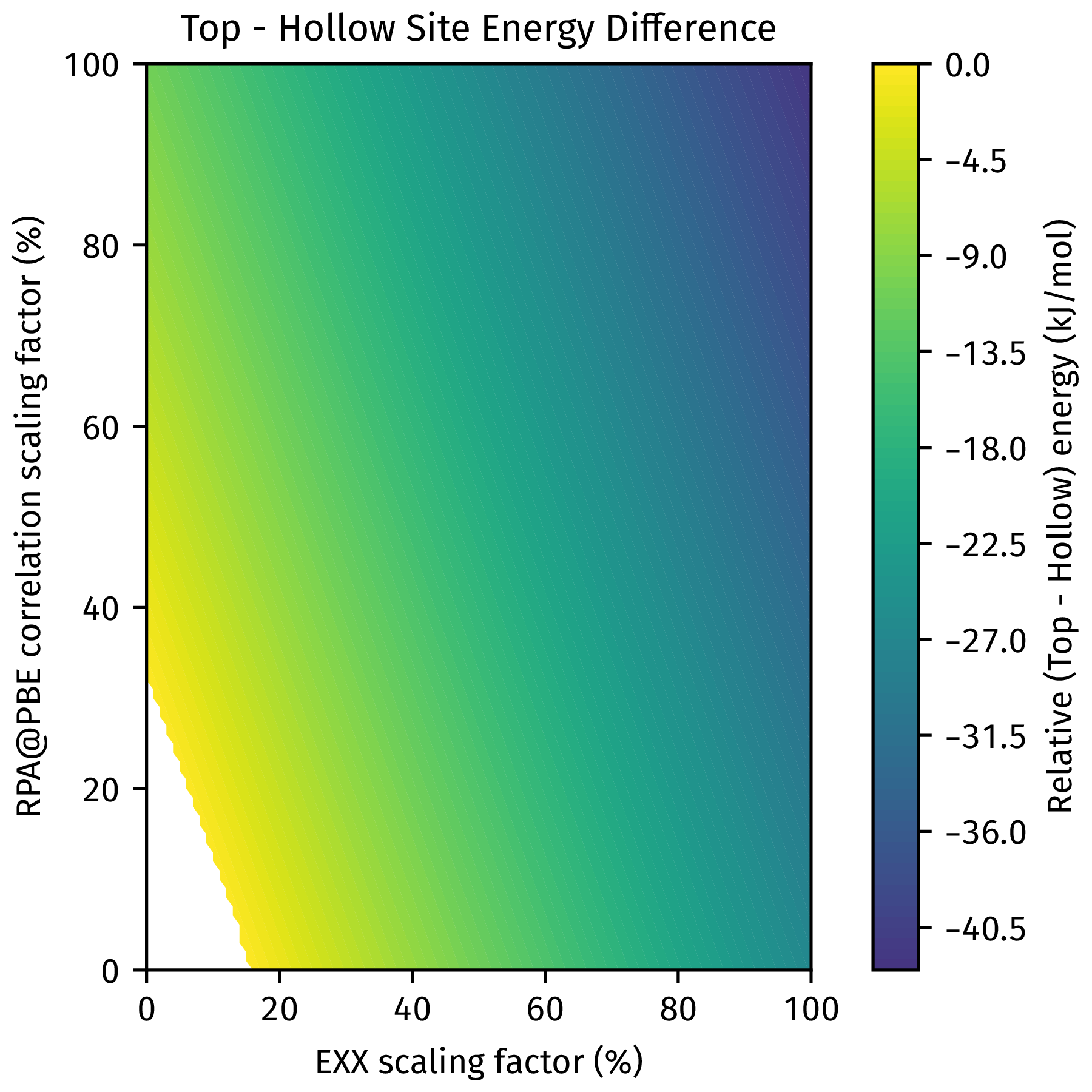}
    \caption{\label{fig:opt_co_ads}Change in relative energy between the top and hollow site for CO adsorbed on Pt(111) as a function of screened exact exchange with $\omega=0.3\,$\AA{}$^{-1}$ and RPA correlation using PBE orbitals. Positive values (i.e., where the hollow site is more stable) are given as white.}
\end{figure}

In Figure~\ref{fig:opt_co_ads}, we show the effect of exact exchange (screened with $\omega=0.3\,$\AA{}$^{-1}$) and RPA correlation energy on the relative energy between the top and hollow sites for CO on Pt(111).
Here, the effect of both EXX and RPA correlation is to increase the stability of the top site.
This is not unexpected given the preference~\cite{schimkaAccurateSurfaceAdsorption2010} of RPA@PBE for the top site.

The most interesting result is that without RPA@PBE correlation, we can stabilize the top site in hBEEF-vdW@BEEF-vdW (corresponding to the x-axis of Figure~\ref{fig:opt_co_ads}).
There is a strong dependence on the amount of EXX, which pushes the stability to the top site at ${\sim}15\,$\% or beyond.
As discussed in the main text, standard self-consistent hybrid DFAs do not stabilize the top site as the de-stablization of the hollow-site from decreased electron back-donation is counteracted by stabilization from decreased steric repulsion.
It arises because self-consistent hybrids localize electrons on the metal atoms much more compared to semilocal DFAs~\cite{stroppaShortcomingsSemilocalHybrid2008}.
We do not have this competing effect within this work as the electron density is kept at the level of the GGA BEEF-vdW.
In addition, as discussed in the next section, the stabilizing effect from decreased back-donation is maintained with hBEEF-vdW@BEEF-vdW.

\subsection{Changes in the density of states for CO on Pt(111)}

The relative stability of the top and hollow CO adsorption sites is often attributed to a competition between several effects~\cite{grinbergCOPt111Puzzle2002,olsenCOPt111Puzzle2003,kresseSignificanceSingleelectronEnergies2003,masonFirstprinciplesExtrapolationMethod2004a,stroppaShortcomingsSemilocalHybrid2008,alaeiCOPt111GGA2008,lazicDensityFunctionalTheory2010,janthonAddingPiecesCO2017,patraRethinkingCOAdsorption2019a}.
The top site is stabilized~\cite{wangSuccessfulPrioriModeling2007a} by $\sigma$ bonds that form between the CO molecule and metal surface due to donation of electrons from the 5$\sigma$ orbital to the empty metal states (predominantly $dz^2$).
The hollow-site is stabilized by back-donation from the metal $d$ orbitals into the anti-bonding 2$\pi^*$ orbitals of the CO molecule, weakening the C--O bond while strengthening the bond between the CO molecule and surrounding three metal atoms.
For this site, due to the close proximity of the CO molecule to several metal atoms, the effect of steric repulsion is also important.

Semilocal DFAs (such as LDA, GGAs and metaGGAs) have a tendency to underestimate the HOMO-LUMO gaps or bandgaps of molecules and materials, respectively.
This tends to place the 2$\pi^*$ orbitals of the CO molecule closer to the Fermi level, thus strengthening the bonding of the hollow-site, hence unphysically favoring the hybrid sites.

\begin{figure}[h]
    \includegraphics[width=0.7\textwidth]{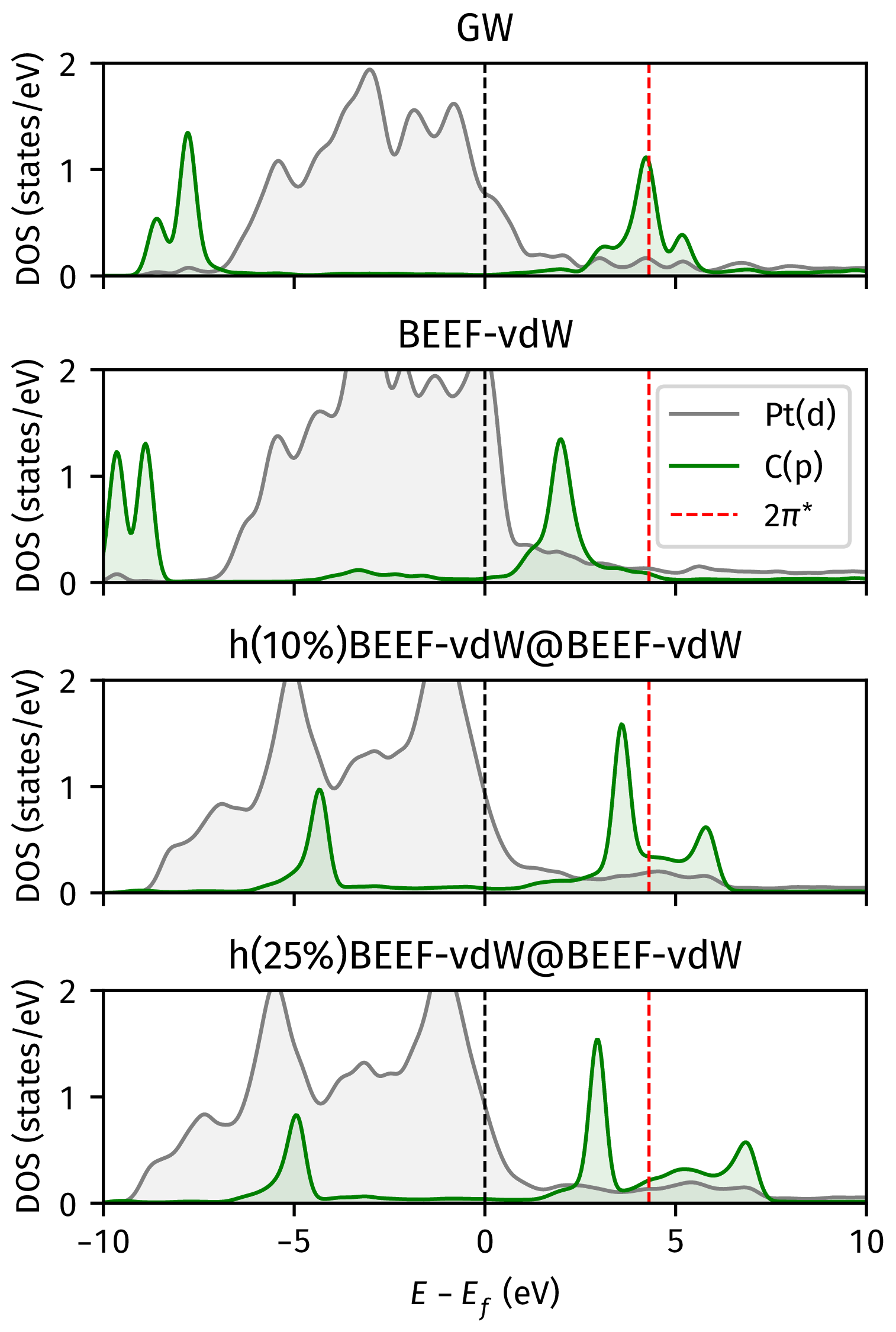}
    \caption{\label{fig:co_ads_dos}Comparison of the electronic density-of-states (DOS) predicted by GW(RPA), BEEF-vdW and hBEEF-vdW@BEEF-vdW. For hBEEF-vdW@BEEF-vdW, the amount of exact-exchange has been varied from 10\% to 25\%. Here, global exact exchange is used in hBEEF-vdW@BEEF-vdW as it was not possible to obtain the DOS when using screened exact exchange. The DOS has been projected on the Pt $d$ orbitals (subsequently normalized by the number of Pt atoms) and on the C $p$ orbitals. The Fermi level is highlighted with a dotted black line while the position of the 2$\pi^*$ orbital from GW(RPA) on the CO atom is indicated with the red line.}
\end{figure}

The effect of hybrid functionals is to shift the 2$\pi^*$ orbitals further above the Fermi level~\cite{stroppaCOAdsorptionMetal2007a,wangSuccessfulPrioriModeling2007a}.
This decreases the amount of back-donation, hence weakening the bond to the hollow site and disfavoring the hollow site.
However, it also localizes the electron density on the metal atoms, and this in turn decreases the amount of steric repulsion they give, favoring the hollow site.
\citet{stroppaShortcomingsSemilocalHybrid2008} showed that while this favors the top site for Cu(111) and Rh(111), it remains unable to resolve this for Pt(111).
They concluded that the ``inclusion of 25\% non-local exchange, bare or long-range screened, hardly improves the description of CO adsorption on metal surfaces.''

The use of our hybrid NSC-DFAs resolves this issue because it uses an electron density coming from a semilocal DFA, hence preventing the localization of electron density on metal atoms.
This means that the main effect is to the weaken the bond between the CO molecule and metal surface for the hollow site.
In Figure~\ref{fig:co_ads_dos}, we show that the position of the 2$\pi^*$ orbital moves upwards as a function of non-local exchange in the system, explaining the steady stabilization of the top site as a function of EXX \%.
While BEEF-vdW by itself predicts the 2$\pi^*$ position to be below that from GW(RPA), the inclusion of exact exchange in hBEEF-vdW@BEEF-vdW brings its position closer to that of the 2$\pi^*$ position.
We note that there remain some qualitative differences in the features of DOS, such as broadening of the $d$-band and also 2$\pi^*$ orbitals.

\subsubsection{Computational details for GW calculations}

The DOS calculations were performed in VASP~\cite{kresseEfficientIterativeSchemes1996a,kresseUltrasoftPseudopotentialsProjector1999b,kresseEfficiencyAbinitioTotal1996a}.
We used the same electronic structure parameters as given in the Methods section of the main text.
For BEEF-vdW, following a self-consistent calculation, an exact diagonalization (\texttt{ALGO = Exact}) is performed, before obtaining the relevant DOS, setting the number of bands to 400 (for this and all DOS calculations).
The hBEEF-vdW@BEEF-vdW DOS required a subsequent recalculation of the one-electron energies, \texttt{ALGO = Eigenval}, from the BEEF-vdW orbitals.
The GW DOS was also obtained in a similar manner, where we used the low-scaling G0W0 algorithm (\texttt{ALGO = EVGW0R} with \texttt{NELMGW = 1}) to update the one-electron energies.
The GW calculations were performed with similar settings to RPA (given in the main text), notably a $310\,$eV energy cutoff with a $12\times12\times1$ $k$-point mesh.

\begin{figure}[h]
    \includegraphics[width=0.7\textwidth]{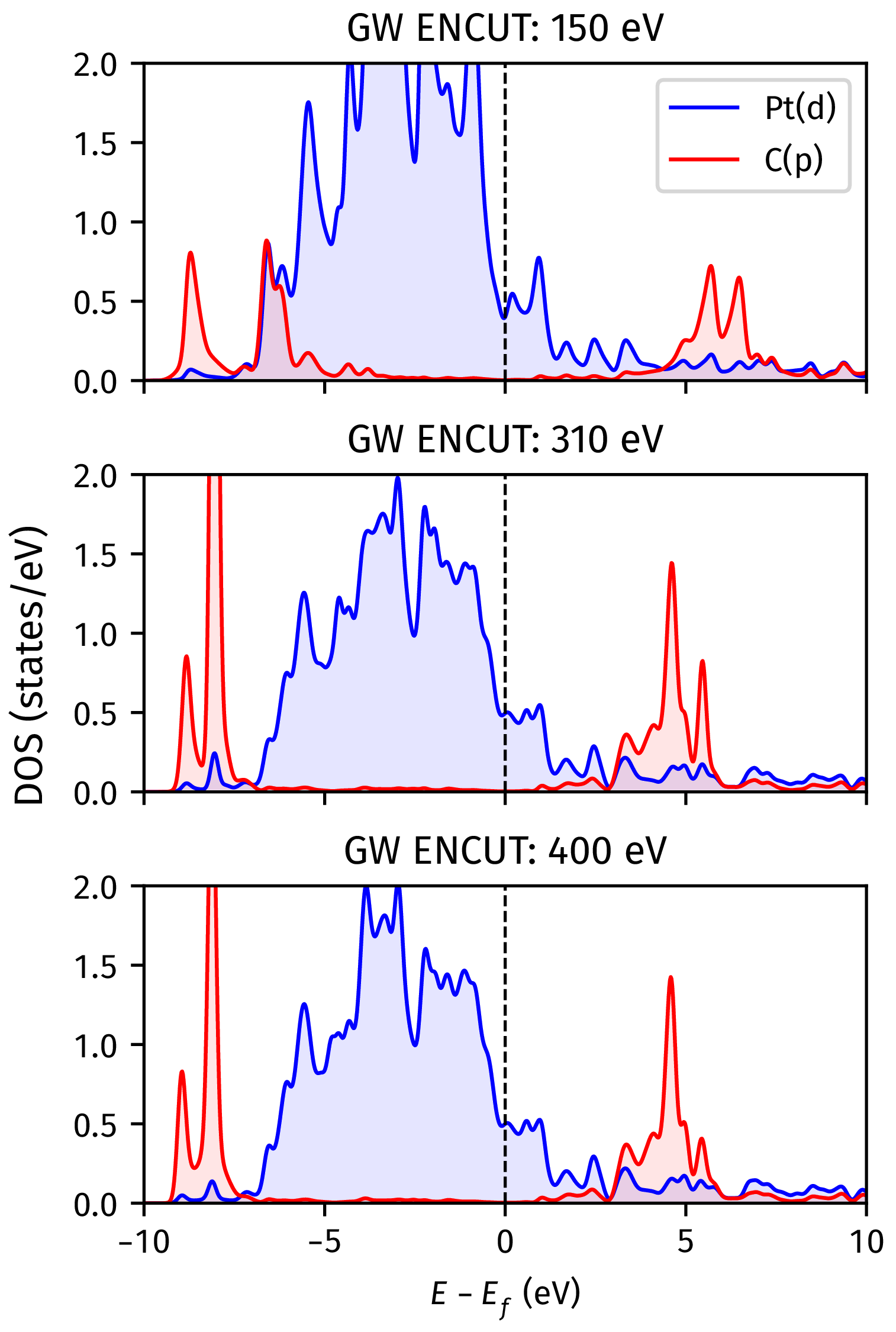}
    \caption{\label{fig:dos_encut_conv}Convergence of the GW(RPA) density of states as a function of energy cutoff for CO adsorbed on the on-top site of a three-layer Pt(111) slab.}
\end{figure}

\begin{figure}[h]
    \includegraphics[width=0.7\textwidth]{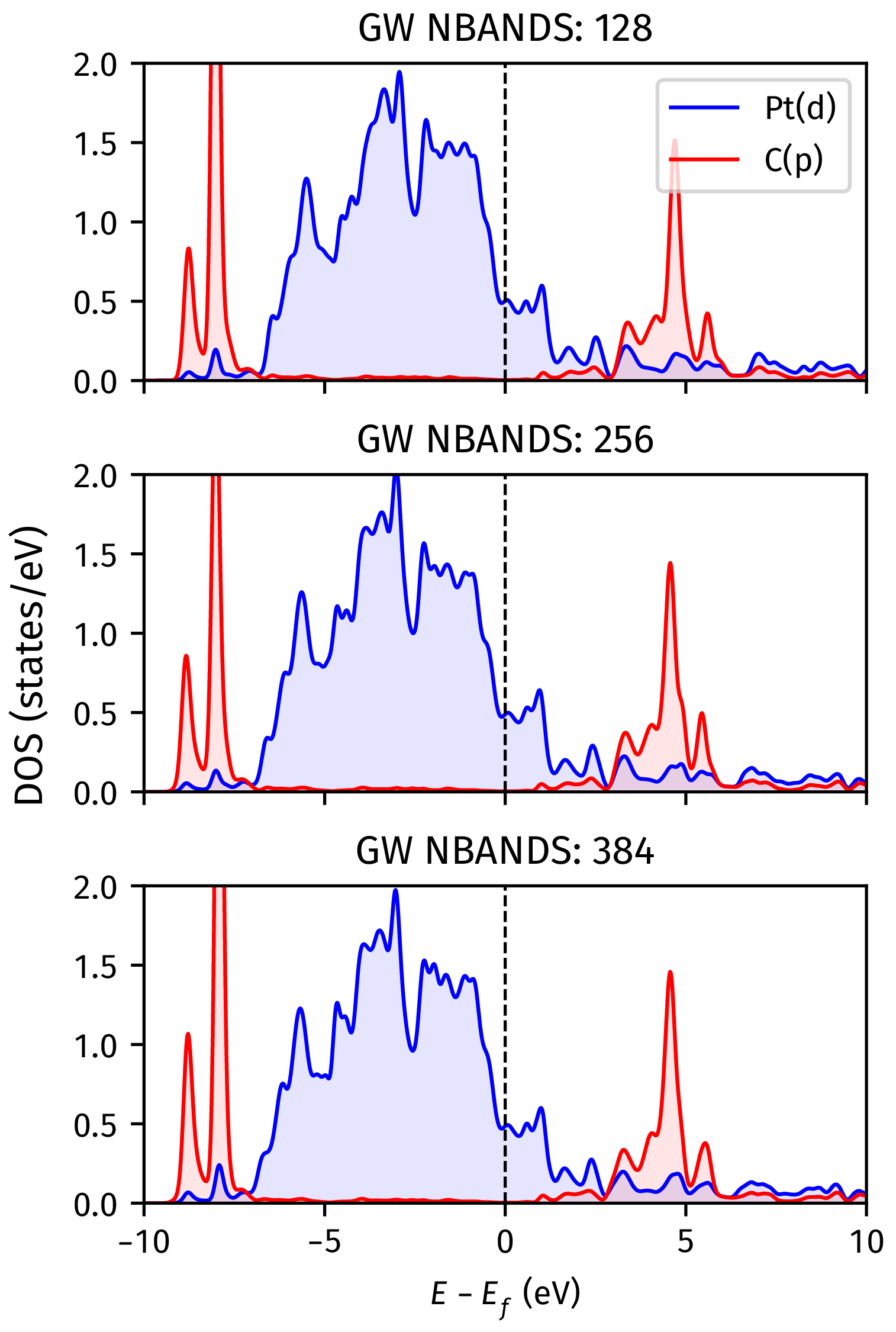}
    \caption{\label{fig:dos_nbands_conv}Convergence of the GW(RPA) density of states as a function of number of one-electron bands (in the GW calculation) for CO adsorbed on the on-top site of a three-layer Pt(111) slab.}
\end{figure}

We have validated our choice of the above parameters as part of this work. In Figure~\ref{fig:dos_encut_conv}, we show that the key features/peaks of the GW(RPA) DOS are already converged by $310\,$eV.
Similarly, we show in in Figure~\ref{fig:dos_encut_conv} that the peaks are already converged by a $9\times9\times1$ $k$-point mesh and 256 (one-electron orbital) bands in Figure~\ref{fig:dos_nbands_conv}.

\begin{figure}[h]
    \includegraphics[width=0.7\textwidth]{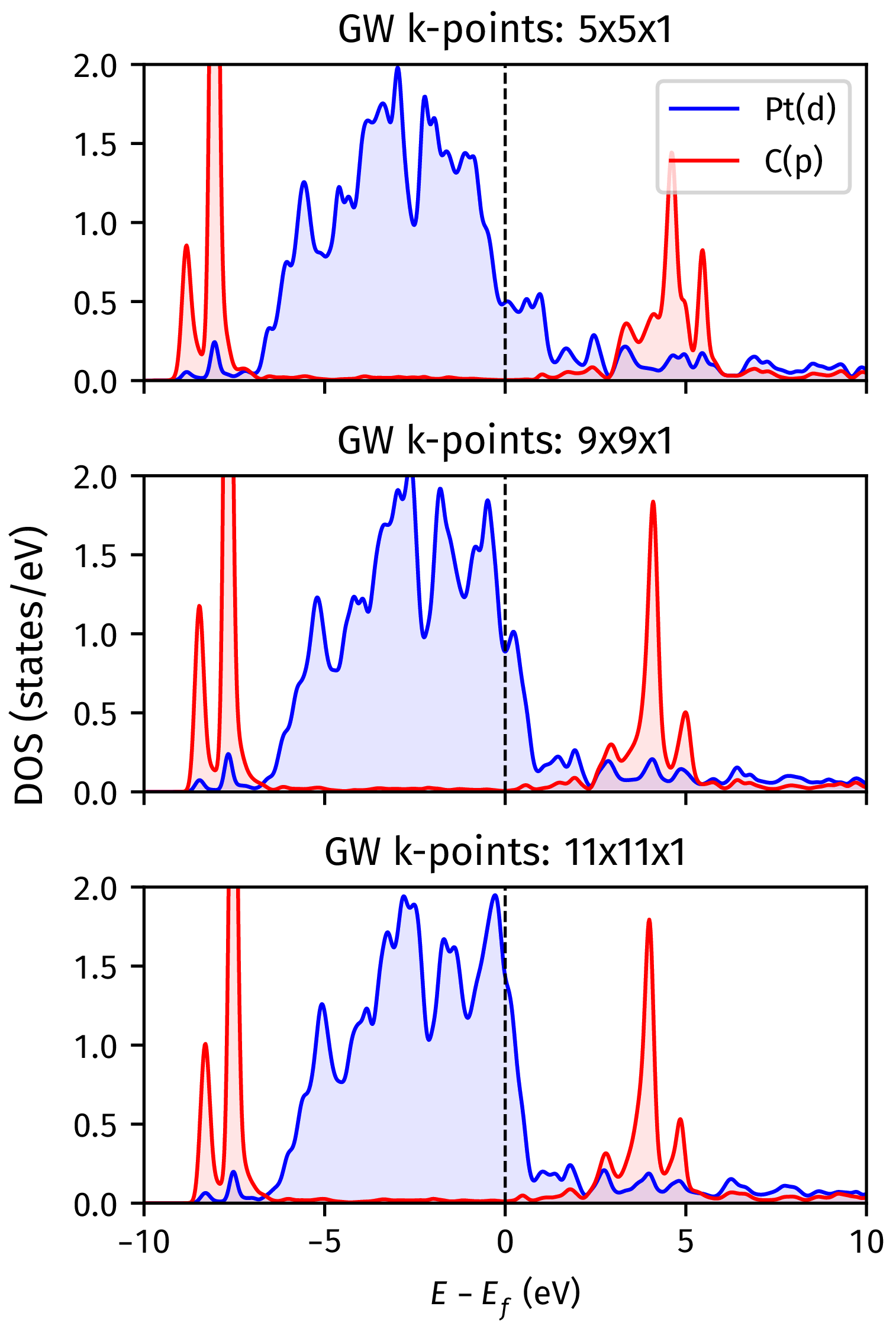}
    \caption{\label{fig:dos_kpoint_conv}Convergence of the GW(RPA) density of states as a function of $k$-point mesh for CO adsorbed on the on-top site of a three-layer Pt(111) slab.}
\end{figure}

\clearpage 

\section{\label{sec:graphene_dft}Challenges with graphene adsorption on Ni(111) for DFAs}

\begin{figure}[h]
    \includegraphics[width=\textwidth]{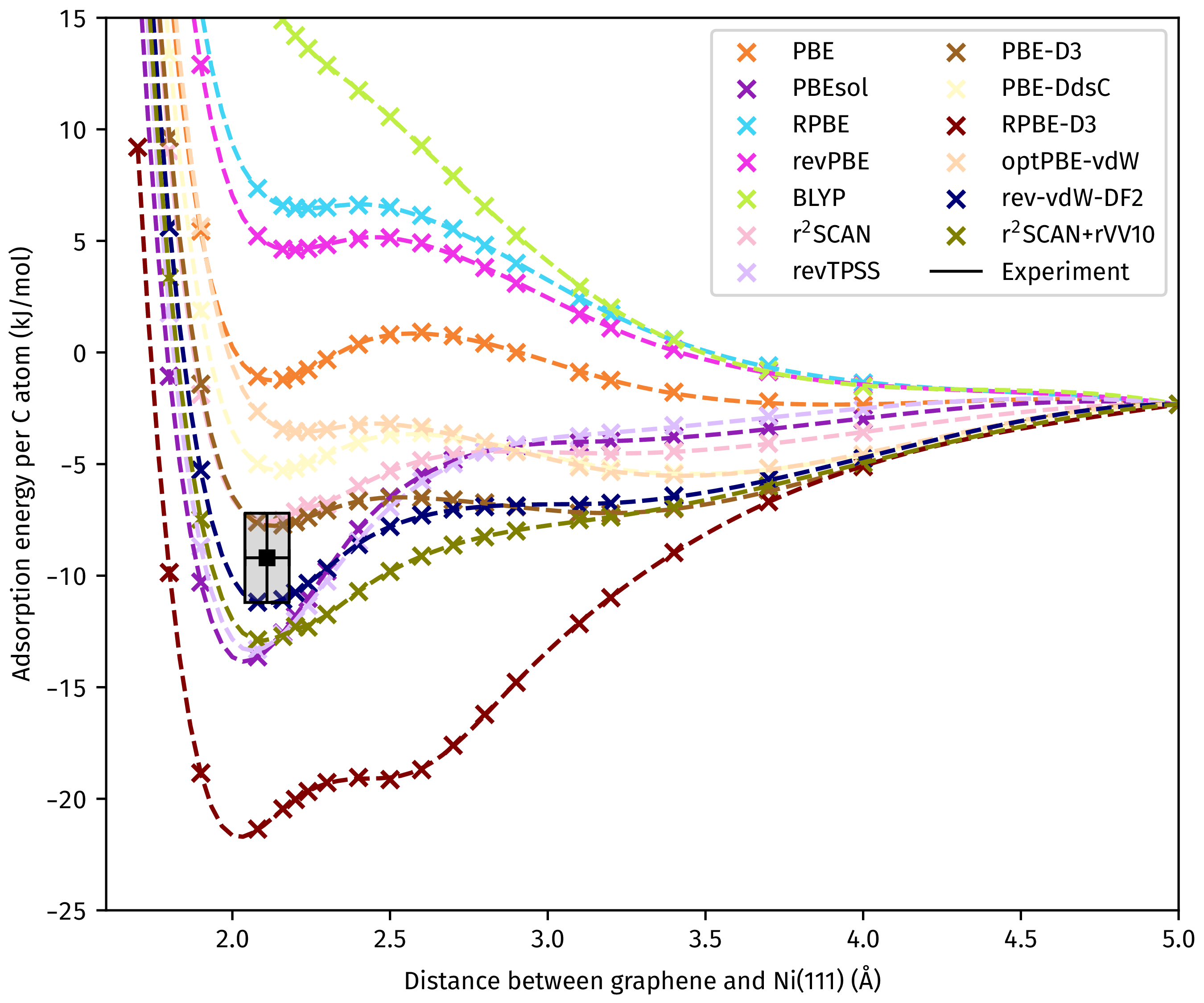}
    \caption{\label{fig:gr_ni111_compare}Comparison of adsorption energy curve of graphene on Ni(111) for several density functional approximations.}
\end{figure}

Within the main text, we have only focused on a small selection of density functional approximations (DFAs).
Here, we highlight a more comprehensive comparison between multiple DFAs in Figure~\ref{fig:gr_ni111_compare}.
It highlights the importance of going beyond non-dispersion-corrected GGAs to stabilize the chemisorption configuration, where most favor a physisorbed configuration.
Upon moving towards meta-GGAs or adding dispersion correction, we find that the chemisorbed state becomes the more stable configuration.
However, achieving experimental agreement remains challenging and only 3 DFAs out of the studied 13 lie within the experimental range.

\clearpage

\section{\label{sec:ct_barrier}Effect of charge transfer on barrier heights}

\begin{figure}[h]
    \includegraphics[width=\textwidth]{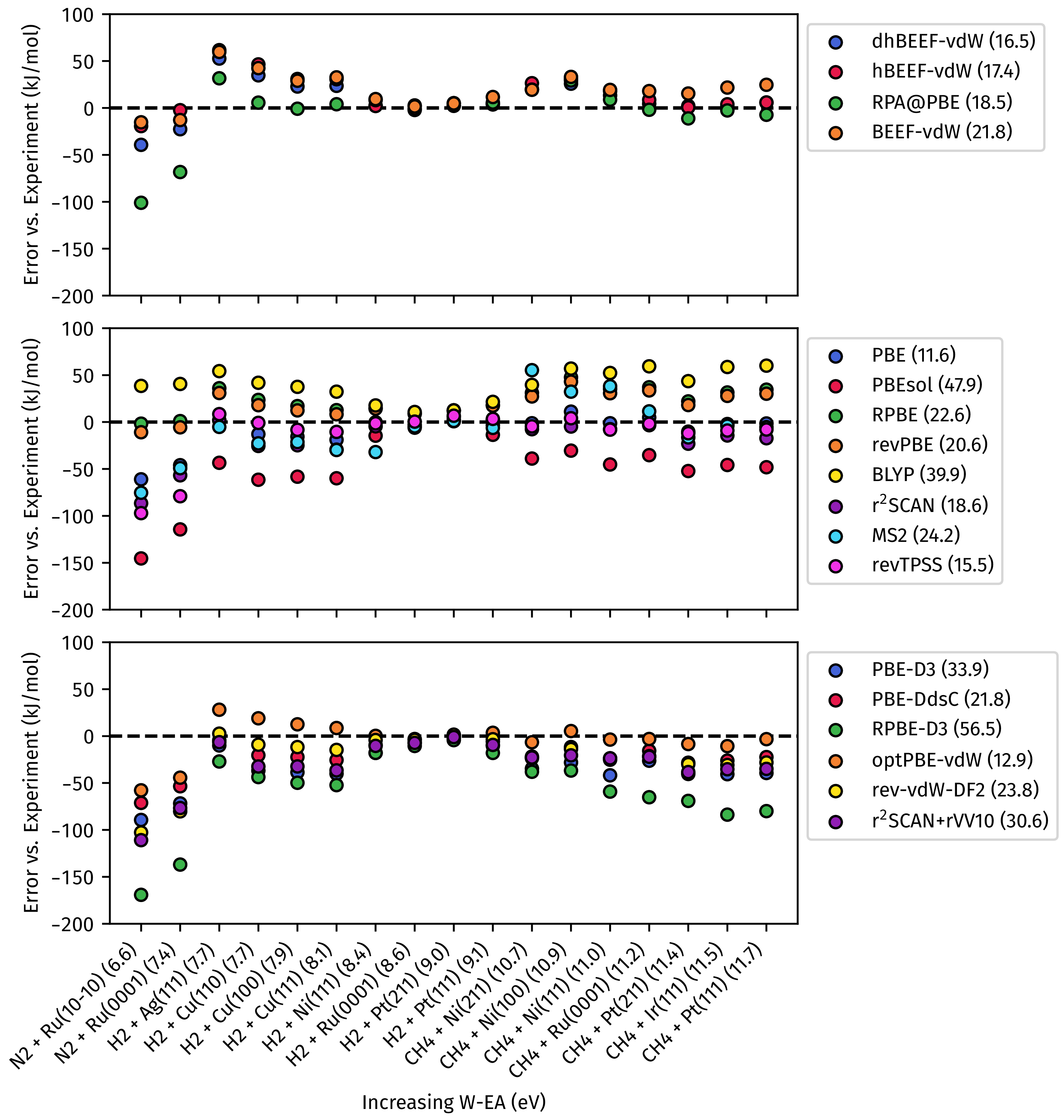}
    \caption{\label{fig:ct_sbh17}Comparison in the errors w.r.t.\ experiments for barrier heights of our NSC-DFAs, RPA (in the top panel) and several DFT functionals in the middle and bottom panels. We give the mean absolute deviation across all systems in the legends given to the right. The systems are ordered based upon the difference of the work function of the metal surface and the electron affinity (value given in parenthesis) of the molecule along the $x$-axis.}
\end{figure}

In Figure~\ref{fig:ct_sbh17}, we assess the errors in the SBH17 dataset as a function of the degree of charge transfer from the metal to the molecule.
We use the difference of the metal work function (W) and the molecule's electron affinity (EA), resulting in W$-$EA, to measure the degree of charge transfer (taken from Ref.~\citenum{gerritsDensityFunctionalTheory2020b}).
A smaller value indicates more significant charge transfer between the molecule and surface, since it either results in a larger EA, meaning the molecule is more likely to gain an electron, or smaller W, where the surface is more likely to give up an electron.

The most challenging systems, involving the highest degree of charge transfer, are the barrier heights for \ce{N2} on the Ru surfaces.
It can be seen that few DFAs can get those systems correct.
Notably, the dispersion-corrected DFAs in the bottom panel of Figure~\ref{fig:ct_sbh17} all suffer average errors that exceed \SI{50}{\kjmol}.
On the other hand, the non-dispersion-corrected DFAs in the middle panel show a wide spread of almost \SI{200}{\kjmol} in the predicted barrier heights.
These high errors of \SI{84.7}{\kjmol} on average for the two \ce{N2} on the Ru barriers are also present in RPA@PBE in the top panel.
In contrast, both hBEEF-vdW@BEEF-vdW and BEEF-vdW perform well for these systems, with average errors of \SI{10.9}{\kjmol}.
Our double-hybrid NSC-DFA, dhBEEF-vdW@BEEF-vdW, also rectifies some of the  errors in RPA@PBE for these two systems, lowering errors below \SI{31.0}{\kjmol} on average.

\clearpage
\section{Timing ratios between RPA, EXX and DFT}

In Table~\ref{tab:ce39_cost_ratio}, we highlight the relative costs to perform the BEEF-vdW calculation (Step 1 in Section~\ref{sec:workflow}), EXX calculation (Step 4) and RPA correlation calculation (Step 8), representing the most costly steps in the hBEEF-vdW@BEEF-vdW and dhBEEF-vdW@BEEF-vdW calculations.
It can be seen that the EXX calculation is on average $20\,\times$ more expensive than the BEEF-vdW calculation, while the RPA correlation calculation is around $2\,\times$ the cost of the EXX calculation.
Both are cost efficient, especially given the fact that a normal self-consistent hybrid DFA would cost ${\sim}50\,\times$ the cost of the hBEEF-vdW@BEEF-vdW since they may require that number of self-consistent field cycles to reach convergence.
\clearpage

\begin{longtable}{lrrr}
\caption{\label{tab:ce39_cost_ratio}Ratio between the computational cost of EXX and RPA calculations compared to BEEF-vdW for the CE39 dataset.} \\
\toprule
 & BEEF-vdW & EXX & RPA \\ 
\endfirsthead

\caption[]{(continued)} \\
\endhead

\multicolumn{4}{r}{{Continued on next page}} \\
\endfoot

\bottomrule
\endlastfoot

Reaction &  &  &  \\
\midrule
\ce{CO + Ni(111) $\rightarrow$ CO/Ni(111)} & 1 & 18 & 24 \\
\ce{CO + Pt(111) $\rightarrow$ CO/Pt(111)} & 1 & 23 & 36 \\
\ce{CO + Pd(111) $\rightarrow$ CO/Pd(111)} & 1 & 25 & 45 \\
\ce{CO + Pd(100) $\rightarrow$ CO/Pd(100)} & 1 & 27 & 56 \\
\ce{CO + Rh(111) $\rightarrow$ CO/Rh(111)} & 1 & 18 & 52 \\
\ce{CO + Ir(111) $\rightarrow$ CO/Ir(111)} & 1 & 15 & 29 \\
\ce{CO + Cu(111) $\rightarrow$ CO/Cu(111)} & 1 & 22 & 35 \\
\ce{CO + Ru(001) $\rightarrow$ CO/Ru(001)} & 1 & 14 & 73 \\
\ce{CO + Co(001) $\rightarrow$ CO/Co(001)} & 1 & 18 & 24 \\
\ce{NO + Ni(100) $\rightarrow$ N/Ni(100) + O/Ni(100)} & 1 & 16 & 28 \\
\ce{NO + Pt(111) $\rightarrow$ NO/Pt(111)} & 1 & 18 & 33 \\
\ce{NO + Pd(111) $\rightarrow$ NO/Pd(111)} & 1 & 22 & 43 \\
\ce{NO + Pd(100) $\rightarrow$ NO/Pd(100)} & 1 & 25 & 58 \\
\ce{O2 + Ni(111) $\rightarrow$ 2O/Ni(111)} & 1 & 17 & 26 \\
\ce{O2 + Ni(100) $\rightarrow$ 2O/Ni(100)} & 1 & 16 & 28 \\
\ce{O2 + Pt(111) $\rightarrow$ 2O/Pt(111)} & 1 & 16 & 30 \\
\ce{O2 + Rh(100) $\rightarrow$ 2O/Rh(100)} & 1 & 16 & 66 \\
\ce{H2 + Pt(111) $\rightarrow$ 2H/Pt(111)} & 1 & 24 & 45 \\
\ce{H2 + Ni(111) $\rightarrow$ 2H/Ni(111)} & 1 & 17 & 27 \\
\ce{H2 + Ni(100) $\rightarrow$ 2H/Ni(100)} & 1 & 16 & 29 \\
\ce{H2 + Rh(111) $\rightarrow$ 2H/Rh(111)} & 1 & 18 & 63 \\
\ce{H2 + Pd(111) $\rightarrow$ 2H/Pd(111)} & 1 & 23 & 50 \\
\ce{I2 + Pt(111) $\rightarrow$ 2I/Pt(111)} & 1 & 24 & 38 \\
\ce{CH2I2 + Pt(111) $\rightarrow$ CH/Pt(111) + H/Pt(111) + 2I/Pt(111)} & 1 & 24 & 44 \\
\ce{CH3I + Pt(111) $\rightarrow$ CH3/Pt(111) + I/Pt(111)} & 1 & 24 & 44 \\
\ce{NH3 + Cu(100) $\rightarrow$ NH3/Cu(100)} & 1 & 20 & 33 \\
\ce{CH3I + Pt(111) $\rightarrow$ CH3I/Pt(111)} & 1 & 23 & 36 \\
\ce{CH3OH + Pt(111) $\rightarrow$ CH3OH/Pt(111)} & 1 & 23 & 42 \\
\ce{CH4 + Pt(111) $\rightarrow$ CH4/Pt(111)} & 1 & 23 & 43 \\
\ce{C2H6 + Pt(111) $\rightarrow$ C2H6/Pt(111)} & 1 & 17 & 31 \\
\ce{C3H8 + Pt(111) $\rightarrow$ C3H8/Pt(111)} & 1 & 18 & 32 \\
\ce{C4H10 + Pt(111) $\rightarrow$ C4H10/Pt(111)} & 1 & 18 & 33 \\
\ce{C6H6 + Pt(111) $\rightarrow$ C6H6/Pt(111)} & 1 & 16 & 28 \\
\ce{C6H6 + Cu(111) $\rightarrow$ C6H6/Cu(111)} & 1 & 25 & 29 \\
\ce{C6H6 + Ag(111) $\rightarrow$ C6H6/Ag(111)} & 1 & 34 & 53 \\
\ce{C6H6 + Au(111) $\rightarrow$ C6H6/Au(111)} & 1 & 31 & 44 \\
\ce{C6H10 + Pt(111) $\rightarrow$ C6H10/Pt(111)} & 1 & 17 & 31 \\
\ce{H2O + Pt(111) $\rightarrow$ H2O/Pt(111)} & 1 & 23 & 38 \\
\ce{H2O + O/Pt(111) $\rightarrow$ H2OOH/Pt(111)} & 1 & 18 & 28 \\
\bottomrule
\end{longtable}

\clearpage
\section{DFT, NSC-DFT and RPA estimates for datasets}

In this section, we provide the supporting data used to generate the mean-absolute deviations highlighted in the main text.
These are separated according to specific datasets and we provide further computational details if it may not have been sufficiently covered in the main text.

\subsection{CE39 adsorption energy dataset}

The calculation of the adsorption energy of the CE39 dataset is discussed in detail in the Methods of the main text.
We highlight the final values (in \SI{}{\kjmol}) for dhBEEF-vdW@BEEF-vdW, hBEEF-vdW@BEEF-vdW, RPA@PBE and BEEF-vdW in Table~\ref{tab:ce39_eads_1} and several DFAs in Table~\ref{tab:ce39_eads_2}.

\begin{longtable}{lrrrrrr}
\caption{\label{tab:ce39_eads_1}Adsorption energies for the CE39 dataset calculated with dhBEEF-vdW@BEEF-vdW, hBEEF-vdW@BEEF-vdW, RPA@PBE, BEEF-vdW compared against experiment. All values are in kJ/mol.} \\
\toprule
\rotatebox{90}{Index} & Reaction & \rotatebox{90}{Experiment} & \rotatebox{90}{dhBEEF-vdW@BEEF-vdW} & \rotatebox{90}{hBEEF-vdW@BEEF-vdW} & \rotatebox{90}{RPA@PBE} & \rotatebox{90}{BEEF-vdW} \\ 
\endfirsthead

\caption[]{(continued)} \\
\endhead

\multicolumn{7}{r}{{Continued on next page}} \\
\endfoot

\bottomrule
\endlastfoot

\midrule
01 & \ce{CO + Ni(111) $\rightarrow$ CO/Ni(111)} & -124 & -140 & -139 & -100 & -151 \\
02 & \ce{CO + Pt(111) $\rightarrow$ CO/Pt(111)} & -124 & -147 & -155 & -135 & -135 \\
03 & \ce{CO + Pd(111) $\rightarrow$ CO/Pd(111)} & -144 & -149 & -164 & -136 & -148 \\
04 & \ce{CO + Pd(100) $\rightarrow$ CO/Pd(100)} & -157 & -153 & -162 & -144 & -150 \\
05 & \ce{CO + Rh(111) $\rightarrow$ CO/Rh(111)} & -142 & -170 & -178 & -140 & -163 \\
06 & \ce{CO + Ir(111) $\rightarrow$ CO/Ir(111)} & -164 & -188 & -196 & -159 & -179 \\
07 & \ce{CO + Cu(111) $\rightarrow$ CO/Cu(111)} & -57 & -52 & -44 & -44 & -50 \\
08 & \ce{CO + Ru(001) $\rightarrow$ CO/Ru(001)} & -161 & -162 & -167 & -142 & -156 \\
09 & \ce{CO + Co(001) $\rightarrow$ CO/Co(001)} & -119 & -134 & -120 & -106 & -137 \\
10 & \ce{NO + Ni(100) $\rightarrow$ N/Ni(100) + O/Ni(100)} & -299 & -393 & -405 & -407 & -385 \\
11 & \ce{NO + Pt(111) $\rightarrow$ NO/Pt(111)} & -119 & -158 & -192 & -149 & -156 \\
12 & \ce{NO + Pd(111) $\rightarrow$ NO/Pd(111)} & -182 & -181 & -225 & -171 & -190 \\
13 & \ce{NO + Pd(100) $\rightarrow$ NO/Pd(100)} & -163 & -165 & -197 & -158 & -173 \\
14 & \ce{O2 + Ni(111) $\rightarrow$ 2O/Ni(111)} & -485 & -444 & -437 & -462 & -430 \\
15 & \ce{O2 + Ni(100) $\rightarrow$ 2O/Ni(100)} & -530 & -510 & -495 & -511 & -476 \\
16 & \ce{O2 + Pt(111) $\rightarrow$ 2O/Pt(111)} & -208 & -221 & -224 & -284 & -208 \\
17 & \ce{O2 + Rh(100) $\rightarrow$ 2O/Rh(100)} & -355 & -371 & -372 & -415 & -366 \\
18 & \ce{H2 + Pt(111) $\rightarrow$ 2H/Pt(111)} & -72 & -68 & -81 & -92 & -48 \\
19 & \ce{H2 + Ni(111) $\rightarrow$ 2H/Ni(111)} & -100 & -66 & -66 & -76 & -66 \\
20 & \ce{H2 + Ni(100) $\rightarrow$ 2H/Ni(100)} & -87 & -58 & -60 & -44 & -53 \\
21 & \ce{H2 + Rh(111) $\rightarrow$ 2H/Rh(111)} & -72 & -76 & -81 & -88 & -58 \\
22 & \ce{H2 + Pd(111) $\rightarrow$ 2H/Pd(111)} & -90 & -46 & -63 & -79 & -39 \\
23 & \ce{I2 + Pt(111) $\rightarrow$ 2I/Pt(111)} & -313 & -343 & -336 & -338 & -282 \\
24 & \ce{CH2I2 + Pt(111) $\rightarrow$ CH/Pt(111) + H/Pt(111) + 2I/Pt(111)} & -455 & -430 & -449 & -464 & -356 \\
25 & \ce{CH3I + Pt(111) $\rightarrow$ CH3/Pt(111) + I/Pt(111)} & -209 & -211 & -210 & -208 & -172 \\
26 & \ce{NH3 + Cu(100) $\rightarrow$ NH3/Cu(100)} & -60 & -37 & -36 & 4 & -35 \\
27 & \ce{CH3I + Pt(111) $\rightarrow$ CH3I/Pt(111)} & -84 & -87 & -81 & -80 & -63 \\
28 & \ce{CH3OH + Pt(111) $\rightarrow$ CH3OH/Pt(111)} & -55 & -40 & -35 & -28 & -33 \\
29 & \ce{CH4 + Pt(111) $\rightarrow$ CH4/Pt(111)} & -14 & -18 & -17 & -8 & -16 \\
30 & \ce{C2H6 + Pt(111) $\rightarrow$ C2H6/Pt(111)} & -27 & -30 & -25 & -27 & -22 \\
31 & \ce{C3H8 + Pt(111) $\rightarrow$ C3H8/Pt(111)} & -39 & -40 & -33 & -36 & -30 \\
32 & \ce{C4H10 + Pt(111) $\rightarrow$ C4H10/Pt(111)} & -48 & -53 & -45 & -42 & -40 \\
33 & \ce{C6H6 + Pt(111) $\rightarrow$ C6H6/Pt(111)} & -162 & -163 & -161 & -213 & -89 \\
34 & \ce{C6H6 + Cu(111) $\rightarrow$ C6H6/Cu(111)} & -66 & -52 & -44 & -45 & -40 \\
35 & \ce{C6H6 + Ag(111) $\rightarrow$ C6H6/Ag(111)} & -61 & -47 & -39 & -42 & -36 \\
36 & \ce{C6H6 + Au(111) $\rightarrow$ C6H6/Au(111)} & -70 & -55 & -46 & -40 & -42 \\
37 & \ce{C6H10 + Pt(111) $\rightarrow$ C6H10/Pt(111)} & -123 & -124 & -118 & -135 & -88 \\
38 & \ce{H2O + Pt(111) $\rightarrow$ H2O/Pt(111)} & -55 & -27 & -24 & -15 & -22 \\
39 & \ce{H2O + O/Pt(111) $\rightarrow$ H2OOH/Pt(111)} & -66 & -67 & -60 & -61 & -55 \\
T & MAD (Total) & 0 & 12 & 17 & 17 & 19 \\
P & MAD (Physisorption) & 0 & 9 & 12 & 21 & 23 \\
C & MAD (Chemisorption) & 0 & 13 & 19 & 15 & 16 \\
\bottomrule
\end{longtable}
\newpage
\begin{longtable}{lrrrrrrrrrrrrrrr}
\caption{\label{tab:ce39_eads_2}Adsorption energies for the CE39 dataset calculated with several density functional approximation compared against experiment. All values are in kJ/mol.} \\
\toprule
\rotatebox{90}{Index}  & \rotatebox{90}{Experiment} & \rotatebox{90}{PBE} & \rotatebox{90}{PBEsol} & \rotatebox{90}{RPBE} & \rotatebox{90}{revPBE} & \rotatebox{90}{BLYP} & \rotatebox{90}{r$^2$SCAN} & \rotatebox{90}{MS2} & \rotatebox{90}{revTPSS} & \rotatebox{90}{PBE-D3} & \rotatebox{90}{PBE-DdsC} & \rotatebox{90}{RPBE-D3} & \rotatebox{90}{optPBE-vdW} & \rotatebox{90}{revvdW-DF2} & \rotatebox{90}{r$^2$SCAN+rVV10} \\ 
\endfirsthead

\caption[]{(continued)} \\
\endhead

\multicolumn{16}{r}{{Continued on next page}} \\
\endfoot

\bottomrule
\endlastfoot

\midrule
01 & -124 & -220 & -229 & -143 & -187 & -126 & -186 & -148 & -174 & -200 & -200 & -206 & -179 & -200 & -202 \\
02 & -124 & -165 & -205 & -133 & -137 & -110 & -182 & -176 & -164 & -197 & -183 & -213 & -160 & -180 & -198 \\
03 & -144 & -182 & -229 & -143 & -148 & -123 & -191 & -172 & -174 & -202 & -197 & -213 & -176 & -199 & -207 \\
04 & -157 & -180 & -220 & -146 & -150 & -132 & -192 & -179 & -170 & -202 & -194 & -208 & -176 & -193 & -206 \\
05 & -142 & -185 & -215 & -157 & -160 & -145 & -197 & -182 & -178 & -204 & -202 & -206 & -184 & -196 & -211 \\
06 & -164 & -198 & -228 & -171 & -174 & -158 & -213 & -189 & -190 & -220 & -212 & -222 & -198 & -209 & -227 \\
07 & -57 & -70 & -99 & -44 & -47 & -33 & -82 & -69 & -70 & -92 & -87 & -109 & -72 & -82 & -96 \\
08 & -161 & -177 & -208 & -150 & -153 & -134 & -186 & -175 & -173 & -199 & -189 & -213 & -176 & -189 & -201 \\
09 & -119 & -161 & -197 & -130 & -134 & -118 & -161 & -188 & -159 & -182 & -178 & -203 & -163 & -178 & -176 \\
10 & -299 & -505 & -508 & -369 & -451 & -382 & -491 & -535 & -450 & -453 & -445 & -442 & -449 & -481 & -511 \\
11 & -119 & -187 & -240 & -145 & -150 & -132 & -198 & -161 & -181 & -212 & -200 & -216 & -191 & -215 & -216 \\
12 & -182 & -221 & -273 & -178 & -184 & -175 & -227 & -201 & -211 & -243 & -236 & -242 & -228 & -251 & -243 \\
13 & -163 & -199 & -245 & -157 & -162 & -156 & -205 & -171 & -191 & -219 & -213 & -223 & -209 & -228 & -222 \\
14 & -485 & -573 & -560 & -437 & -524 & -476 & -566 & -547 & -500 & -512 & -511 & -490 & -528 & -552 & -584 \\
15 & -530 & -608 & -595 & -481 & -560 & -514 & -632 & -709 & -560 & -550 & -546 & -533 & -569 & -588 & -650 \\
16 & -208 & -295 & -371 & -233 & -242 & -239 & -350 & -327 & -320 & -325 & -307 & -330 & -312 & -348 & -371 \\
17 & -355 & -435 & -495 & -379 & -386 & -400 & -495 & -466 & -469 & -457 & -444 & -480 & -466 & -490 & -516 \\
18 & -72 & -101 & -146 & -68 & -72 & -45 & -114 & -108 & -86 & -116 & -111 & -117 & -71 & -97 & -121 \\
19 & -100 & -185 & -151 & -74 & -157 & -67 & -116 & -160 & -103 & -129 & -123 & -115 & -90 & -112 & -123 \\
20 & -87 & -169 & -141 & -65 & -142 & -48 & -105 & -207 & -96 & -117 & -104 & -121 & -75 & -97 & -112 \\
21 & -72 & -107 & -151 & -75 & -79 & -54 & -119 & -120 & -102 & -122 & -120 & -133 & -82 & -106 & -127 \\
22 & -90 & -94 & -140 & -59 & -64 & -45 & -101 & -108 & -80 & -107 & -104 & -110 & -68 & -93 & -108 \\
23 & -313 & -291 & -376 & -225 & -236 & -159 & -344 & -348 & -350 & -401 & -344 & -659 & -324 & -369 & -397 \\
24 & -455 & -396 & -522 & -305 & -321 & -210 & -472 & -453 & -471 & -529 & -467 & -793 & -420 & -489 & -538 \\
25 & -209 & -174 & -231 & -125 & -131 & -89 & -215 & -214 & -208 & -259 & -223 & -403 & -212 & -236 & -257 \\
26 & -60 & -38 & -56 & -17 & -18 & -16 & -51 & -40 & -41 & -69 & -59 & -65 & -54 & -56 & -65 \\
27 & -84 & -50 & -83 & -14 & -16 & 2 & -71 & -67 & -60 & -123 & -94 & -192 & -95 & -102 & -107 \\
28 & -55 & -19 & -30 & -3 & -1 & -1 & -32 & -27 & -20 & -55 & -47 & -61 & -50 & -46 & -52 \\
29 & -14 & -3 & -5 & 1 & 3 & 5 & -8 & -7 & -2 & -24 & -20 & -25 & -23 & -16 & -18 \\
30 & -27 & -12 & -21 & -1 & 1 & 4 & -23 & -22 & -13 & -50 & -39 & -60 & -42 & -37 & -41 \\
31 & -39 & -6 & -15 & 8 & 10 & 11 & -19 & -14 & -5 & -55 & -44 & -61 & -47 & -39 & -44 \\
32 & -48 & -9 & -22 & 11 & 14 & 14 & -27 & -20 & -7 & -75 & -60 & -83 & -64 & -54 & -60 \\
33 & -162 & -115 & -232 & -29 & -43 & 54 & -179 & -184 & -174 & -233 & -190 & -336 & -162 & -219 & -247 \\
34 & -66 & -5 & -17 & 13 & 15 & 21 & -26 & -17 & -7 & -76 & -65 & -80 & -66 & -52 & -61 \\
35 & -61 & -7 & -18 & 10 & 12 & 15 & -26 & -20 & -8 & -63 & -57 & -71 & -60 & -48 & -58 \\
36 & -70 & -7 & -21 & 15 & 16 & 19 & -30 & -24 & -9 & -75 & -61 & -86 & -69 & -59 & -67 \\
37 & -123 & -76 & -132 & -24 & -29 & -2 & -120 & -101 & -91 & -178 & -148 & -204 & -142 & -156 & -176 \\
38 & -55 & -19 & -30 & -4 & -4 & -3 & -30 & -26 & -21 & -41 & -34 & -46 & -34 & -34 & -41 \\
39 & -66 & -59 & -81 & -31 & -33 & -39 & -76 & -76 & -62 & -88 & -79 & -102 & -76 & -81 & -91 \\
T & 0 & 37 & 47 & 33 & 35 & 39 & 32 & 31 & 28 & 36 & 28 & 52 & 22 & 32 & 40 \\
P & 0 & 36 & 25 & 62 & 61 & 72 & 19 & 25 & 31 & 22 & 11 & 40 & 9 & 15 & 18 \\
C & 0 & 37 & 59 & 16 & 20 & 21 & 39 & 35 & 26 & 44 & 37 & 58 & 29 & 42 & 52 \\
\bottomrule
\end{longtable}
\clearpage
\subsection{CO adsorption relative energy}

In Table~\ref{tab:co_ads}, we compare the relative energy between the on-top and FCC site for our NSC-DFAs, RPA@PBE and several standard DFAs.
This is compared for the Cu(111), Pt(111), Rh(111) and Pd(111) surfaces.
We calculate the relative energy between the on-top and FCC site as follows:
\begin{equation}
    E_\text{rel} = E[\text{Top}] - E[\text{FCC}] + \Delta_\text{layer},
\end{equation}
where a negative value indicates that the on-top site is more stable.
These calculations were performed on a 4L slab, with a subsequent layer correction to a 5L slab with 2L fixed at the bottom.
The correction is overall very small, below \SI{1}{\kjmol} for each system.

As shown in the main text, the on-top site is most stable for the Cu(111), Pt(111) and Rh(111) surfaces.
In contrast, Pd(111) is included (only) as additional supplementary here to highlight a system where the FCC hollow site is more stable.
Out of all the studed semilocal DFAs, none can get all three systems correctly identified.
This can only be obtained correctly when going beyond to hBEEF-vdW@BEEF-vdW or dhBEEF-vdW@BEEF-vdW.

In addition to previous DFT work in Section~\ref{sec:co_ads}, we also add a brief discussion here on some of the previous beyond-DFT work in this area.
Our RPA@PBE estimates for Cu(111) are in good agreement with previous RPA@PBE estimates~\cite{renExploringRandomPhase2009a,schimkaAccurateSurfaceAdsorption2010} ($10-21\,$\SI{}{\kjmol}) as well as DMC~\cite{hsingQuantumMonteCarlo2019b,fantaResolutionSelectivitySteps2025} ($37-39$\SI{}{\kjmol}).
For Pt(111), these estimates also lie within the range ($24-73\,$\SI{}{\kjmol}) from previous DMC estimates~\cite{hsingQuantumMonteCarlo2019b,iyerFiniteSizeErrorCancellation2022a} and close to an estimate of $40\,$\SI{}{\kjmol} from coupled cluster theory~\cite{carboneCOAdsorptionPt1112024}.
It lies outside of a previous RPA@PBE estimate~\cite{schimkaAccurateSurfaceAdsorption2010}, which indicates much weaker stability of the on-top site ($8\,$\SI{}{\kjmol}) compared to the DMC, CCSD(T) and our RPA@PBE estimates.

\begin{table}[h]

\caption{\label{tab:co_ads}Relative adsorption energies of CO at the atop and fcc sites on Cu(111), Pt(111), Rh(111), and Pd(111) calculated with hBEEF-vdW@BEEF-vdW, dhBEEF-vdW@BEEF-vdW, RPA@PBE, BEEF-vdW and selection of density functional approximations. All values are in kJ/mol.}
\begin{tabular}{lrrrr}
\toprule
 & Cu(111) & Pt(111) & Rh(111) & Pd(111) \\ 
\midrule
hBEEF-vdW & -8.1 & -0.6 & -18.9 & 50.4 \\
dhBEEF-vdW & -1.7 & -5.3 & -11.9 & 49.1 \\
RPA@PBE & -14.5 & -47.8 & -27.6 & 87.7 \\
BEEF-vdW & 2.7 & 4.9 & -11.1 & 44.7 \\
PBE & 15.6 & 8.1 & -6.9 & 53.2 \\
PBEsol & 25.8 & 14.6 & 7.5 & 64.2 \\
RPBE & 8.3 & 2.9 & -15.9 & 45.7 \\
revPBE & 9.7 & 3.5 & -14.2 & 46.9 \\
BLYP & 1.1 & -1.9 & -28.4 & 36.9 \\
r$^2$SCAN & 13.9 & 3.6 & -9.2 & 56.2 \\
MS2 & 7.4 & -16.5 & -20.7 & 48.1 \\
revTPSS & 11.8 & 3.1 & -5.3 & 49.0 \\
PBE-D3 & 13.1 & 1.5 & -5.9 & 51.1 \\
PBE-DdsC & 14.3 & 6.5 & -7.5 & 51.9 \\
RPBE-D3 & 10.0 & -7.6 & 8.7 & 47.6 \\
optPBE-vdW & 9.5 & 5.5 & -11.7 & 47.3 \\
revvdW-DF2 & 15.7 & 8.3 & -2.5 & 53.6 \\
r$^2$SCAN+rVV10 & 15.4 & 4.7 & -6.4 & 57.5 \\
\bottomrule
\end{tabular}

\end{table}

\clearpage

\subsection{SBH17 barrier heights}

The calculation of the adsorption energy of the SBH17 dataset is discussed in detail in the Methods of the main text.
We highlight the final values (in \SI{}{\kjmol}) for dhBEEF-vdW@BEEF-vdW, hBEEF-vdW@BEEF-vdW, RPA@PBE and BEEF-vdW in Table~\ref{tab:sbh17_1} and several DFAs in Table~\ref{tab:sbh17_2}.

\begin{table}[h]

\caption{\label{tab:sbh17_1}Barrier heights of the SBH17 benchmark set calculated with hBEEF-vdW, dhBEEF-vdW, RPA@PBE, BEEF-vW. These are compared to experimental estimates from the SBH17 dataset~\cite{tchakouaSBH17BenchmarkDatabase2023}. All values are in kJ/mol.}
\begin{tabular}{lrrrrr}
\toprule
System & \rotatebox{90}{Experiment} & \rotatebox{90}{hBEEF-vdW} & \rotatebox{90}{dhBEEF-vdW} & \rotatebox{90}{RPA@PBE} & \rotatebox{90}{BEEF-vdW} \\ 
\midrule
\ce{H2 + Cu(111)} & 60.6 & 91.2 & 84.2 & 64.6 & 93.2 \\
\ce{H2 + Cu(100)} & 71.4 & 102.3 & 94.3 & 70.6 & 100.4 \\
\ce{H2 + Cu(110)} & 76.1 & 122.6 & 110.9 & 81.8 & 118.6 \\
\ce{H2 + Pt(111)} & -0.8 & 3.0 & 3.9 & 5.0 & 10.9 \\
\ce{H2 + Pt(211)} & -8.0 & -5.3 & -5.7 & -2.8 & -3.2 \\
\ce{H2 + Ru(0001)} & 0.4 & -0.3 & -1.8 & 3.1 & 2.0 \\
\ce{H2 + Ni(111)} & 2.3 & 4.4 & 4.9 & 10.7 & 11.8 \\
\ce{H2 + Ag(111)} & 104.4 & 166.0 & 157.1 & 136.0 & 164.1 \\
\ce{N2 + Ru(0001)} & 177.5 & 175.2 & 154.9 & 109.3 & 164.7 \\
\ce{N2 + Ru(10-10)} & 38.6 & 19.3 & -0.7 & -62.4 & 23.3 \\
\ce{CH4 + Ni(111)} & 97.9 & 115.8 & 111.3 & 106.8 & 117.2 \\
\ce{CH4 + Ni(100)} & 73.3 & 104.5 & 99.2 & 102.7 & 106.5 \\
\ce{CH4 + Ni(211)} & 67.4 & 93.9 & 87.5 & 87.0 & 86.8 \\
\ce{CH4 + Pt(111)} & 78.6 & 84.5 & 84.3 & 71.4 & 103.3 \\
\ce{CH4 + Pt(211)} & 53.9 & 54.6 & 56.1 & 42.8 & 69.3 \\
\ce{CH4 + Ir(111)} & 80.7 & 84.6 & 81.8 & 78.0 & 102.4 \\
\ce{CH4 + Ru(0001)} & 77.2 & 85.7 & 81.7 & 75.3 & 95.2 \\
MAD (Total) & - & 17.4 & 16.5 & 18.5 & 21.8 \\
MAD (H2) & - & 22.4 & 18.2 & 8.0 & 23.9 \\
MAD (CH4) & - & 13.5 & 10.4 & 11.5 & 21.7 \\
MAD (N2) & - & 10.9 & 31.0 & 84.7 & 14.1 \\
\bottomrule
\end{tabular}

\end{table}
\begin{table}[h]

\caption{\label{tab:sbh17_2}Barrier heights of the SBH17 benchmark set calculated with various DFT functionals. These are compared to experimental estimates from the SBH17 dataset~\cite{tchakouaSBH17BenchmarkDatabase2023}. All values are in kJ/mol.}
\begin{adjustbox}{max width=1\textwidth}
\begin{tabular}{lrrrrrrrrrrrrrr}
\toprule
System & \rotatebox{90}{PBE} & \rotatebox{90}{PBEsol} & \rotatebox{90}{RPBE} & \rotatebox{90}{revPBE} & \rotatebox{90}{BLYP} & \rotatebox{90}{r$^2$SCAN} & \rotatebox{90}{MS2} & \rotatebox{90}{revTPSS} & \rotatebox{90}{PBE-D3} & \rotatebox{90}{PBE-DdsC} & \rotatebox{90}{RPBE-D3} & \rotatebox{90}{optPBE-vdW} & \rotatebox{90}{rev-vdW-DF2} & \rotatebox{90}{r$^2$SCAN+rVV10} \\ 
\midrule
\ce{H2 + Cu(111)} & 41.7 & 0.7 & 73.3 & 68.9 & 93.0 & 31.2 & 30.8 & 50.1 & 20.5 & 35.1 & 8.3 & 69.2 & 45.9 & 24.1 \\
\ce{H2 + Cu(100)} & 55.8 & 13.1 & 88.4 & 83.8 & 108.9 & 46.6 & 50.1 & 63.1 & 32.8 & 49.1 & 21.6 & 84.0 & 59.7 & 39.1 \\
\ce{H2 + Cu(110)} & 63.2 & 14.6 & 99.7 & 94.1 & 118.0 & 50.6 & 53.4 & 75.2 & 39.4 & 55.7 & 32.5 & 95.1 & 67.0 & 43.7 \\
\ce{H2 + Pt(111)} & 1.6 & -14.4 & 17.5 & 16.7 & 20.7 & -4.3 & -7.0 & 2.6 & -11.2 & -6.6 & -18.7 & 2.8 & -4.1 & -10.2 \\
\ce{H2 + Pt(211)} & -2.7 & -7.2 & 3.8 & 4.5 & 4.4 & -5.3 & -6.8 & -1.3 & -10.4 & -8.3 & -12.3 & -7.0 & -6.5 & -9.1 \\
\ce{H2 + Ru(0001)} & 0.7 & -5.6 & 9.0 & 9.6 & 11.1 & -2.5 & -4.3 & 0.8 & -7.5 & -5.8 & -10.1 & -2.7 & -3.7 & -6.8 \\
\ce{H2 + Ni(111)} & 2.0 & -12.2 & 17.3 & 16.6 & 20.0 & -2.3 & -29.7 & 0.9 & -15.6 & -7.2 & -15.6 & 2.6 & -1.9 & -8.1 \\
\ce{H2 + Ag(111)} & 106.2 & 61.0 & 140.4 & 135.3 & 158.7 & 106.2 & 99.2 & 112.8 & 94.4 & 99.9 & 77.3 & 132.5 & 107.0 & 97.9 \\
\ce{N2 + Ru(0001)} & 131.5 & 63.1 & 178.5 & 171.9 & 218.1 & 120.7 & 128.3 & 98.5 & 105.8 & 124.0 & 40.6 & 133.2 & 97.3 & 100.9 \\
\ce{N2 + Ru(10-10)} & -22.5 & -106.8 & 36.9 & 27.7 & 77.1 & -48.0 & -36.8 & -58.5 & -50.8 & -32.6 & -130.6 & -19.2 & -64.1 & -72.3 \\
\ce{CH4 + Ni(111)} & 96.6 & 52.7 & 133.2 & 128.6 & 150.2 & 90.3 & 136.1 & 89.8 & 56.1 & 74.2 & 38.8 & 94.2 & 72.8 & 74.3 \\
\ce{CH4 + Ni(100)} & 84.4 & 42.8 & 121.0 & 116.4 & 130.3 & 68.5 & 105.7 & 77.4 & 45.2 & 61.7 & 36.6 & 78.7 & 59.5 & 52.8 \\
\ce{CH4 + Ni(211)} & 66.2 & 28.5 & 98.6 & 94.7 & 107.0 & 59.8 & 122.7 & 62.8 & 32.8 & 43.9 & 29.4 & 60.9 & 45.6 & 44.6 \\
\ce{CH4 + Pt(111)} & 77.1 & 30.5 & 113.1 & 108.8 & 138.7 & 61.4 & 73.2 & 70.6 & 39.3 & 56.3 & -1.3 & 75.4 & 50.5 & 43.7 \\
\ce{CH4 + Pt(211)} & 43.7 & 1.7 & 75.9 & 72.0 & 97.5 & 30.8 & 37.6 & 42.0 & 13.6 & 25.5 & -15.1 & 45.5 & 24.0 & 15.6 \\
\ce{CH4 + Ir(111)} & 78.5 & 34.9 & 112.2 & 108.5 & 139.3 & 66.3 & 76.6 & 71.6 & 40.1 & 54.5 & -3.1 & 70.0 & 49.8 & 45.3 \\
\ce{CH4 + Ru(0001)} & 82.1 & 41.9 & 114.3 & 110.9 & 136.6 & 73.8 & 88.6 & 75.4 & 50.9 & 61.6 & 12.0 & 74.1 & 55.8 & 55.1 \\
MAD (Total) & 11.6 & 47.9 & 22.6 & 20.6 & 39.9 & 18.6 & 24.2 & 15.5 & 33.9 & 21.8 & 56.5 & 12.9 & 23.8 & 30.6 \\
MAD (H2) & 7.2 & 32.3 & 17.9 & 15.4 & 28.5 & 11.9 & 15.4 & 5.0 & 20.5 & 11.8 & 27.9 & 9.5 & 6.4 & 17.0 \\
MAD (CH4) & 4.6 & 42.3 & 34.2 & 30.1 & 52.9 & 11.2 & 23.3 & 6.8 & 35.9 & 21.6 & 61.7 & 5.9 & 24.5 & 28.2 \\
MAD (N2) & 53.5 & 129.9 & 1.4 & 8.2 & 39.6 & 71.7 & 62.3 & 88.1 & 80.6 & 62.3 & 153.1 & 51.1 & 91.5 & 93.8 \\
\bottomrule
\end{tabular}
\end{adjustbox}

\end{table}

\clearpage
\subsection{NBH56 gas-phase barrier heights}

The NBH56 dataset are a set of 56 barrier heights obtained from the BH76 dataset~\cite{zhaoMulticoefficientExtrapolatedDensity2005,zhengDBH2408Database2009}, chosen from the neutral subset sampled from the recent Gold-Standard Chemical Database 137 (GSCDB137)~\cite{liangGoldstandardChemicalDatabase2025}.
The specific reaction labels from the GSCDB137 are given in Table~\ref{tab:nbh56_1}.
We highlight the final values (in \SI{}{\kjmol}) for dhBEEF-vdW@BEEF-vdW, hBEEF-vdW@BEEF-vdW, RPA@PBE and BEEF-vdW in Table~\ref{tab:nbh56_1} and several DFAs in Table~\ref{tab:nbh56_2}.
To calculate the barrier heights, the set of systems for a specific reaction label is given alongside the stoichiometries.
For example, for a reaction involving the following systems $\{\text{h}, \text{n2o}, \text{n2ohts}\}$ with stoichiometric coefficients $-1$, $-1$, and $+1$, respectively, the barrier height ($E_b$) is obtained by forming the linear combination of their total energies using these coefficients:
\begin{equation}
E_b = -1\,E_{\text{h}} - 1\,E_{\text{n2o}} + 1\,E_{\text{n2ohts}}
\end{equation}

\begin{longtable}{lrrrrr}
\caption{\label{tab:nbh56_1}Reaction energies of the NBH56 benchmark set calculated with hBEEF-vdW, dhBEEF-vdW, RPA@PBE, BEEF-vdW. These are compared to CCSD(T) estimates from Gold-Standard Chemical Database 137~\cite{liangGoldstandardChemicalDatabase2025} (GSCDB137).  All values are in kJ/mol.} \\
\toprule
Label & \rotatebox{90}{CCSD(T)} & \rotatebox{90}{hBEEF-vdW} & \rotatebox{90}{dhBEEF-vdW} & \rotatebox{90}{RPA@PBE} & \rotatebox{90}{BEEF-vdW} \\ 
\endfirsthead

\caption[]{(continued)} \\
\endhead

\multicolumn{6}{r}{{Continued on next page}} \\
\endfoot

\bottomrule
\endlastfoot

\midrule
BH76\_3 & 176.1 & 132.4 & 129.0 & 161.8 & 117.3 \\
BH76\_7 & 127.6 & 89.1 & 85.2 & 115.6 & 71.1 \\
BH76\_8 & 238.1 & 204.5 & 195.6 & 223.5 & 185.1 \\
BH76\_9 & 6.3 & -28.4 & -30.9 & 1.4 & -43.7 \\
BH76\_10 & 438.5 & 406.6 & 391.0 & 430.6 & 358.6 \\
BH76\_31 & 13.4 & 1.0 & 1.2 & 18.5 & -3.9 \\
BH76\_32 & 95.4 & 109.6 & 107.5 & 98.8 & 106.9 \\
BH76\_35 & 26.8 & 21.4 & 17.5 & 33.0 & 16.8 \\
BH76\_36 & 138.1 & 131.4 & 128.2 & 136.3 & 118.1 \\
BH76\_39 & 25.5 & -1.7 & -1.7 & 15.0 & -8.0 \\
BH76\_40 & 33.5 & -29.8 & -49.7 & 48.2 & 4.3 \\
BH76\_41 & 21.8 & 19.0 & 13.7 & 28.5 & 4.6 \\
BH76\_42 & 90.4 & 54.0 & 52.6 & 75.1 & 44.1 \\
BH76\_43 & 49.8 & 48.8 & 45.9 & 53.9 & 43.6 \\
BH76\_44 & 62.8 & 40.5 & 40.8 & 60.4 & 33.5 \\
BH76\_47 & 40.6 & 27.6 & 28.2 & 41.4 & 23.3 \\
BH76\_49 & 14.2 & -0.0 & -9.6 & 5.8 & -23.9 \\
BH76\_50 & 57.3 & 37.0 & 28.1 & 41.3 & 17.7 \\
BH76\_51 & 7.5 & -0.7 & -4.4 & 6.4 & -7.3 \\
BH76\_52 & 28.5 & -37.0 & -57.5 & 46.2 & -5.1 \\
BH76\_53 & 14.6 & 5.0 & -1.5 & 22.1 & -14.0 \\
BH76\_54 & 85.4 & 65.3 & 59.3 & 74.1 & 54.1 \\
BH76\_55 & 6.7 & -9.9 & -15.4 & 12.1 & -31.1 \\
BH76\_56 & 141.4 & 87.1 & 84.8 & 111.7 & 76.8 \\
BH76\_57 & 60.2 & 41.7 & 36.8 & 62.8 & 22.0 \\
BH76\_58 & 37.2 & 21.6 & 17.0 & 34.0 & 11.8 \\
BH76\_59 & 12.1 & -0.6 & -0.5 & 14.1 & -6.1 \\
BH76\_60 & 103.3 & 109.1 & 106.4 & 111.6 & 106.2 \\
BH76\_65 & 43.5 & 24.3 & 8.9 & 31.8 & -12.8 \\
BH76\_66 & 41.4 & -32.2 & -63.9 & 42.7 & -20.7 \\
BH76\_67 & 37.2 & 29.8 & 25.0 & 37.3 & 20.7 \\
BH76\_68 & 92.0 & 78.9 & 74.3 & 93.5 & 65.4 \\
BH76\_69 & 41.0 & 34.9 & 29.8 & 40.7 & 26.6 \\
BH76\_70 & 81.2 & 67.0 & 62.3 & 85.0 & 52.8 \\
BH76\_71 & 47.3 & 42.0 & 36.3 & 53.0 & 29.2 \\
BH76\_72 & 74.5 & 65.2 & 59.4 & 69.5 & 55.6 \\
BH76\_73 & 58.2 & 54.0 & 48.6 & 61.8 & 42.0 \\
BH76\_74 & 70.7 & 60.3 & 54.9 & 66.3 & 49.9 \\
BH76\_75 & 166.1 & 159.4 & 155.2 & 166.3 & 149.8 \\
DBH24\_1 & 71.9 & 47.4 & 44.9 & 66.1 & 38.8 \\
DBH24\_2 & 73.6 & 57.7 & 55.8 & 73.1 & 49.8 \\
DBH24\_3 & 28.9 & -4.7 & -11.2 & 12.4 & -22.5 \\
DBH24\_7 & 59.9 & 30.6 & 30.5 & 59.8 & 22.7 \\
DBH24\_8 & 7.7 & 0.5 & 1.1 & 13.2 & -1.4 \\
DBH24\_9 & 201.4 & 196.8 & 195.1 & 193.6 & 194.8 \\
DBH24\_10 & 25.6 & 18.5 & 12.4 & 31.2 & 1.1 \\
DBH24\_11 & 44.6 & 16.9 & 15.8 & 37.4 & 8.0 \\
DBH24\_12 & 15.8 & 2.2 & 2.2 & 16.6 & -3.4 \\
DBH24\_13 & 343.7 & 307.9 & 297.2 & 354.6 & 266.8 \\
DBH24\_15 & 248.2 & 145.9 & 118.5 & 238.1 & 156.9 \\
DBH24\_19 & 44.8 & 54.5 & 52.1 & 45.4 & 47.3 \\
DBH24\_20 & 175.2 & 184.3 & 179.4 & 176.3 & 174.3 \\
DBH24\_21 & 137.7 & 141.8 & 138.8 & 133.1 & 136.3 \\
DBH24\_22 & 81.0 & 61.8 & 56.4 & 71.3 & 50.6 \\
DBH24\_23 & 54.0 & 45.4 & 40.7 & 59.7 & 28.3 \\
DBH24\_24 & 71.5 & 78.0 & 75.6 & 86.1 & 73.3 \\
MAD & - & 21.0 & 26.1 & 6.8 & 29.9 \\
\bottomrule
\end{longtable}
\begin{longtable}{lrrrrrrrrrrrrrr}
\caption{\label{tab:nbh56_2}Reaction energies of the NBH56 benchmark set calculated with various DFT functionals. All values are in kJ/mol.} \\
\toprule
Label & \rotatebox{90}{PBE} & \rotatebox{90}{PBEsol} & \rotatebox{90}{RPBE} & \rotatebox{90}{revPBE} & \rotatebox{90}{BLYP} & \rotatebox{90}{r$^2$SCAN} & \rotatebox{90}{MS2} & \rotatebox{90}{revTPSS} & \rotatebox{90}{PBE-D3} & \rotatebox{90}{PBE-DdsC} & \rotatebox{90}{RPBE-D3} & \rotatebox{90}{optPBE-vdW} & \rotatebox{90}{rev-vdW-DF2} & \rotatebox{90}{r$^2$SCAN+rVV10} \\ 
\endfirsthead

\caption[]{(continued)} \\
\endhead

\multicolumn{15}{r}{{Continued on next page}} \\
\endfoot

\bottomrule
\endlastfoot

\midrule
BH76\_3 & 105.1 & 91.1 & 114.9 & 114.6 & 100.8 & 112.4 & 120.4 & 108.7 & 104.9 & 104.8 & 112.5 & 101.3 & 95.4 & 111.7 \\
BH76\_7 & 70.1 & 64.1 & 74.3 & 75.1 & 59.9 & 71.6 & 85.4 & 69.1 & 68.7 & 68.9 & 71.8 & 59.1 & 59.3 & 70.9 \\
BH76\_8 & 164.1 & 148.9 & 182.1 & 181.8 & 168.6 & 188.4 & 166.2 & 162.5 & 162.0 & 162.1 & 175.0 & 167.2 & 160.4 & 186.4 \\
BH76\_9 & -44.5 & -51.3 & -37.3 & -37.0 & -52.7 & -44.3 & -38.2 & -46.6 & -44.9 & -44.7 & -39.9 & -51.9 & -53.6 & -44.9 \\
BH76\_10 & 326.3 & 308.5 & 347.6 & 346.5 & 332.2 & 367.0 & 342.7 & 330.9 & 325.8 & 326.0 & 344.0 & 336.0 & 325.7 & 365.3 \\
BH76\_31 & -7.8 & -16.9 & -2.3 & -2.7 & -9.0 & -14.3 & -3.5 & -21.5 & -8.6 & -8.4 & -9.6 & -13.8 & -15.8 & -14.8 \\
BH76\_32 & 105.6 & 108.4 & 104.1 & 104.4 & 100.0 & 102.3 & 110.9 & 92.7 & 105.0 & 105.2 & 100.8 & 105.2 & 105.6 & 102.3 \\
BH76\_35 & 6.9 & -5.9 & 19.8 & 19.8 & 19.8 & 9.3 & 9.4 & 13.0 & 2.0 & 0.5 & -0.7 & 5.7 & 2.5 & 6.8 \\
BH76\_36 & 124.8 & 134.3 & 119.0 & 120.1 & 103.4 & 129.6 & 134.6 & 117.7 & 123.9 & 125.0 & 120.6 & 114.7 & 120.6 & 130.1 \\
BH76\_39 & 2.8 & -0.1 & 3.5 & 4.1 & -9.7 & -2.9 & -3.2 & -20.9 & 2.4 & 2.7 & -1.5 & -16.7 & -12.9 & -3.3 \\
BH76\_40 & -4.8 & -22.6 & 10.9 & 9.9 & 10.4 & 8.9 & 12.8 & 10.1 & -5.2 & -4.9 & 1.3 & 10.5 & 0.7 & 8.0 \\
BH76\_41 & -25.6 & -46.1 & -6.6 & -8.0 & -13.7 & -5.0 & 2.6 & -1.1 & -26.2 & -26.0 & -12.6 & -10.3 & -22.3 & -5.8 \\
BH76\_42 & 53.9 & 52.4 & 52.3 & 53.4 & 40.0 & 48.6 & 30.7 & 25.1 & 53.3 & 53.5 & 48.0 & 34.4 & 39.2 & 48.2 \\
BH76\_43 & 17.0 & -0.3 & 32.2 & 31.3 & 30.8 & 34.2 & 30.7 & 33.3 & 15.8 & 15.7 & 22.7 & 31.8 & 22.5 & 33.0 \\
BH76\_44 & 38.2 & 32.0 & 38.9 & 39.3 & 31.2 & 31.3 & 43.4 & 20.1 & 37.0 & 36.8 & 32.6 & 23.7 & 24.9 & 30.4 \\
BH76\_47 & 15.3 & 4.6 & 20.6 & 20.1 & 12.2 & 10.6 & 19.5 & 3.5 & 15.2 & 15.3 & 15.0 & 12.6 & 9.0 & 10.3 \\
BH76\_49 & -48.8 & -68.7 & -29.9 & -31.2 & -38.1 & -28.2 & -14.5 & -25.1 & -50.7 & -51.8 & -39.3 & -41.6 & -51.6 & -29.8 \\
BH76\_50 & -4.9 & -21.9 & 12.0 & 11.1 & 4.9 & 15.1 & 11.8 & 13.1 & -6.9 & -7.9 & 1.5 & 1.3 & -6.7 & 13.4 \\
BH76\_51 & -21.7 & -34.9 & -8.6 & -9.0 & -11.9 & -9.6 & -8.1 & -12.9 & -24.2 & -25.0 & -21.6 & -19.3 & -24.4 & -11.3 \\
BH76\_52 & -8.1 & -25.1 & 5.5 & 4.8 & 8.6 & -0.7 & 20.6 & 4.9 & -10.5 & -11.4 & -8.9 & -0.1 & -8.3 & -2.5 \\
BH76\_53 & -36.2 & -55.9 & -18.1 & -19.1 & -24.5 & -17.2 & -0.4 & -12.4 & -39.2 & -40.0 & -27.9 & -30.3 & -40.1 & -19.1 \\
BH76\_54 & 43.9 & 34.0 & 54.7 & 55.0 & 50.0 & 58.7 & 31.3 & 44.9 & 40.1 & 38.8 & 42.5 & 42.1 & 40.1 & 56.5 \\
BH76\_55 & -52.8 & -69.1 & -35.9 & -36.8 & -47.6 & -29.1 & -24.9 & -31.8 & -53.4 & -53.1 & -39.2 & -43.9 & -53.5 & -29.7 \\
BH76\_56 & 95.3 & 100.6 & 89.5 & 91.5 & 76.5 & 90.1 & 53.6 & 53.5 & 94.7 & 95.1 & 86.6 & 70.5 & 79.5 & 89.8 \\
BH76\_57 & -0.4 & -15.5 & 16.2 & 15.5 & 8.5 & 12.6 & 26.1 & 34.0 & -2.0 & -2.9 & 9.4 & 6.8 & -0.7 & 11.2 \\
BH76\_58 & -3.8 & -19.3 & 8.7 & 8.4 & 5.1 & 13.6 & 11.4 & 2.3 & -5.5 & -6.3 & -0.3 & 0.2 & -6.3 & 12.2 \\
BH76\_59 & -7.1 & -15.4 & -3.1 & -3.2 & -10.7 & -13.0 & 0.2 & -19.8 & -8.0 & -7.9 & -10.4 & -16.2 & -17.6 & -13.5 \\
BH76\_60 & 77.2 & 60.6 & 93.1 & 91.9 & 92.2 & 84.7 & 84.8 & 85.2 & 76.1 & 76.0 & 79.0 & 94.9 & 85.2 & 83.7 \\
BH76\_65 & -40.1 & -58.9 & -20.1 & -21.3 & -33.2 & -13.6 & -0.9 & -1.6 & -40.4 & -42.2 & -29.1 & -33.7 & -43.5 & -15.1 \\
BH76\_66 & -30.0 & -52.8 & -13.5 & -14.6 & -16.1 & -3.7 & 13.1 & -15.5 & -30.3 & -32.1 & -26.0 & -21.2 & -33.0 & -5.3 \\
BH76\_67 & 3.8 & -13.0 & 16.8 & 16.3 & 15.4 & 20.9 & 25.6 & 13.4 & 1.6 & 0.4 & 5.6 & 8.6 & 1.3 & 19.4 \\
BH76\_68 & 44.8 & 30.6 & 59.3 & 59.0 & 55.9 & 54.9 & 66.1 & 76.2 & 42.6 & 41.4 & 48.9 & 52.0 & 45.7 & 53.4 \\
BH76\_69 & 13.0 & -3.1 & 25.7 & 25.5 & 25.4 & 27.7 & 31.7 & 20.7 & 9.4 & 7.5 & 12.8 & 14.8 & 8.6 & 25.4 \\
BH76\_70 & 32.2 & 16.9 & 47.7 & 47.3 & 44.7 & 42.3 & 55.9 & 65.7 & 29.4 & 28.0 & 36.2 & 38.6 & 32.0 & 40.2 \\
BH76\_71 & 6.9 & -12.9 & 24.1 & 23.2 & 22.5 & 23.7 & 40.6 & 30.6 & 3.4 & 2.0 & 11.9 & 13.6 & 4.2 & 21.5 \\
BH76\_72 & 43.1 & 30.1 & 55.0 & 55.1 & 54.1 & 56.3 & 46.1 & 49.7 & 38.9 & 36.9 & 41.5 & 43.2 & 39.4 & 53.9 \\
BH76\_73 & 19.0 & 0.1 & 35.5 & 34.6 & 33.8 & 36.2 & 51.9 & 41.5 & 16.4 & 15.1 & 24.4 & 27.1 & 17.8 & 34.4 \\
BH76\_74 & 33.5 & 19.5 & 45.8 & 45.6 & 44.2 & 49.5 & 41.0 & 42.8 & 30.8 & 29.6 & 34.4 & 37.1 & 31.9 & 47.7 \\
BH76\_75 & 130.4 & 114.3 & 141.1 & 139.1 & 150.3 & 137.9 & 156.9 & 144.8 & 129.8 & 126.9 & 127.0 & 140.2 & 130.9 & 136.3 \\
DBH24\_1 & 37.1 & 26.5 & 43.6 & 43.9 & 31.6 & 37.7 & 37.0 & 33.7 & 36.3 & 36.4 & 36.5 & 25.5 & 23.4 & 36.8 \\
DBH24\_2 & 41.3 & 28.7 & 48.8 & 48.2 & 42.1 & 38.0 & 51.1 & 32.0 & 41.1 & 41.1 & 42.8 & 37.3 & 33.5 & 37.3 \\
DBH24\_3 & -24.0 & -28.9 & -13.9 & -12.8 & -27.9 & -14.1 & -15.6 & -19.9 & -27.0 & -26.8 & -21.8 & -31.9 & -31.4 & -16.0 \\
DBH24\_7 & 17.6 & 5.3 & 24.1 & 23.7 & 18.7 & 14.4 & 22.3 & 12.5 & 17.4 & 17.4 & 17.2 & 11.9 & 8.7 & 13.7 \\
DBH24\_8 & -0.8 & -8.5 & 3.5 & 3.5 & -3.3 & -14.8 & -2.1 & -19.4 & -2.3 & -2.0 & -5.5 & -10.7 & -11.2 & -15.5 \\
DBH24\_9 & 190.4 & 186.1 & 189.9 & 190.2 & 196.8 & 191.8 & 190.6 & 195.6 & 190.5 & 190.7 & 189.3 & 193.5 & 191.6 & 191.6 \\
DBH24\_10 & -22.2 & -41.6 & -4.5 & -5.6 & -10.5 & -3.3 & 13.4 & 0.9 & -24.5 & -25.3 & -13.3 & -14.8 & -24.7 & -4.9 \\
DBH24\_11 & 12.3 & 4.2 & 13.7 & 14.0 & 3.3 & 9.3 & 18.7 & -11.4 & 11.9 & 12.2 & 9.5 & -1.7 & -1.9 & 8.9 \\
DBH24\_12 & -4.5 & -12.7 & -0.3 & -0.4 & -8.5 & -10.8 & 3.4 & -19.2 & -5.3 & -4.9 & -7.9 & -13.7 & -14.9 & -11.3 \\
DBH24\_13 & 218.6 & 179.3 & 249.5 & 246.0 & 258.3 & 262.8 & 303.0 & 248.7 & 217.6 & 216.0 & 235.6 & 245.5 & 221.3 & 260.4 \\
DBH24\_15 & 166.3 & 161.7 & 174.5 & 175.7 & 166.5 & 183.9 & 179.5 & 167.0 & 164.0 & 164.1 & 165.1 & 162.6 & 162.9 & 182.0 \\
DBH24\_19 & 42.2 & 42.5 & 43.2 & 42.9 & 38.3 & 44.3 & 49.2 & 34.0 & 42.0 & 42.0 & 41.4 & 42.7 & 41.9 & 44.2 \\
DBH24\_20 & 169.2 & 169.1 & 171.9 & 171.7 & 159.5 & 181.3 & 184.8 & 158.8 & 168.3 & 169.0 & 167.0 & 167.4 & 165.2 & 181.2 \\
DBH24\_21 & 131.0 & 129.3 & 130.3 & 130.5 & 136.3 & 135.5 & 129.4 & 132.6 & 131.1 & 131.1 & 131.3 & 133.9 & 133.3 & 135.3 \\
DBH24\_22 & 36.1 & 24.7 & 47.8 & 47.7 & 42.8 & 53.2 & 28.8 & 40.4 & 33.8 & 33.0 & 37.4 & 38.0 & 34.4 & 51.6 \\
DBH24\_23 & -5.4 & -24.4 & 14.5 & 13.0 & 6.3 & 11.2 & 20.7 & 33.6 & -5.8 & -5.6 & 9.2 & 13.0 & 1.3 & 10.4 \\
DBH24\_24 & 39.9 & 19.7 & 57.7 & 56.2 & 60.2 & 51.3 & 58.9 & 57.0 & 39.0 & 39.4 & 43.6 & 62.6 & 50.0 & 50.2 \\
MAD & 41.6 & 53.8 & 31.1 & 31.4 & 36.7 & 32.0 & 28.5 & 36.0 & 43.0 & 43.5 & 39.0 & 40.3 & 45.3 & 33.1 \\
\bottomrule
\end{longtable}

\clearpage
\subsection{S66 non-covalent interactions}

The S66 is a well-studied databse of dimer interaction energies covering a subset of H-bonding, dispersion-bonding and mixed interactions.
The interaction energies $E_\text{int}$ are calculated as:
\begin{equation}
    E_\text{int} = E[\text{AB}] - E[\text{A}] - E[\text{B}],
\end{equation}
where AB is the dimer complex formed from the A and B monomers.
We highlight the final values (in \SI{}{\kjmol}) for dhBEEF-vdW@BEEF-vdW, hBEEF-vdW@BEEF-vdW, RPA@PBE and BEEF-vdW in Table~\ref{tab:s66_1} and several DFAs in Table~\ref{tab:s66_2}.

\begin{longtable}{lrrrrr}
\caption{\label{tab:s66_1}Interaction energies of the S66 benchmark set calculated with hBEEF-vdW, dhBEEF-vdW, RPA@PBE, BEEF-vdW. These are compared to CCSD(T) estimates from Gold-Standard Chemical Database 137~\cite{liangGoldstandardChemicalDatabase2025} (GSCDB137).  All values are in kJ/mol.} \\
\toprule
Label & \rotatebox{90}{CCSD(T)} & \rotatebox{90}{hBEEF-vdW} & \rotatebox{90}{dhBEEF-vdW} & \rotatebox{90}{RPA@PBE} & \rotatebox{90}{BEEF-vdW} \\ 
\endfirsthead

\caption[]{(continued)} \\
\endhead

\multicolumn{6}{r}{{Continued on next page}} \\
\endfoot

\bottomrule
\endlastfoot

\midrule
S66\_1 & -20.8 & -19.8 & -22.1 & -18.4 & -18.7 \\
S66\_2 & -23.7 & -22.3 & -24.9 & -21.0 & -21.0 \\
S66\_3 & -29.2 & -28.1 & -31.0 & -25.9 & -27.3 \\
S66\_4 & -34.3 & -32.1 & -36.1 & -30.7 & -30.0 \\
S66\_5 & -24.3 & -23.3 & -26.0 & -21.6 & -22.0 \\
S66\_6 & -31.8 & -31.2 & -34.4 & -28.3 & -30.2 \\
S66\_7 & -34.8 & -32.7 & -36.7 & -30.9 & -30.7 \\
S66\_8 & -21.2 & -20.5 & -22.8 & -18.6 & -19.4 \\
S66\_9 & -12.9 & -13.5 & -14.7 & -11.0 & -12.8 \\
S66\_10 & -17.4 & -16.1 & -17.6 & -14.7 & -15.0 \\
S66\_11 & -22.8 & -20.3 & -22.7 & -19.3 & -18.3 \\
S66\_12 & -30.7 & -29.1 & -32.2 & -27.3 & -28.0 \\
S66\_13 & -26.1 & -25.7 & -28.8 & -23.5 & -23.7 \\
S66\_14 & -31.4 & -30.7 & -34.1 & -28.3 & -29.1 \\
S66\_15 & -36.4 & -36.4 & -40.8 & -33.1 & -33.8 \\
S66\_16 & -21.6 & -22.1 & -24.7 & -19.5 & -20.6 \\
S66\_17 & -73.2 & -72.0 & -81.3 & -67.5 & -67.4 \\
S66\_18 & -29.0 & -28.3 & -31.4 & -25.9 & -27.6 \\
S66\_19 & -31.2 & -30.7 & -34.0 & -27.9 & -29.9 \\
S66\_20 & -81.3 & -74.1 & -84.3 & -76.3 & -69.1 \\
S66\_21 & -69.1 & -61.4 & -70.2 & -65.2 & -57.3 \\
S66\_22 & -83.0 & -76.6 & -87.2 & -77.1 & -71.6 \\
S66\_23 & -81.6 & -75.5 & -86.1 & -76.2 & -70.7 \\
S66\_24 & -11.6 & -10.1 & -10.6 & -8.5 & -9.0 \\
S66\_25 & -16.1 & -13.5 & -14.6 & -12.6 & -11.9 \\
S66\_26 & -41.1 & -34.7 & -39.5 & -35.8 & -30.1 \\
S66\_27 & -14.2 & -12.0 & -12.9 & -10.8 & -10.6 \\
S66\_28 & -23.7 & -18.7 & -20.8 & -18.9 & -15.7 \\
S66\_29 & -28.4 & -23.0 & -25.9 & -23.5 & -19.9 \\
S66\_30 & -5.7 & -5.3 & -5.2 & -3.6 & -4.8 \\
S66\_31 & -13.9 & -12.3 & -13.6 & -11.2 & -10.6 \\
S66\_32 & -15.5 & -14.3 & -15.9 & -12.8 & -12.5 \\
S66\_33 & -7.5 & -6.7 & -6.8 & -5.2 & -6.0 \\
S66\_34 & -15.5 & -16.5 & -17.3 & -11.3 & -14.4 \\
S66\_35 & -10.8 & -11.9 & -12.4 & -7.6 & -10.7 \\
S66\_36 & -7.3 & -9.4 & -9.5 & -4.8 & -8.8 \\
S66\_37 & -9.9 & -10.9 & -11.2 & -6.8 & -9.7 \\
S66\_38 & -12.3 & -13.1 & -13.7 & -8.9 & -11.7 \\
S66\_39 & -14.7 & -13.2 & -14.2 & -11.3 & -11.4 \\
S66\_40 & -11.9 & -11.4 & -12.2 & -9.3 & -10.0 \\
S66\_41 & -20.1 & -17.4 & -19.0 & -15.9 & -14.4 \\
S66\_42 & -17.1 & -15.2 & -16.5 & -13.4 & -12.9 \\
S66\_43 & -15.4 & -14.6 & -15.9 & -12.4 & -12.7 \\
S66\_44 & -8.2 & -9.0 & -9.3 & -5.9 & -8.1 \\
S66\_45 & -7.1 & -8.7 & -9.0 & -5.0 & -8.1 \\
S66\_46 & -17.7 & -16.6 & -17.9 & -13.7 & -14.2 \\
S66\_47 & -11.8 & -10.3 & -11.3 & -9.5 & -8.9 \\
S66\_48 & -14.6 & -12.4 & -13.9 & -12.2 & -10.7 \\
S66\_49 & -13.7 & -11.6 & -12.9 & -11.3 & -10.1 \\
S66\_50 & -11.9 & -11.2 & -12.4 & -10.0 & -9.7 \\
S66\_51 & -6.4 & -7.3 & -7.9 & -5.3 & -6.9 \\
S66\_52 & -19.6 & -17.1 & -19.3 & -17.7 & -14.8 \\
S66\_53 & -18.3 & -16.4 & -18.4 & -16.2 & -14.4 \\
S66\_54 & -13.7 & -12.4 & -13.8 & -11.9 & -11.0 \\
S66\_55 & -17.3 & -15.6 & -17.2 & -14.7 & -14.0 \\
S66\_56 & -13.3 & -12.0 & -13.1 & -10.9 & -10.6 \\
S66\_57 & -22.0 & -19.1 & -21.4 & -19.0 & -16.7 \\
S66\_58 & -17.6 & -13.6 & -15.5 & -15.3 & -12.0 \\
S66\_59 & -12.2 & -12.3 & -13.8 & -10.7 & -11.4 \\
S66\_60 & -20.7 & -19.0 & -21.2 & -18.0 & -17.0 \\
S66\_61 & -12.0 & -12.9 & -13.6 & -9.2 & -11.4 \\
S66\_62 & -14.6 & -14.2 & -15.2 & -11.4 & -12.4 \\
S66\_63 & -15.6 & -14.0 & -15.4 & -12.9 & -12.0 \\
S66\_64 & -12.4 & -12.4 & -13.4 & -10.1 & -11.0 \\
S66\_65 & -17.1 & -16.7 & -18.6 & -15.1 & -15.9 \\
S66\_66 & -16.5 & -14.6 & -16.0 & -13.5 & -13.4 \\
MAD & - & 1.8 & 1.5 & 3.1 & 3.4 \\
\bottomrule
\end{longtable}
\begin{longtable}{lrrrrrrrrrrrrrr}
\caption{\label{tab:s66_2}Interaction energies of the S66 benchmark set calculated with various DFT functionals. All values are in kJ/mol.} \\
\toprule
Label & \rotatebox{90}{PBE} & \rotatebox{90}{PBEsol} & \rotatebox{90}{RPBE} & \rotatebox{90}{revPBE} & \rotatebox{90}{BLYP} & \rotatebox{90}{r$^2$SCAN} & \rotatebox{90}{MS2} & \rotatebox{90}{revTPSS} & \rotatebox{90}{PBE-D3} & \rotatebox{90}{PBE-DdsC} & \rotatebox{90}{RPBE-D3} & \rotatebox{90}{optPBE-vdW} & \rotatebox{90}{rev-vdW-DF2} & \rotatebox{90}{r$^2$SCAN+rVV10} \\ 
\endfirsthead

\caption[]{(continued)} \\
\endhead

\multicolumn{15}{r}{{Continued on next page}} \\
\endfoot

\bottomrule
\endlastfoot

\midrule
S66\_1 & -21.1 & -24.3 & -15.5 & -14.6 & -17.8 & -22.4 & -22.3 & -18.8 & -22.9 & -22.7 & -20.2 & -20.7 & -21.0 & -23.4 \\
S66\_2 & -22.2 & -26.1 & -15.6 & -14.6 & -18.4 & -24.8 & -24.0 & -19.5 & -25.4 & -25.7 & -22.8 & -24.0 & -23.8 & -26.3 \\
S66\_3 & -30.7 & -35.3 & -23.5 & -23.0 & -25.9 & -31.3 & -30.2 & -27.5 & -33.9 & -33.8 & -31.0 & -30.5 & -31.1 & -32.8 \\
S66\_4 & -31.3 & -36.7 & -22.2 & -21.2 & -25.9 & -35.9 & -34.6 & -28.1 & -35.8 & -36.5 & -33.1 & -34.2 & -34.0 & -38.2 \\
S66\_5 & -21.9 & -25.9 & -15.2 & -14.2 & -17.7 & -24.7 & -23.7 & -19.1 & -26.1 & -26.6 & -23.5 & -25.6 & -24.8 & -26.8 \\
S66\_6 & -30.8 & -36.2 & -22.7 & -21.6 & -24.4 & -32.9 & -30.9 & -27.3 & -36.7 & -37.3 & -33.9 & -34.5 & -34.0 & -35.4 \\
S66\_7 & -30.8 & -36.7 & -21.2 & -20.2 & -24.6 & -35.4 & -33.8 & -27.5 & -37.2 & -38.0 & -34.1 & -36.0 & -35.3 & -38.6 \\
S66\_8 & -20.7 & -24.0 & -15.1 & -14.2 & -17.2 & -22.4 & -22.0 & -18.4 & -23.2 & -23.3 & -20.5 & -21.9 & -21.7 & -23.7 \\
S66\_9 & -9.5 & -11.6 & -5.0 & -3.3 & -5.1 & -11.0 & -10.2 & -7.2 & -14.8 & -14.8 & -13.2 & -15.1 & -12.8 & -13.1 \\
S66\_10 & -13.1 & -16.9 & -6.1 & -4.5 & -7.0 & -15.3 & -13.6 & -10.5 & -19.0 & -20.0 & -17.8 & -18.7 & -17.1 & -17.9 \\
S66\_11 & -14.3 & -19.1 & -5.3 & -3.6 & -7.0 & -19.4 & -17.1 & -11.5 & -23.2 & -24.1 & -21.4 & -24.2 & -21.5 & -23.5 \\
S66\_12 & -30.6 & -35.8 & -22.5 & -21.8 & -25.1 & -32.8 & -30.9 & -27.2 & -34.4 & -34.9 & -32.0 & -31.8 & -32.2 & -34.6 \\
S66\_13 & -19.7 & -23.8 & -12.1 & -10.6 & -14.0 & -23.4 & -21.8 & -16.9 & -26.8 & -27.5 & -23.8 & -27.1 & -25.0 & -26.7 \\
S66\_14 & -26.2 & -31.3 & -17.7 & -16.4 & -18.8 & -29.2 & -26.3 & -23.3 & -34.4 & -35.5 & -32.0 & -33.5 & -31.7 & -32.8 \\
S66\_15 & -27.5 & -33.0 & -17.8 & -16.2 & -19.8 & -32.9 & -30.5 & -24.6 & -37.9 & -38.8 & -34.9 & -38.1 & -35.3 & -37.8 \\
S66\_16 & -18.8 & -21.6 & -13.4 & -12.1 & -14.8 & -20.5 & -19.9 & -16.6 & -22.6 & -22.6 & -19.8 & -22.1 & -20.9 & -22.4 \\
S66\_17 & -66.6 & -77.1 & -51.1 & -50.8 & -58.2 & -72.7 & -69.1 & -63.1 & -75.8 & -76.5 & -70.4 & -72.2 & -73.0 & -78.0 \\
S66\_18 & -29.4 & -33.7 & -22.4 & -21.7 & -25.2 & -30.1 & -29.1 & -26.2 & -33.2 & -33.1 & -31.0 & -30.7 & -30.8 & -31.9 \\
S66\_19 & -29.8 & -34.6 & -22.3 & -21.5 & -24.7 & -31.4 & -29.6 & -26.5 & -35.5 & -35.7 & -33.4 & -34.1 & -33.4 & -34.1 \\
S66\_20 & -79.5 & -93.1 & -61.3 & -62.0 & -69.7 & -88.5 & -85.4 & -75.3 & -86.0 & -87.8 & -79.3 & -79.9 & -83.4 & -92.6 \\
S66\_21 & -63.7 & -73.8 & -48.6 & -48.4 & -55.3 & -68.9 & -66.3 & -60.5 & -70.9 & -71.7 & -65.6 & -66.4 & -67.9 & -72.9 \\
S66\_22 & -78.2 & -90.3 & -61.1 & -61.4 & -69.2 & -85.8 & -82.8 & -74.3 & -85.9 & -87.1 & -79.8 & -80.8 & -83.2 & -90.3 \\
S66\_23 & -75.2 & -86.0 & -59.4 & -59.2 & -66.5 & -81.4 & -78.1 & -71.5 & -83.4 & -84.2 & -78.0 & -79.1 & -80.6 & -86.0 \\
S66\_24 & 4.8 & 1.3 & 12.5 & 14.7 & 14.4 & -3.9 & -1.3 & 6.8 & -11.1 & -11.6 & -10.7 & -16.5 & -10.4 & -11.5 \\
S66\_25 & 1.8 & -2.6 & 10.6 & 12.8 & 12.1 & -8.0 & -5.1 & 3.8 & -15.0 & -15.2 & -15.0 & -20.2 & -14.4 & -16.0 \\
S66\_26 & -10.7 & -19.7 & 6.1 & 8.8 & 4.3 & -30.6 & -22.6 & -8.1 & -37.0 & -36.4 & -37.1 & -45.6 & -38.3 & -43.7 \\
S66\_27 & 3.2 & -0.9 & 11.5 & 13.8 & 13.2 & -6.2 & -3.5 & 5.1 & -13.3 & -13.8 & -13.1 & -18.6 & -12.7 & -14.1 \\
S66\_28 & 1.2 & -5.5 & 14.0 & 16.4 & 14.3 & -13.6 & -8.9 & 3.2 & -20.4 & -20.8 & -20.2 & -27.3 & -20.7 & -24.1 \\
S66\_29 & -3.4 & -10.1 & 9.6 & 12.1 & 9.3 & -18.3 & -13.5 & -1.3 & -24.7 & -24.9 & -24.6 & -31.9 & -25.4 & -28.8 \\
S66\_30 & 2.9 & 0.7 & 8.2 & 10.0 & 9.7 & -2.1 & -0.9 & 4.6 & -6.5 & -6.7 & -5.5 & -9.2 & -5.2 & -6.3 \\
S66\_31 & -1.7 & -5.3 & 5.9 & 7.9 & 6.5 & -9.2 & -7.0 & 0.3 & -13.5 & -13.6 & -12.9 & -16.7 & -12.3 & -14.5 \\
S66\_32 & -4.6 & -8.0 & 2.6 & 4.5 & 2.9 & -11.7 & -9.9 & -2.8 & -14.2 & -14.6 & -14.4 & -17.9 & -14.1 & -16.4 \\
S66\_33 & 1.6 & -1.0 & 7.5 & 9.3 & 8.6 & -3.9 & -2.5 & 3.4 & -8.1 & -8.3 & -7.4 & -10.8 & -6.8 & -8.2 \\
S66\_34 & 1.6 & -3.5 & 13.2 & 16.9 & 13.9 & -6.0 & -2.8 & 5.3 & -18.0 & -24.4 & -16.2 & -21.9 & -14.5 & -14.4 \\
S66\_35 & 0.4 & -2.8 & 8.2 & 11.0 & 9.4 & -3.6 & -2.0 & 3.2 & -13.5 & -17.8 & -11.6 & -16.0 & -10.0 & -9.7 \\
S66\_36 & 0.1 & -1.7 & 5.1 & 7.5 & 6.9 & -1.6 & -1.3 & 2.3 & -10.3 & -13.2 & -8.4 & -12.2 & -6.9 & -6.3 \\
S66\_37 & 0.7 & -2.5 & 8.1 & 10.7 & 9.1 & -2.9 & -1.6 & 3.3 & -12.6 & -16.5 & -10.8 & -14.6 & -9.2 & -8.8 \\
S66\_38 & 0.3 & -3.3 & 9.0 & 12.0 & 9.6 & -3.9 & -2.2 & 3.5 & -14.7 & -18.8 & -13.2 & -17.0 & -11.5 & -10.5 \\
S66\_39 & 1.0 & -3.4 & 10.1 & 12.5 & 11.2 & -6.6 & -4.1 & 3.1 & -15.3 & -18.3 & -14.6 & -18.7 & -13.2 & -14.1 \\
S66\_40 & 0.2 & -3.0 & 7.2 & 9.4 & 8.8 & -5.3 & -3.7 & 2.1 & -13.2 & -15.7 & -11.8 & -15.9 & -10.6 & -11.5 \\
S66\_41 & 2.0 & -4.3 & 15.0 & 18.2 & 15.1 & -9.9 & -5.2 & 4.8 & -19.9 & -23.0 & -18.4 & -24.6 & -17.4 & -19.8 \\
S66\_42 & 2.2 & -2.9 & 13.1 & 16.1 & 13.3 & -7.6 & -3.8 & 4.9 & -16.9 & -19.3 & -15.7 & -21.4 & -14.9 & -16.6 \\
S66\_43 & 0.2 & -3.8 & 8.9 & 11.6 & 9.7 & -8.0 & -4.8 & 2.7 & -15.8 & -18.0 & -14.6 & -19.9 & -13.6 & -15.5 \\
S66\_44 & -0.1 & -2.6 & 6.1 & 8.5 & 7.2 & -3.5 & -2.3 & 2.4 & -9.8 & -12.2 & -9.6 & -11.4 & -7.2 & -7.5 \\
S66\_45 & -0.5 & -2.3 & 4.1 & 6.3 & 5.9 & -3.7 & -2.7 & 1.7 & -8.6 & -10.0 & -8.2 & -11.0 & -6.9 & -7.1 \\
S66\_46 & 0.1 & -5.4 & 12.0 & 15.3 & 12.0 & -9.3 & -5.5 & 3.4 & -18.3 & -22.6 & -16.8 & -22.3 & -15.7 & -17.4 \\
S66\_47 & -0.9 & -4.0 & 5.2 & 7.0 & 6.5 & -5.7 & -4.4 & 0.6 & -11.7 & -12.7 & -12.1 & -13.6 & -9.6 & -10.7 \\
S66\_48 & -3.3 & -6.8 & 3.8 & 5.6 & 4.2 & -8.3 & -6.8 & -1.4 & -14.0 & -14.8 & -14.4 & -15.7 & -12.0 & -13.3 \\
S66\_49 & -2.6 & -6.0 & 3.8 & 5.6 & 4.9 & -7.6 & -6.3 & -1.1 & -13.5 & -14.5 & -14.1 & -15.1 & -11.4 & -12.6 \\
S66\_50 & -5.2 & -7.9 & -0.3 & 1.1 & 0.8 & -8.7 & -8.0 & -4.2 & -11.9 & -12.7 & -13.2 & -12.7 & -10.2 & -11.6 \\
S66\_51 & -5.2 & -6.0 & -3.0 & -1.9 & -2.1 & -5.3 & -5.9 & -4.1 & -7.7 & -7.4 & -8.0 & -7.6 & -6.1 & -6.4 \\
S66\_52 & -9.9 & -14.5 & -2.2 & -0.8 & -1.5 & -17.1 & -15.2 & -8.7 & -19.8 & -19.9 & -20.2 & -20.3 & -17.4 & -21.7 \\
S66\_53 & -9.8 & -13.5 & -2.8 & -1.2 & -2.6 & -14.7 & -13.1 & -7.8 & -18.8 & -18.9 & -18.1 & -19.0 & -16.1 & -18.7 \\
S66\_54 & -8.4 & -11.4 & -2.9 & -1.6 & -2.3 & -13.2 & -11.9 & -6.8 & -15.1 & -14.8 & -14.9 & -14.2 & -12.3 & -15.8 \\
S66\_55 & -8.0 & -12.1 & -0.5 & 1.3 & 0.3 & -14.7 & -12.7 & -6.2 & -18.4 & -19.1 & -18.8 & -19.1 & -16.0 & -19.1 \\
S66\_56 & -3.4 & -6.9 & 3.4 & 5.3 & 4.5 & -9.4 & -7.4 & -1.8 & -14.3 & -15.1 & -14.2 & -15.4 & -11.8 & -13.9 \\
S66\_57 & -7.2 & -12.2 & 2.0 & 4.1 & 3.3 & -15.5 & -12.7 & -5.5 & -22.2 & -23.5 & -22.4 & -23.9 & -19.3 & -22.1 \\
S66\_58 & -10.5 & -13.8 & -4.0 & -2.6 & -5.0 & -11.5 & -11.5 & -8.4 & -16.5 & -16.7 & -15.9 & -16.2 & -14.4 & -14.6 \\
S66\_59 & -11.3 & -12.7 & -8.1 & -7.0 & -8.6 & -11.4 & -12.0 & -9.9 & -13.1 & -12.7 & -11.7 & -12.3 & -11.5 & -12.3 \\
S66\_60 & -17.6 & -21.9 & -10.3 & -9.3 & -11.2 & -21.2 & -21.0 & -16.0 & -22.5 & -22.3 & -22.2 & -20.9 & -19.9 & -23.6 \\
S66\_61 & 0.2 & -3.2 & 8.2 & 11.2 & 9.2 & -6.2 & -3.6 & 3.2 & -13.6 & -15.7 & -12.2 & -16.7 & -10.9 & -12.1 \\
S66\_62 & -1.2 & -5.5 & 8.4 & 11.3 & 8.7 & -8.1 & -5.4 & 1.9 & -16.1 & -18.7 & -14.2 & -18.7 & -13.1 & -14.6 \\
S66\_63 & -2.3 & -6.4 & 5.8 & 7.9 & 6.8 & -10.2 & -7.8 & -0.4 & -15.1 & -16.1 & -15.0 & -18.4 & -13.7 & -16.1 \\
S66\_64 & -4.1 & -6.9 & 2.3 & 4.5 & 2.8 & -8.5 & -7.1 & -1.5 & -13.3 & -14.0 & -12.5 & -14.8 & -11.1 & -12.3 \\
S66\_65 & -15.5 & -17.8 & -11.3 & -10.3 & -11.9 & -15.1 & -15.3 & -14.1 & -19.0 & -18.7 & -18.2 & -18.0 & -16.9 & -16.8 \\
S66\_66 & -7.4 & -11.3 & 0.3 & 2.2 & -0.2 & -12.5 & -10.2 & -5.0 & -16.9 & -17.8 & -16.5 & -18.6 & -15.5 & -16.8 \\
MAD & 9.0 & 6.7 & 17.3 & 19.0 & 16.6 & 4.3 & 5.8 & 11.4 & 1.7 & 2.6 & 1.2 & 2.5 & 1.3 & 1.6 \\
\bottomrule
\end{longtable}

\clearpage
\subsection{SE20 surface energies}

The SE20 dataset is a subset of (metal) surface energies taken from Ref.~\citenum{lundgaardMBEEFvdWRobustFitting2016}, shown in Table~\ref{tab:se20_1}.
We calculate the surface energy ($\gamma$) as follows:
\begin{equation}
\gamma = \frac{E_{\text{slab}} - N E_{\text{bulk}}}{2A},
\end{equation}
where $E_{\text{slab}}$ is the total energy of the slab and $E_{\text{bulk}}$ is the energy per atom, while $A$ is the surface area of one surface.
We use a 4L slab, with the bottom 2L fixed, which we expect to be converged~\cite{sheldonAdsorptionCH4Pt1112021} to \SI{0.05}{\joule\per\meter\squared}.
We highlight the final values (in \SI{}{\joule\per\meter\squared}) for dhBEEF-vdW@BEEF-vdW, hBEEF-vdW@BEEF-vdW, RPA@PBE and BEEF-vdW in Table~\ref{tab:se20_1} and several DFAs in Table~\ref{tab:se20_2}.

\begin{table}[h]

\caption{\label{tab:se20_1}Surface energies of the SE20 benchmark set calculated with hBEEF-vdW, dhBEEF-vdW, RPA@PBE, BEEF-vdW. These are compared to experimental estimates Lundgaard~\etal{}~\cite{lundgaardMBEEFvdWRobustFitting2016}.}
\begin{tabular}{lrrrrr}
\toprule
Label & \rotatebox{90}{Experiment} & \rotatebox{90}{hBEEF-vdW} & \rotatebox{90}{dhBEEF-vdW} & \rotatebox{90}{RPA@PBE} & \rotatebox{90}{BEEF-vdW} \\ 
\midrule
Li(110) & 0.52 & 0.55 & 0.56 & 0.49 & 0.56 \\
Na(110) & 0.26 & 0.25 & 0.26 & 0.22 & 0.24 \\
K(110) & 0.14 & 0.12 & 0.13 & 0.14 & 0.12 \\
Rb(110) & 0.11 & 0.10 & 0.11 & 0.11 & 0.09 \\
Ba(110) & 0.38 & 0.36 & 0.39 & 0.41 & 0.36 \\
Ca(111) & 0.50 & 0.52 & 0.53 & 0.52 & 0.52 \\
Sr(111) & 0.41 & 0.39 & 0.41 & 0.41 & 0.38 \\
Nb(110) & 2.68 & 2.35 & 2.40 & 2.48 & 2.17 \\
Ta(110) & 3.03 & 2.76 & 2.80 & 2.79 & 2.56 \\
Mo(110) & 2.95 & 3.33 & 3.27 & 3.50 & 3.01 \\
W(110) & 3.47 & 3.87 & 3.80 & 3.95 & 3.47 \\
Al(111) & 1.15 & 1.08 & 1.10 & 1.06 & 1.03 \\
Ni(111) & 2.42 & 2.01 & 2.07 & 2.30 & 1.92 \\
Cu(111) & 1.81 & 1.26 & 1.40 & 1.57 & 1.23 \\
Rh(111) & 2.68 & 2.33 & 2.33 & 2.59 & 2.03 \\
Pd(111) & 2.03 & 1.26 & 1.41 & 1.76 & 1.13 \\
Ag(111) & 1.25 & 0.68 & 0.81 & 1.09 & 0.59 \\
Ir(111) & 3.02 & 2.64 & 2.63 & 3.06 & 2.24 \\
Pt(111) & 2.48 & 1.57 & 1.68 & 2.01 & 1.31 \\
Au(111) & 1.50 & 0.63 & 0.77 & 1.15 & 0.41 \\
MAD & 0.00 & 0.32 & 0.27 & 0.17 & 0.38 \\
\bottomrule
\end{tabular}

\end{table}
\begin{table}[h]

\caption{\label{tab:se20_2}Surface energies of the SE20 benchmark set calculated with various DFT functionals. All values are in J/m$^2$.}
\begin{tabular}{lrrrrrrrrrrrrrr}
\toprule
Label & \rotatebox{90}{PBE} & \rotatebox{90}{PBEsol} & \rotatebox{90}{RPBE} & \rotatebox{90}{revPBE} & \rotatebox{90}{BLYP} & \rotatebox{90}{r$^2$SCAN} & \rotatebox{90}{MS2} & \rotatebox{90}{revTPSS} & \rotatebox{90}{PBE-D3} & \rotatebox{90}{PBE-DdsC} & \rotatebox{90}{RPBE-D3} & \rotatebox{90}{optPBE-vdW} & \rotatebox{90}{rev-vdW-DF2} & \rotatebox{90}{r$^2$SCAN+rVV10} \\ 
\midrule
Li(110) & 0.49 & 0.52 & 0.46 & 0.46 & 0.34 & 0.51 & 0.47 & 0.52 & 0.60 & 0.52 & 0.88 & 0.52 & 0.55 & 0.55 \\
Na(110) & 0.21 & 0.23 & 0.19 & 0.19 & 0.11 & 0.24 & 0.24 & 0.24 & 0.30 & 0.25 & 0.44 & 0.24 & 0.25 & 0.27 \\
K(110) & 0.11 & 0.12 & 0.08 & 0.08 & 0.04 & 0.12 & 0.12 & 0.12 & 0.15 & 0.13 & 0.20 & 0.13 & 0.13 & 0.14 \\
Rb(110) & 0.08 & 0.10 & 0.06 & 0.06 & 0.03 & 0.09 & 0.09 & 0.09 & 0.12 & 0.11 & 0.17 & 0.11 & 0.11 & 0.11 \\
Ba(110) & 0.31 & 0.36 & 0.26 & 0.27 & 0.20 & 0.34 & 0.35 & 0.36 & 0.37 & 0.49 & 0.64 & 0.38 & 0.40 & 0.40 \\
Ca(111) & 0.47 & 0.51 & 0.43 & 0.43 & 0.31 & 0.50 & 0.51 & 0.53 & 0.55 & 0.51 & 0.79 & 0.52 & 0.54 & 0.56 \\
Sr(111) & 0.34 & 0.39 & 0.30 & 0.31 & 0.21 & 0.38 & 0.41 & 0.40 & 0.41 & 0.39 & 0.64 & 0.39 & 0.41 & 0.43 \\
Nb(110) & 2.08 & 2.40 & 1.88 & 1.93 & 1.44 & 2.21 & 2.42 & 2.52 & 2.82 & 2.34 & 5.21 & 2.17 & 2.40 & 2.49 \\
Ta(110) & 2.37 & 2.67 & 2.21 & 2.25 & 1.67 & 2.62 & 2.76 & 2.83 & 3.16 & 2.68 & 5.58 & 2.45 & 2.69 & 2.89 \\
Mo(110) & 2.89 & 3.28 & 2.68 & 2.73 & 2.11 & 3.11 & 3.34 & 3.41 & 3.68 & 3.24 & 5.94 & 2.98 & 3.27 & 3.44 \\
W(110) & 3.35 & 3.73 & 3.15 & 3.20 & 2.45 & 3.67 & 3.86 & 3.93 & 4.26 & 3.72 & 6.92 & 3.40 & 3.72 & 4.00 \\
Al(111) & 0.83 & 0.98 & 0.73 & 0.75 & 0.29 & 0.98 & 1.00 & 1.06 & 1.08 & 0.94 & 2.18 & 0.85 & 0.99 & 1.13 \\
Ni(111) & 1.94 & 2.38 & 1.66 & 1.73 & 1.27 & 2.14 & 2.06 & 2.59 & 2.68 & 2.47 & 4.02 & 2.03 & 2.31 & 2.46 \\
Cu(111) & 1.29 & 1.70 & 1.01 & 1.08 & 0.65 & 1.61 & 1.79 & 1.87 & 2.22 & 1.75 & 4.15 & 1.35 & 1.60 & 1.90 \\
Rh(111) & 2.03 & 2.54 & 1.74 & 1.82 & 1.24 & 2.29 & 2.60 & 2.68 & 2.94 & 2.52 & 4.84 & 2.16 & 2.51 & 2.67 \\
Pd(111) & 1.25 & 1.78 & 0.92 & 1.00 & 0.52 & 1.52 & 1.75 & 1.89 & 2.18 & 1.62 & 4.27 & 1.38 & 1.72 & 1.88 \\
Ag(111) & 0.66 & 1.11 & 0.37 & 0.44 & 0.09 & 0.91 & 1.11 & 1.20 & 1.36 & 1.05 & 3.08 & 0.78 & 1.04 & 1.20 \\
Ir(111) & 2.10 & 2.63 & 1.83 & 1.89 & 1.12 & 2.40 & 2.81 & 2.82 & 3.20 & 2.52 & 5.49 & 2.20 & 2.59 & 2.81 \\
Pt(111) & 1.29 & 1.88 & 0.98 & 1.06 & 0.34 & 1.58 & 2.01 & 2.00 & 2.46 & 1.71 & 4.92 & 1.38 & 1.80 & 1.97 \\
Au(111) & 0.53 & 1.11 & 0.18 & 0.27 & -0.34 & 0.79 & 1.20 & 1.20 & 1.45 & 0.91 & 3.66 & 0.61 & 1.01 & 1.15 \\
MAD & 0.41 & 0.18 & 0.58 & 0.54 & 0.93 & 0.28 & 0.17 & 0.14 & 0.18 & 0.22 & 1.56 & 0.34 & 0.20 & 0.15 \\
\bottomrule
\end{tabular}

\end{table}

\clearpage

\bibliography{references}